\documentclass[apj,numberedappendix]{emulateapj}
\usepackage{lscape}
\usepackage{apjfonts}
\usepackage{epsf}

\slugcomment{to appear in The Astrophysical Journal Supplement Series,
2006 Sept., vol.~166 no.~1}

\shorttitle{HST Proper Motions in the Core of 47 Tuc}
\shortauthors{D.~E.~McLaughlin et al.}

\begin{document}

\title{HST Proper Motions and Stellar Dynamics
       in the Core of the Globular Cluster 47 Tucanae\altaffilmark{1}}

\author{Dean E. McLaughlin,\altaffilmark{2}
Jay Anderson,\altaffilmark{3}
Georges Meylan,\altaffilmark{4}
Karl Gebhardt,\altaffilmark{5}
Carlton Pryor,\altaffilmark{6}
Dante Minniti,\altaffilmark{7}
and Sterl Phinney\altaffilmark{8}}

\altaffiltext{1}{Based on observations made with the NASA/ESA {\it Hubble
Space Telescope}, obtained at the Space Telescope Science Institute,
which is operated by the Association of Universities for Research in
Astronomy, INC., under NASA contract NAS 5-26555.}
\altaffiltext{2}{University of Leicester, Department of Physics and Astronomy,
  University Road, Leicester, UK LE1 7RH;
  {\tt dean.mclaughlin@astro.le.ac.uk}}
\altaffiltext{3}{Rice University, Dept. of Physics and Astronomy, MS 61,
6100 South Main, Houston, TX 77005; {\tt jay@eeyore.rice.edu}}
\altaffiltext{4}{Laboratoire d'Astrophysique, Ecole Polytechnique F\'ed\'erale
de Lausanne (EPFL), Observatoire, CH-1290 Sauverny, Switzerland}
\altaffiltext{5}{University of Texas at Austin, Dept. of Astronomy, C1400,
Austin, TX 78712}
\altaffiltext{6}{Rutgers, The State University of New Jersey, Dept. of
Physics and Astronomy, 136 Frelinghuysen Road, Piscataway, NJ 08854}
\altaffiltext{7}{Department of Astronomy, Pontificia Universidad Cat\'olica,
  Casilla 306, Santiago 22, Chile}
\altaffiltext{8}{Caltech, Theoretical Astrophysics, MS 130-33, Pasadena, CA
91125}

\begin{abstract}
We have used HST imaging of the central regions of the globular cluster 47
Tucanae (= NGC 104), taken with the WFPC2 and ACS cameras between 1995 and
2002, to derive
proper motions and $U$- and $V$-band magnitudes for 14,366 stars within
100\arcsec\ (about 5 core radii) of the cluster center. This represents the
largest set of member velocities collected for any globular cluster. The stars
involved range in brightness from just fainter than the horizontal branch of
the cluster, to more than 2.5 mag below the main-sequence turn-off.
In the course of obtaining these
kinematical data, we also use a recent set of ACS images to define a
list of astrometrically calibrated positions (and F475W magnitudes) for nearly
130,000 stars in a larger, $\simeq3\arcmin\times3\arcmin$ central area.
We describe our data-reduction procedures in some detail and provide
the full position, photometry, and velocity data in the form of
downloadable electronic tables. We have used the star counts to obtain a new
estimate for the position of the cluster center and to define the density
profile of main-sequence turn-off and giant-branch stars into essentially
zero radius,
thus constraining the global spatial structure of the cluster better than
before. A single-mass, isotropic King-model fit to it is then used as a rough
point of reference against which to compare the gross characteristics of our
proper-motion data. We search in particular for any evidence of very
fast-moving stars, in significantly greater numbers than expected
for the extreme tails of the velocity distribution in a sample of our
size. We find that likely fewer than 0.1\%, and no more than about 0.3\%, of
stars with measured proper motions have total speeds above the nominal central
escape velocity of the cluster. At lower speeds, the proper-motion velocity
distribution very closely matches that of a regular King model (which is
itself nearly Gaussian given the high stellar density) at all observed radii.
Considerations of only the velocity dispersion then lead to
a number of results.
(1) Blue stragglers in the core of 47 Tuc have a velocity dispersion
$\sigma_\mu$ smaller
than that of the cluster giants by a factor of $\sqrt{2}$, consistent with the
former being on average twice as massive as normal, main-sequence turn-off
stars.
(2) The velocity distribution in the inner five core radii of the cluster is
essentially isotropic, and the detailed dependence of $\sigma_\mu$ on
$R$ for the brighter stars suggests that heavy remnants contribute only a
fraction of a percent to the total cluster mass.
Both of these results are in keeping with earlier, more realistic multimass
and anisotropic models of 47 Tuc.
(3) Using a sample of 419 line-of-sight velocities measured for bright giants
within $R\le 105\arcsec$, we obtain a kinematic distance to the cluster:
$D=4.0\pm0.35$ kpc, formally some 10\%--20\% lower than
recent estimates based on
standard CMD fitting, and more consistent with the value implied by fitting
to the white-dwarf cooling sequence.
And (4) by fitting simple models of isotropic, single-mass
stellar clusters with central point masses to our observed $\sigma_\mu(R)$
profile, we infer a 
1-$\sigma$ upper limit of $M_\bullet\la1000$--$1500\,M_\odot$ for any
intermediate-mass black hole in 47 Tuc. The formal best-fit hole mass
ranges from 0 if only the kinematics of stars near the main-sequence
turn-off mass are modeled, to $\sim$700--$800\,M_\odot$ if fainter, less
massive stars are also used.
We can neither confirm nor refute the hypothesis that
47 Tuc might lie on an extension of the $M_\bullet-\sigma$
relation observed for galaxy bulges.

\noindent
{\bf Note:} all online material is also available at
{\tt {\bf http://www.astro.le.ac.uk/$\sim$dm131/47tuc.html}}

\end{abstract}

\keywords{globular clusters: individual
(\objectname{NGC 104})---astrometry---stellar dynamics}

\vspace*{0.2truein}

\section{Introduction}
\label{sec:intro}

Galactic globular clusters, which are ancient building blocks of the halo,
represent an interesting family of ``hot'' stellar systems in which some
fundamental dynamical processes have taken place on time scales comparable
to the age of the universe. Intermediate in mass between galaxies and open
clusters, globulars are  unique  laboratories for learning about two-body
relaxation, mass segregation and equipartition of energy, stellar collisions
and mergers, and core collapse.

The whole concept of core collapse, linked to the gravothermal
instability which may develop due to the negative specific heat of
self-gravitating systems, was first investigated theoretically in the 
1960s and observed indirectly in the 1980s
(see \citealt{meyheg97} for a review). The stellar density in the core
may increase by up to six orders of magnitudes during the phases
of deep collapse. This significantly increases the frequency of interactions
and collisions between stars.

Binary stars play an essential role during
these late phases of the dynamical evolution of a globular cluster, as they
transfer energy to passing stars and can thus strongly influence the cluster
evolution---enough to delay, halt, and even reverse core collapse.
At the same time, stellar collisions  are effective in destroying binaries
(the outcome of most binary-binary interactions being the destruction of one
participant), in hardening those that remain, and in ejecting stars towards
the outer parts of the cluster. Observational evidence of possible products
of stellar encounters include blue stragglers, X-ray sources, pulsars, and
high-velocity stars.

In their pioneering radial-velocity study of the globular M3 $\equiv$ NGC~5272,
\citet{gg79} noted the puzzling presence of two stars that they called
``interlopers''.  These are two high-velocity stars located in the core 
of the cluster, both about 20\arcsec\ from the centre. They have radial
velocities relative to the cluster mean of +17.0 km~s$^{-1}$ and
$-22.9$ km~s$^{-1}$, corresponding to 3.5 and 4.5 times the velocity
dispersion in the core. These radial velocities are still close enough to
the mean radial velocity of the cluster
($\langle v_r\rangle \simeq -147\ {\rm km s}^{-1}$, high enough to make
contamination by field stars very unlikely) to carry a strong implication
of membership.

Similarly, \citet{mdm91} discovered two high-velocity stars
in the core of the globular cluster \object{47 Tucanae}. Located respectively
at about 3\arcsec\ and 38\arcsec\ from the centre, they have line-of-sight
velocities relative to the cluster of $-36.7$ km~s$^{-1}$\ and
$+32.4$ km~s$^{-1}$, corresponding to $\ga 3$ times the core velocity
dispersion but appearing in a total sample of only 50 radial velocities.
Repeated observations over 1.5 years indicated that neither of these
two stars is a binary or a pulsating star, and high-resolution echelle spectra
confirmed their luminosity classes and, consequently, their membership in
47 Tucanae.

Prompted in 1995 by the presence of these four unusually fast-moving stars in
two different globulars, and their potential link to the extreme
dynamical processes in high-density environments, we decided to investigate
further the core of 47 Tucanae, the closest of the two clusters.
The capability limit of the radial velocity observations
which could be obtained from the ground in the crowded core of 47 Tuc having
already been reached, we concluded that if more progress was
to be made in the search for high-velocity stars, it had to be made by
obtaining proper motions---a task for which only HST is suitable.

We thus used WFPC2 to obtain  images of the core of 47 Tucanae, at three
different epochs over four years between 1995--1999, in order to perform
precise astrometry and obtain a complete census of high-velocity stars.
Choosing the F300W ($\approx U$-band) filter allowed stars to be measured
over the whole color-magnitude diagram, from the red-giant branch to well
down the main sequence, ultimately yielding a velocity database of
unprecedented size for a globular cluster. Meanwhile, subsequent observations
of the center of 47 Tuc, by unrelated WFPC2 and ACS imaging programs between
1999--2002, have provided extremely useful supplements to our original
dataset.

In this paper, we present our analysis of these HST data. Not surprisingly,
we have found it possible to address a number of issues beyond simply
characterizing the stellar velocity distribution \citep[e.g., see][]{min97}.
But the latter does remain our primary focus here, and our look
at other questions (estimating the distance to 47 Tuc; defining the internal
velocity-dispersion profile as a function of stellar luminosity/mass;
assessing the possibility of a compact central mass concentration) is not as
comprehensive. However, we are also providing full details of the data
themselves, including extensive tables of star-by-star astrometry, photometry,
and proper-motion solutions. This thorough census of the
stellar distribution and kinematics in 47 Tuc, used together
with sophisticated modeling techniques, will ultimately allow for unique and
precise constraints to be placed on relaxation processes, stellar collision
and ejection rates, and many other aspects of the dynamical structure and
evolution of globular cluster cores.

It is perhaps worth noting that previous studies of internal
globular-cluster dynamics
using HST-based proper motions \citep{druketal03,mcn03} have employed
samples of $\sim$1000 member stars and tended to focus on deriving the
stellar velocity dispersion very near the cluster centers.
The largest sample of ground-based proper motions comes from
the analysis by \citet{vanl00} of 9847 stars in
NGC 5139 $\equiv$ $\omega$ Centauri, using observations over a
$\sim$50-year baseline. \citet{vandeven06} have used a high-quality subset 
of 2295 of these stars to explore the internal dynamics of this large cluster
and estimate a distance to it.
Our full velocity sample for 47 Tuc includes 14,366 stars, and the majority
of these prove useful for a variety of precise kinematics analyses.

\subsection{Outline of the Paper}
\label{subsec:overview}

We begin in \S\ref{subsec:hstdataoutline} by giving the basic details of the
WFPC2 and ACS image sets that we have used to derive proper motions for stars
in 47 Tuc. Section \ref{subsec:master} then focuses on the construction
and astrometric calibration of a
comprehensive catalogue of positions and F475W magnitudes for nearly 130,000
stars in one central ACS field measuring about 3\arcmin\ on a side. This
``master'' star list is presented in Table \ref{tab:master}, and both it and
an associated image that we have made of the field are available from the
online edition of the {\it Astrophysical Journal}.
We then use this list to
re-evaluate the coordinates of the center of 47 Tuc. In
\S\ref{subsec:multiepoch} we describe our procedures for performing
{\it local} coordinate transformations of the data at other epochs into the
master reference frame, in order to obtain {\it relative} proper motions for
as many stars as possible in a rather smaller area ($R<100\arcsec$) of the
sky.

Section \ref{sec:velsamples} discusses our derivation of the velocities
themselves, focusing on statistical properties such as goodness-of-fit and
error distributions in order to identify a useful working sample for
kinematics analyses. A catalogue of $U$, $V$, and F475W photometry,
epoch-by-epoch displacements, and associated proper motions for 14,366 stars
is given in Table \ref{tab:posdata}.
This is also available
electronically, along with an SM code which extracts and plots the data for
any given star's position vs.\ time. In \S\ref{subsec:rvsample}, we also
describe a set of line-of-sight velocities that we ultimately use to estimate
a kinematic distance to 47 Tuc.

\begin{deluxetable*}{lll}
\tabletypesize{\scriptsize}
\tablecaption{Basic Data on 47 Tucanae = NGC~104 \label{tab:basic}}
\tablewidth{0pt}
\tablecolumns{3}
\tablehead{
\multicolumn{2}{l}{Property}     & \multicolumn{1}{l}{Reference}
}
\startdata
Cluster Center (J2000)  & $\alpha=0^{\rm h}24^{\rm m}05\fs67$,
                          $\delta=-72^{\circ}04^{\prime}52\farcs62$
     & this paper, \S\ref{subsubsec:center}             \\
Galactic Coordinates    & $\ell=305\fdg9$, $b=-44\fdg9$
     & \citet{har96}                                    \\
Apparent Magnitude      & $V_{\rm tot}=3.95$
     & \citet{har96}                                    \\
Integrated Colors       & $(B-V)=0.88$, $(U-V)=1.25$
     & \citet{har96}                                    \\
Main-Sequence Turn-off   & $V_{\rm TO}=17.65$
     & \citet{zoc01}; \citet{perc02}                    \\
Metallicity             & [Fe/H] = $-0.76$
     & \citet{har96}                                    \\
Central Surface Brightness & $\mu_{V,0}=14.26\pm0.26$ mag arcsec$^{-2}$
     & this paper, \S\ref{subsec:density}               \\
King (1966) Core Radius ($V\le V_{\rm TO}$) & $r_0=20\farcs84 \pm5\farcs05$
     & this paper, \S\ref{subsec:density}               \\
King (1966) concentration  & $c\equiv\log\,(r_t/r_0)=2.01\pm0.12$\ \
                             ($W_0=8.6\pm0.4$)
     & this paper, \S\ref{subsec:density}               \\
Foreground Reddening    & $E(B-V)=0.04$
     & \citet{har96}                                    \\
Field Contamination
($V\le 21$)             & $\Sigma_{\rm fore}=(0.8\pm0.2)$ stars arcmin$^{-2}$
     & \citet{rat85}                                    \\
\sidehead{{\it Heliocentric Distance:}}
Main-Sequence Fitting   & $D=4.85\pm0.18$ kpc
     & \citet{grat03}                                   \\
Main-Sequence Fitting   & $D=4.45\pm0.15$ kpc
     & \citet{perc02}                                   \\
White Dwarf             & $D=4.15\pm0.27$ kpc
     & \citet{zoc01}                                    \\
Kinematic               & $D=4.02\pm0.35$ kpc
     & this paper, \S\ref{subsec:distance}              \\
\sidehead{\it {Central Velocity Dispersion ($m_{*}\simeq0.85\ M_{\odot}$):}}
Line-of-sight           & $\sigma_z(R=0)=11.6\pm0.8$ km s$^{-1}$
     & this paper, \S\ref{subsec:bhmods}                 \\
Plane-of-Sky            & $\sigma_\mu(R=0)=0.609\pm0.010$ mas yr$^{-1}$
     & this paper, \S\ref{subsec:bhmods}                 \\
\enddata
\end{deluxetable*}

In \S\ref{subsec:density} we use our master star list to construct the
number-density profile at $R<100\arcsec$ for stars brighter than the
main-sequence turn-off. This provides a direct extension of a wider-field,
ground-based $V$-band surface-brightness profile already in the literature.
Combining these data, we fit a standard single-mass and isotropic
\citet{king66} model to the cluster, to give a rough framework for the
physical interpretation of some of our results. In particular, in
\S\ref{subsec:modvel} (supplemented by Appendix \ref{sec:kingveldist}) we
describe the calculation of projected, two-dimensional proper-motion
velocity distributions for generic \citeauthor{king66} models.

Sections \ref{sec:veldist} and \ref{sec:veldisp} then examine various
aspects of the stellar kinematics in the central 5 core radii of 47 Tuc.
Some preliminary results from earlier stages of this work have been
presented in conference proceedings by \citet{ka2001} and \citet{mcl03}, which
naturally are superseded here.

In \S\S\ref{subsec:onedvel} and
\ref{subsec:twodvel}, we construct the one- and two-dimensional distributions
of proper motion and compare them both to Gaussians and to
\citeauthor{king66} models with finite escape velocities. We look especially
for evidence of stars with total speeds on the plane of the sky exceeding
the nominal central escape velocity of 47 Tuc, but find only a few dozen
potential candidates. Section \ref{subsec:highpm} summarizes the overall
properties of these high-velocity stars.

In \S\ref{subsec:strag} we go on to
compare the velocity dispersion of blue stragglers in our field to that of
similarly bright stars on the cluster's giant branch. Section
\ref{subsec:kinmags} considers the run of velocity dispersion with
clustercentric radius, as a function of stellar magnitude, and obtains an
estimate of the average velocity anisotropy in the central regions. Section
\ref{subsec:distance} then compares the velocity dispersion
profile of the brighter stars in our proper-motion sample to that of
our much smaller radial-velocity sample, to derive a kinematic estimate of the
distance to 47 Tuc. In \S\ref{subsec:bhmods} we focus on the kinematics
at the smallest projected radii, to fit the proper-motion velocity
dispersions there with models based on those of \citet{king66} but allowing
for the possible presence of a dark central point mass.

We should emphasize from the start that, although 47 Tuc is known to be
rotating \citep{mey86,ak2003a}, we do not attempt to include this complication
in any of our kinematics analyses. The justification for this is essentially
that we are working here only on relatively small scales, in regions of the
cluster for which rotation is indeed dynamically dominated to a large extent
by random stellar motions. This point has been made previously by 
\citet{mey86}, and we illustrate it again, quantitatively, in
\S\ref{subsec:norotation} of this paper.

For reference throughout the paper, some basic data on 47 Tuc are provided in
Table \ref{tab:basic}. Some of the numbers there rely on new analyses of the
HST data that we have collected.
Note up front the small field contamination predicted by the Galaxy model of
\citet{rat85}, which for the area covered by our proper-motion sample
(very roughly, $\approx3.5$--4 arcmin$^{2}$) amounts to of order 3
($\pm$2) interloping field stars brighter than $V\le21$. For the majority of
our work, this is evidently a negligible effect.

\section{HST Astrometry and Photometry}
\label{sec:hstdata}

\subsection{The Available Data and General Approach}
\label{subsec:hstdataoutline}

This project began with a series of WFPC2 exposures of the center of 
47Tuc in 1995, 1997, and 1999 (GO-5912, GO-6467, and GO-7503, PI Meylan).
The goal of these observations was to search for the proper-motion analogues
of the high-velocity ``cannonball'' stars that \citet{mdm91} had found
with CORAVEL radial velocities from the ground (and which have
well-known counterparts in the Galactic globular cluster M3; \citealt{gg79}). 
These original exposures are confined within the inner $\sim$4--5 core radii
($R\la100\arcsec$) of the cluster.
They were taken with the F300W ($\sim U$-band) filter in order to suppress the
background from the red giants and thus allow better position measurements
of the more numerous stars at the main-sequence turnoff and fainter. As a
result, we found it possible to measure accurate motions not only for the
fast-moving stars in the cluster, but for many thousands of the average
members as well. Furthermore, subsequent (unrelated) HST observations of
the core of 47 Tuc have nearly doubled our original four-year time
baseline and allowed the derivation of more precise proper motions.

Additional WFPC2 images of the same central field as the Meylan pointings
were obtained in 1999 and 2001 (GO-8267 and GO-9266, PI Gilliland) 
through the similarly short-wavelength filter, F336W.  More recently, 
ACS images that cover the same area were obtained in 2002 for various 
calibration programs (PIs Meurer, King, and Bohlin), all through the 
slightly redder F475W filter. We have reduced all of these images from the
HST archive and included them in our analysis. The ACS data in particular have
proven extremely useful in providing much higher-precision position
measurements than are possible with the lower-resolution WF chips of the
WFPC2. In addition, the even coverage of these ACS-WFC data
(as contrasted with the lop-sided WFPC2 footprint) allows us to construct an
accurate, uniform, and nearly complete census of stars---independently of any
proper-motion goals---within about $\pm 1.5$ arcminutes of the cluster center.

\begin{deluxetable}{lcccr}
\tablecaption{WFPC2 and ACS Observations of 47 Tucanae \label{tab:datasets}}
\tablewidth{0pt}
\tablecolumns{5}
\tablehead{
\colhead{} & \colhead{Program} & \colhead{} & \colhead{} & \colhead{} \\
\colhead{Data set} & \colhead{ID}      & \colhead{$N_{\rm obs}$}
                  & \colhead{Filter}  & \colhead{Date}
}
\startdata
MEYLANe1 & 5912 & 15 & F300W & 25 Oct 1995 = 1995.82 \\
MEYLANe2 & 6467 & 16 & F300W & 03 Nov 1997 = 1997.84 \\
GILLILU1 & 8267 & 28 & F336W & 05 Jul 1999 = 1999.51 \\
MEYLANe3 & 7503 & 16 & F300W & 28 Oct 1999 = 1999.82 \\
GILLILU2 & 9266 & 11 & F336W & 13 Jul 2001 = 2001.53 \\
WFC-MEUR & 9028 & 20 & F475W & 05 Apr 2002 = 2002.26 \\
HRC-MEUR & 9028 & 40 & F475W & 05 Apr 2002 = 2002.26 \\
HRC-BOHL & 9019 & 10 & F475W & 13 Apr 2002 = 2002.28 \\
WFC-KING & 9443 &  6 & F475W & 07 Jul 2002 = 2002.52 \\
HRC-KING & 9443 & 20 & F475W & 24 Jul 2002 = 2002.56 \\
\enddata
\end{deluxetable}

All these data sets and their basic attributes are listed
in Table \ref{tab:datasets}. There are 182 independent exposures taken as
parts of ten distinct sets, which we refer to loosely as ten ``epochs''
spanning nearly seven years in total. Our general approach to collating these
for analysis is first to combine all the exposures for each epoch so that
we have a single position and flux for each star measured in a frame natural
to that epoch.  This intra-epoch 
averaging also gives an empirical estimate of the error in position 
and flux for each star at each epoch.  We then compare the positions of stars
measured at the different epochs to derive proper motions.

The proper motions we measure are, of course, simply changes in the
relative positions of stars over time.  The many observations are taken at
different times, at different pointings and orientations, through
different filters, and with different instruments.  Each observation
therefore has a different (and a priori unknown) mapping of
the chip coordinates to the sky.  Before we can compare relative positions
of stars measured in different images, we must transform all our positions
into a common reference frame.  
We have chosen to use the WFC images of the GO-9028 data set to construct this 
``master frame,'' since these images have a large and very even spatial
coverage. The large majority of stars found in any of our data sets will be
found in the GO-9028 data set.

In \S\ref{subsec:master}, then, we define this master frame and discuss
its astrometric and photometric calibration. We also use it to find a new
estimate for the cluster center, which will be useful for our later analyses.
After this, we briefly describe the process by which
we transform multi-epoch observations into the reference frame
(\S\ref{subsec:multiepoch}). In \S\ref{sec:velsamples} we detail our
derivation of the proper motions themselves and define a working sample for
investigation of the cluster kinematics in the rest of the paper.

\subsection{The Master Star List}
\label{subsec:master}

Special care is required in constructing the star list for the reference
frame, since even if a star is not optimally measured in the GO-9028 data set, 
we still want to allow for the possibility that it might be found well in
other epochs.  Our primary goal is therefore to make the master
list as complete as possible.

The GO-9028 data set has some small dithers and some large dithers about
a central pointing.  We do not want the master star list to be affected by 
the location of the inter-chip gap in the centeral pointing, so we made use
only of the central-pointing image and the pointings that had ditherings
larger than
the  inter-chip gap to generate an unbiased master list.  This amounts to 
13 images out of the 20 in GO-9028.  We first fit a PSF to every peak in each 
of these 13 images, and then corrected each peak's raw pixel position 
(from the {\tt \_flt} images) for distortion according to the prescription 
in Anderson (2006, in preparation). We next determined the transformation
from 
each image into the frame of the central pointing and identified a star
wherever a peak at the same master-frame location was found in 7 or more
of the 13 individual images.

This gave us a list of 129,733 coincident peaks, covering an area 
of about 202\arcsec$\times$202\arcsec, with positions in the 
distortion-corrected frame of the central-pointing image ({\tt j8cd01a9q}). 
Plotting the star list on a stacked image that we made of the field 
shows that no obvious stars are missing from the list.  There are, however,
a (relatively small) number of PSF artifacts that occurred in the same place
in all images and which were therefore misidentified as stars.  

\subsubsection{Purging of Non-stellar Artifacts}
\label{subsubsec:purge}

\begin{figure}[!t]
\vspace*{+0.1truecm}
\centerline{\resizebox{90mm}{!}{\includegraphics{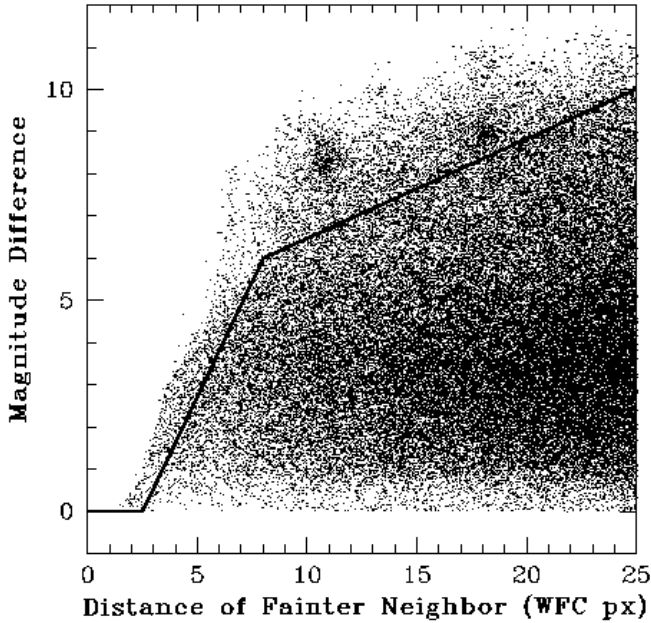}\hfil}}
\caption{Distribution of neighbors
found around saturated stars in the GO-9028 F475W exposures used to
construct the master frame.  Each point represents a
fainter neighbor of a saturated star and is plotted at the
appropriate distance and magnitude offset from the main star.
PSF artifacts show up as very faint objects relatively close to
the star. All points above this line are considered to be
potential artifacts. Some
real stars are necessarily rejected as well, but in a very regular
way so that completeness can be defined well.
\label{fig:PSFartifacts}}
\end{figure}

In order to isolate PSF artifacts in the master list, we went through the
list star by star.  For each star we found every fainter neighboring
``star'' within 25 pixels. In Figure~\ref{fig:PSFartifacts} we show the
distribution of neighbors for all saturated stars only (about 3000
altogether). On the horizontal axis we plot the distance in WFC pixels from
the bright star to the fainter neighbor, and on the vertical axis we show the
magnitude difference. PSF artifacts occupy a clearly recognizable 
region in this parameter space (e.g., the clump of points at a
$\sim$10-pixel distance and $\sim$8-magnitude flux difference).

We therefore define the discriminating line shown in
Figure \ref{fig:PSFartifacts} to
distinguish real stars from (possible) artifacts. For every star in turn in
the full master list, we have flagged all neighbors that fall above this
line. These correspond to objects that are too close to brighter stars to be
considered trustworthy, and they should not be used in any detailed
analyses. This procedure is bound to reject some real stars along with true
artifacts, but it does so in a quantifiable (and therefore correctable) way.
The end result is a more robust and better-defined star list, consisting
ultimately of 114,973 reliable stars. The full star list is presented in
\S\ref{subsubsec:masterlist} below, after we have described the astrometric
and photometric calibration of the data.

\subsubsection{Finding the Cluster Center}
\label{subsubsec:center}

The ACS/WFC images provide us the widest, the most uniform, and the deepest 
survey to date of the central regions of 47 Tuc.  Previous images taken by
WFPC2 have a very asymmetric and non-uniform coverage, due to the lop-sided
shape of the WFPC2 footprint and the gaps between the chips. By contrast, the
dithered set of WFC images have essentially uniform coverage out to a radius
of about 100\arcsec, which is nearly five core radii. These data therefore
permit us the best determination to date of the cluster center. Indeed,
given the high degree of completeness in the present star counts, it is
difficult to imagine any significant improvement in the near future.

Note that a determination of the center from star counts should not depend on 
incompleteness corrections, provided of course that incompleteness 
is a function only of stellar magnitude and clustercentric radius.
Nor should mass segregation 
enter the problem, so long as the stellar distribution is radially symmetric.
Thus, in determining the center we work directly with our master star 
list, making no attempt to correct for either of these effects.

\begin{figure*}
\centerline{\resizebox{135mm}{!}{\includegraphics{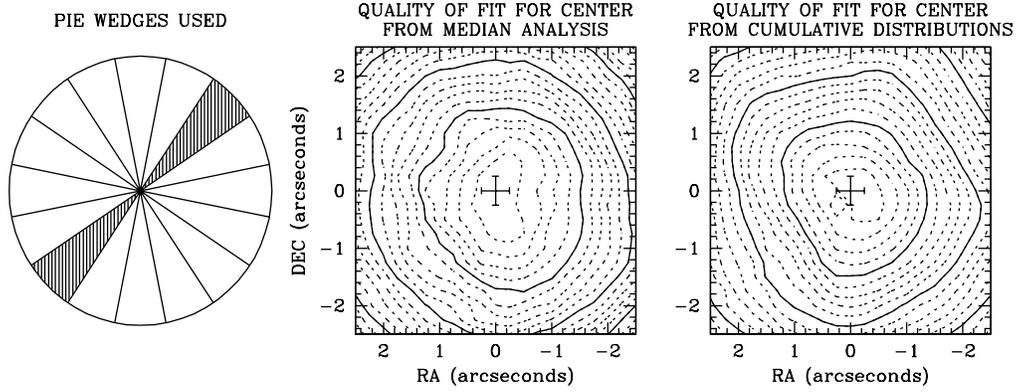}\hfil}}
\caption{
{\it Left}:  a shematic illustrating the wedges used to
home-in on a cluster center; see text.
{\it Middle}:  contours of equal quality-of-fit statistic for
determining the center using a ``median'' analysis, i.e., by
comparing the total numbers of stars in pairs of opposing wedges.  
{\it Right}:   contours of equal quality-of-fit statistic for
determining the center using the cumulative radial distributions of
stars in opposing wedges. In the {\it middle} and {\it right}
panels, dotted contours represent changes of 20\% in the quantity
plotted. Errorbars represent the estimated uncertainty in the center.
\label{fig:findcenA}}
\end{figure*}

We begin with our list of 114,973 stars after culling faint neighbors
(possible PSF artifacts). For each 
star we have a position $(x,y)$ in the reference-frame coordinate
system, rotated in order to align the $y$ axis roughly with North.
Next, an array of trial centers $(x_0,y_0)$ is defined, and for each trial
center in turn we find all the stars within a radius of 1500 WFC pixels
(75\arcsec) from that center.  We then divide these stars into sixteen,
$22.5^{\circ}$-wide pie wedges---shown schematically in the left
panel of Figure \ref{fig:findcenA}---and use two tests to compare the
distributions of stars in the eight distinct pairs of opposing wedges.

In the first test, we look at the difference in the total number of stars
between the members of the opposing-wedge pairs defined by each $(x_0,y_0)$
in our pre-defined array of trial centers, and we form a ``quality-of-fit''
statistic as the sum of the differences over the eight wedge pairs.
The coordinates which minimize this statistic then define a ``median''
estimate of the cluster center. The middle panel of Figure 
\ref{fig:findcenA} shows a contour plot of the sum of differences for a
3\arcsec$\times$3\arcsec\ subgrid of trial centers, with $(0,0)$
corresponding to the best center ultimately implied by this method (the
conversion to calibrated right ascension and declination is discussed below).

The second test is to generate the cumulative radial distribution for stars
within each of the sixteen wedges for any specified $(x_0,y_0)$. We then find
the absolute value of the integrated difference between the radial
distributions in any two opposing wedges, and define a quality-of-fit
statistic as the sum of these absolute differences over the eight wedge
pairs. The right panel of Figure \ref{fig:findcenA} shows a contour plot of
this statistic over a grid of trial centers, with $(0,0)$ again corresponding
to the best-estimate coordinates which minimize the statistic.

\begin{deluxetable*}{ccccrrccc}
\tabletypesize{\scriptsize}
\tablecaption{Stars Used for Astrometric Calibration, and the Resulting
              Cluster Center   \label{tab:center}}
\tablewidth{0pt}
\tablecolumns{9}
\tablehead{
\colhead{Point/Star}    & \colhead{Master Frame ID} & \colhead{F475W} &
\colhead{$x_{\rm raw}$} & \colhead{$y_{\rm raw}$}   &
\colhead{$\Delta$RA}    & \colhead{$\Delta$Dec}     &
\colhead{RA (J2000)}    & \colhead{Dec (J2000)}     \\ 
\colhead{}              & \colhead{}                & \colhead{}      &
\multicolumn{2}{c}{[pixels]}                        &
\multicolumn{2}{c}{[arcsec]}                        &
\colhead{[hh:mm:ss]}    & \colhead{[dd:mm:ss]}      \\
\colhead{(1)} & \colhead{(2)} & \colhead{(3)} & \colhead{(4)} & 
\colhead{(5)} & \colhead{(6)} & \colhead{(7)} & \colhead{(8)} & 
\colhead{(9)}
}
\startdata
 $\alpha$   & M052296 & 13.6 & 1991 &  415  
            & $-4.532$ & $-5.983$
            & 00:24:04.69$\pm$0.10 & $-$72:04:58.58$\pm$0.12 \\
 $\beta$    & M060833 & 14.3 & 1961 &  239  
            & $-4.924$ & $+2.935$
            & 00:24:04.61$\pm$0.10 & $-$72:04:49.69$\pm$0.11 \\
 $\gamma$   & M061604 & 13.8 & 2163 &  182  
            & $+5.180$ & $+3.720$
            & 00:24:06.79$\pm$0.10 & $-$72:04:48.90$\pm$0.11 \\
 $\delta$   & M056400 & 13.6 & 2208 &  281  
            & $+6.809$ & $-1.665$
            & 00:24:07.15$\pm$0.11 & $-$72:04:54.25$\pm$0.11 \\
 $\epsilon$ & M056630 & 12.9 & 2013 &  319  
            & $-2.903$ & $-1.413$
            & 00:24:05.04$\pm$0.11 & $-$72:04:54.08$\pm$0.11 \\
\hline
 $\zeta$    & M004529 & 14.2 & 2113 &  812  
            & $ -6.731$& $-74.669$
            & 00:24:04.21$\pm$0.11 & $-$72:06:07.29$\pm$0.10 \\
 $\eta$     & M034527 & 13.4 & 3380 &  545  
            & $+62.662$& $-26.247$
            & 00:24:19.25$\pm$0.10 & $-$72:05:18.77$\pm$0.12 \\
\hline
 Center     & \nodata & \nodata & 2066 &  277   
            & $ 0.000$ & $ 0.000$
            & 00:24:05.67$\pm$0.07 & $-$72:04:52.62$\pm$0.26 \\
\enddata

\tablecomments{Key to columns:   \hfill\break
{\bf Column (1)}---Label for the star in Fig.~\ref{fig:findcenB}.
See text for the distinction between stars $\alpha$, $\beta$, $\gamma$,
$\delta$, and $\epsilon$ vs.\ stars $\zeta$ and $\eta$.  \hfill\break
{\bf Column (2)}---Stellar ID on the sequential numbering
system for the full master-frame star list of Table \ref{tab:master}.
\hfill\break
{\bf Column (3)}---Calibrated F475W magnitude. \hfill\break
{\bf Column (4)}---``Raw'' $x$ position, in pixels, in the top chip (WFC1)
of the {\tt j8cd01a9q\_flt} frame from program GO-9028.  \hfill\break
{\bf Column (5)}---``Raw'' $y$ position, in pixels, in the top chip (WFC1)
of the {\tt j8cd01a9q\_flt} frame from program GO-9028.  \hfill\break
{\bf Column (6)}---RA offset in arcsec (positive Eastward) from the cluster
center in the master-frame system.  \hfill\break
{\bf Column (7)}---Dec offset in arcsec (positive Northward) from the cluster
center in the master-frame system.  \hfill\break
{\bf Column (8)}---Calibrated, absolute right ascension.  \hfill\break
{\bf Column (9)}---Calibrated, absolute declination.  \hfill\break
}

\end{deluxetable*}

\begin{figure*}
\centerline{\resizebox{135mm}{!}{\includegraphics{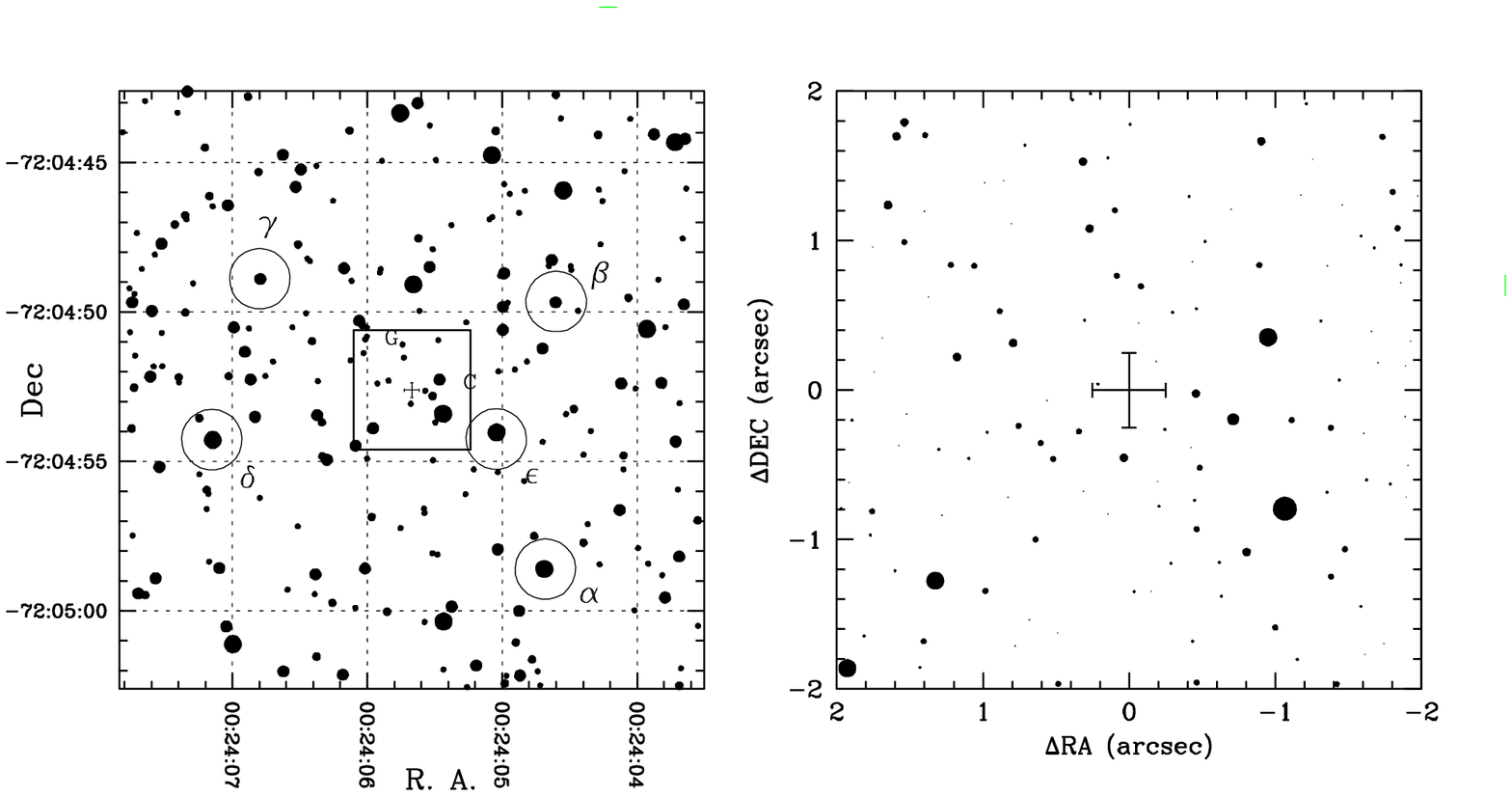}\hfil}}
\caption{
{\it Left}: stars in a 20\arcsec$\times$20\arcsec\ field  
about our cluster center.  Point size is correlated
with F475W magnitude.  The 5 stars used
in the absolute calibration are indicated (see Table
\ref{tab:center}), as are the cluster centers estimated by
\citet{guh92} and \citet{cal93} .
{\it Right}: a close up of the inner 4\arcsec$\times$4\arcsec.
The uncertainty in the center is about $\pm0\farcs25$ in each coordinate.
\label{fig:findcenB}}
\end{figure*}

These two approaches give quite consistent positions for the cluster center.
To get an idea of the accuracy of our center, we again take eight pairs
of opposing wedges. For each pair, we find the location along the wedge axis
that minimizes the difference between the cumulative radial distributions for 
the two wedges independently of any others.  This yields
eight estimates of the center along different axes using independent samples
of stars.  From the scatter among these independent estimates, we find that
our final center is good (in the master-frame coordinate system) to about
$\pm$5 WFC pixels (or $\simeq0\farcs25$) in both directions. This
uncertainty is indicated by the errorbars at $(0,0)$ in the middle and
right-hand panels of Figure \ref{fig:findcenA}.

\subsubsection{Astrometric Calibration}
\label{subsubsec:astrocal}

We now have a position for the cluster center in the reference frame, 
which is based on the distortion-corrected and rotated frame of the 
first image of GO-9028.  In order to transform the Master-frame positions 
into absolute RA and Dec, we used the image header information from several 
WFPC2 images ({\tt u2ty0201t}, {\tt u2vo0101t}, {\tt u4f40101r}, and 
{\tt u5jm120dr}) to obtain absolute positions for seven stars---five stars 
at the center and two stars in the outskirts.  These four images were 
taken at different pointings and orientations, so they should all use 
different guide stars and give independent estimates of the absolute 
coordinates.  

Table \ref{tab:center} gives details of these seven stars. First are their IDs
in our master star list and calibrated F475W magnitudes (both of which items
are described in general below). Following this are the stars' locations in the
master frame, both in terms of pixel positions and in terms of relative RA and
Dec offsets from the cluster center determined in \S\ref{subsubsec:center}.
Then we list the average absolute RA and Dec (J2000) obtained from the header
information in the four WFPC2 images. Combining the absolute positions of
the five central stars with their relative offsets in the reference frame
then sets the absolute astrometric zeropoint of our master-frame system.
The two outer stars $\zeta$ and $\eta$ are used to fix the orientation angle.
As is also stated in the bottom line of Table \ref{tab:center}, the absolute
position of the cluster center is
\begin{equation}
\begin{array}{rcl}
{\rm RA(J2000)}  & = & ~~00^{\rm h}\, 24^{\rm m}\, 05\fs67\,
               \pm 0\fs07   \\
{\rm Dec(J2000)} & = & -72^{\circ}\, 04^{\prime}\, 52\farcs62\,
               \pm 0\farcs26 \ ,   
\end{array}
\label{eq:center}
\end{equation}
where the uncertainties come from the averaging of the five stars and
essentially reflect the $\pm0\farcs25$ internal uncertainty in the uncalibrated
master-frame coordinates of the cluster center.
This absolute calibration should be good to about
0.1 arcsecond throughout our $\sim$3\arcmin$\times$3\arcmin\ master field,
although we note that it may ultimately suffer from a small ($\sim$1\arcsec)
inaccuracy if the positional errors of the HST guide stars for the WFPC2
frames we have used are correlated \citep[see][]{taff90}.

The position of our adopted center is intermediate to those determined by
\citet{guh92} and \citet{cal93}. This
is illustrated in Figure \ref{fig:findcenB}, the left panel of which shows
a 20\arcsec$\times$20\arcsec\ region about our adopted center with the 5
central reference stars in Table \ref{tab:center} marked. 
Our star $\epsilon$ is the star that \citeauthor{guh92} used as a reference
position (their star E).  Our absolute coordinate for this star differs from
theirs by about 1.5 arcseconds.  The center position as estimated by
\citeauthor{guh92} is labeled with a ``G,'' placed at the position they report
relative to star $\epsilon$ (not at the absolute coordinates given in their
paper). We also mark the center found by \citeauthor{cal93}
with a ``C.'' Finally, the right panel of Figure \ref{fig:findcenB} shows a
close-up of the 4\arcsec$\times$4\arcsec\ box around our center.

\subsubsection{F475W Photometry}
\label{subsubsec:photo}

In the course of fitting PSFs to find positions, we also computed the average
F475W fluxes of stars in the master frame. These fluxes include a
spatially-dependent correction, of order $\sim3\%$, for the fact that the ACS
flat fields were designed to preserve surface-brightness rather than flux
(Anderson 2006, ACS ISR in preparation).
However, they were calculated from only the
inner 5$\times$5 pixels around each stellar peak, and they correspond to a
60-second exposure.
To calibrate the photometry, we must turn this flux into that which 
would be measured through the standard ``infinite'' aperture of 
5 arcseconds (100 WFC pixels) in 1 second. We first compute the ratio of
our measured flux to the flux contained within 10 pixels, or 0\farcs5.
This ratio is 1.238. The encircled energy curves in the ACS Handbook
\citep{pav05} then show that 92.5\% of the light should be contained
within this radius.  Thus, we scaled all our fluxes 
by a net factor of 1.339 upward and then divided by 60 (seconds) before
adding the VEGAMAG zeropoint of 26.168 \citep{dem04} to obtain a final,
calibrated F475W magnitude for each star in the master list.

\subsubsection{Completeness Fractions}
\label{subsubsec:compfrac}

The broad, uniform coverage of our WFC master frame also makes this
a useful data set for constructing an accurate surface density
profile for the cluster. We can detect {\it almost} all the 
stars there are, so it should be possible to come up with the
definitive radial profile from the center out to nearly 5 core radii.
There are, however, two issues that complicate the construction 
of any density profile from star counts in globular clusters: incompleteness
and mass segregation. Mass segregation really just means that the density
profile can differ, at least in principle, for stars in different mass
(magnitude) ranges. It is a physical effect, separate from any instrumentation
or data-reduction issues, and we discuss it briefly in
\S\ref{sec:space} below, where we actually derive a density profile for the
innermost parts of the cluster.
Incompleteness, on the other hand, is a technical limitation of the
observations themselves.

Incompleteness is always a joint function of both the images 
and the algorithm used to find stars in the images.  It can 
arise from several sources.  A particular star might not be found 
because (1) it is too close to a brighter neighbor and is not 
bright enough to generate its own peak in the image; (2) it could 
land on a defect in the chip or it could be hit by a cosmic ray;
or (3) it could be close to the background and not bright enough 
to generate a peak above the noise.  The fact that we have 
generated our list from many observations all at different 
pointings saves us from (2).  And since the stars we analyze in this paper
are always several magnitudes brighter than the faintest stars
that can be detected in the master frame, issue (3) is not important
for us. The first source of incompleteness (bright-star crowding) is the
only thing we need to concern ourselves with here.  

The usual strategy for evaluating incompleteness involves applying
a set data-reduction procedure to a large number of artificially generated
data sets. This is a rather daunting prospect in our case, and we have instead
addressed the problem directly from the algorithm we used in
\S\ref{subsubsec:purge} to identify (likely) non-stellar artifacts in our
original list of 129,733 PSF peaks in the master frame.

First, we used the ACS images from program GO-9028 to create a circular image
of the region $R\le 150\arcsec$, with a uniform pixel size of
$0\farcs05$ pixel$^{-1}$. (Note that this reference image completely
contains the master field itself, which is non-circular and measures only
$\simeq3\arcmin$ on a side. This is because we required a point to be
covered by at least 7 ACS pointings to contribute to the master star list, but
only one was sufficient to build the circular mosaic.) We then went
pixel by pixel through the intersection of this circular frame with our master
field, made a list of all stars found by our PSF-fitting within a
25-pixel ($1\farcs25$) radius from each point, and recorded the faintest
magnitude, $m_{\rm max}$, of the stars which could have
survived the artifact-purging procedure of \S\ref{subsubsec:purge} and
Figure \ref{fig:PSFartifacts}. This yields an estimate of the limiting
magnitude at every pixel in our master frame.

Given the limiting $m_{\rm max}$ at every point in our field, we
calculated a ``completeness fraction'' for every star in the master list
individually. Knowing the position $(x_*,y_*)$ and magnitude $m_*$ of each
star in the list, we looked at the values of $m_{\rm max}$ at all points
within 100 pixels ($=5\arcsec$) of $(x_*,y_*)$ and counted the number
of pixels for which $m_{\rm max}<m_*$. That is, we computed the fraction
$f \in [0,1]$
of the local area around each star where an identical star could fall but be
discounted as unreliable by our criterion in \S\ref{subsubsec:purge} (or not
be detected at all). The local completeness fraction at $(x_*,y_*,m_*)$ is
then just $c \equiv (1-f)$. 

With $c$ defined pointwise in this way, every star in our master list is
interpreted as representing $1/c$ actual stars, and corrected density profiles
follow simply from the sums of $1/c$ over all points within specified areas on
the sky. Again, we actually construct such a profile in \S\ref{sec:space}.

\subsubsection{Image of the Central Regions and the Final Star List}
\label{subsubsec:masterlist}

\begin{figure}[!b]
\centerline{\resizebox{100mm}{!}{\includegraphics{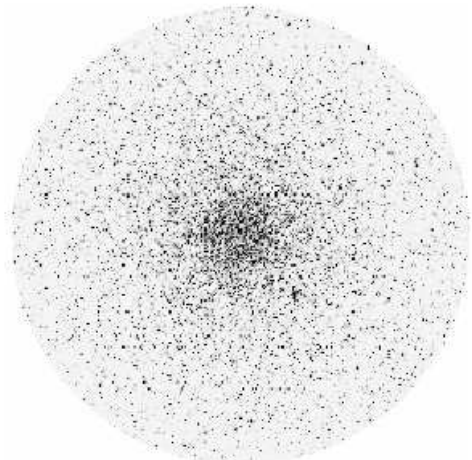}\hfil}}
\caption{
ACS-based image of the central $R\le 150\arcsec$ in 47 Tuc, inside of which
we identify stars for the unbiased master list of Table \ref{tab:master}.
The image is available in {\tt .fits} format from the online edition of
the {\it Astrophysical Journal}. Pixel size in the online image is 0\farcs05
per pixel, and coordinates are defined with the cluster center at
$(x,y)=(3001,3001)$ px. North is up and East is to the left.
\label{fig:metaimage}}
\end{figure}

As was mentioned just above, as part of our estimation of completeness
fractions we constructed a meta-image of the innermost $R\le150\arcsec$ from
the center of 47 Tuc. This convenient reference image,
{\tt 47TucMaster.fits}, oriented in the usual
North-up, East-left way, is available through the online edition of
the {\it Astrophysical Journal}. A low-resolution version of it is shown in
Figure \ref{fig:metaimage}.

Table \ref{tab:master} finally presents all important information on the
129,733 coincident PSF peaks in our master field, which is fully contained in
the circular area of the meta-image.
(A sample of Table \ref{tab:master} can be found at the end of this preprint.)
Only 114,973 of these peaks can be
said confidently to be bona fide stars, but to be comprehensive, we have here
retained (and flagged) the 14,760 peaks which could be PSF artifacts according
to \S\ref{subsubsec:purge}. The table gives the offset of each detection from
the cluster center in arcseconds;
the calibrated F475W  magnitude; the absolute RA and Dec in real and
celestial formats; the faintest magnitude a star at each position could have
and still be found; a flag indicating whether the star survives the artifact
purging; the local completeness fraction for the brightness and position of
each star; a serial ID number; and $(x,y)$ coordinates in both our meta-image
and the original image of the ACS/GO-9028 central pointing.

The absolute astrometric calibration of Table \ref{tab:master} should be good
to about 0\farcs1, but the relative positions should be much better than
this---particularly for bright, unsaturated stars and those with small
separations, for which we estimate an accuracy of $\sim$0\farcs001.
Of course, all the positions refer specifically to the epoch (2002.26) of
the GO-9028 data set.

\subsection{Reducing Images from Multiple Epochs}
\label{subsec:multiepoch}

With a well-defined reference frame in hand, the next step toward deriving
proper motions is to measure the positions of stars in individual images taken
at different epochs and transform these into the master coordinate system.
We began this task by measuring each star in each data set in
Table \ref{tab:datasets} with an appropriate PSF, derived according to the
method of \citet{ak2000}. For the ACS-WFC
observations, we used a single PSF to treat the entire 2-chip association. 
We also used a single PSF for the entire ACS-HRC chip. The PSF does vary
significantly with position over the ACS, but our data were reduced
before the methods of \citet{ak2006} were developed to deal with this.
Nevertheless, \citet{and02} shows that the biggest effect of this variation
on astrometry is a small bias of 0.01 pixel in the positions.
Since all our images are well-dithered, this error averages out and is
included in the internal uncertainties for the positions. The
constant-PSF assumption can also introduce systematic errors of up to 0.03
magnitudes in the photometry, but this is not of concern to us here.

All the raw measured positions were corrected for distortion 
using the prescriptions in \citet{ak2003b} for WFPC2 data and 
Anderson (2006, in preparation) for ACS data. These corrections include the
global and fine-scale distortion corrections, as well as a correction for the
68th-row defect. Charge-transfer inefficiency should not be an issue,
as the background is relatively high in the images that we have used.

It was then necessary to combine all the observations of each star
in the multiple pointings at each epoch. In all cases we adopted the
centermost pointing as the ``fiducial'' frame for each epoch, and used general
6-parameter linear transformations (along the lines of those described in
eq.~[\ref{eq:transform}] below) to transform positions
from the individual, distortion-corrected pointings for that epoch into the
central frame. In finding these transformations for WFPC2 images, we treated
the PC chip and the three WF chips independently of the others; for the ACS
images, the HRC chip and two WF chips were likewise considered independently.
In this way, we found an average position for each star in each of 27
chip-epoch combinations. The standard deviation of a star's position in the
independent pointings at each epoch defines the uncertainty in $x$ and $y$
locations.

\begin{figure}[!t]
\centerline{\resizebox{85mm}{!}{\includegraphics{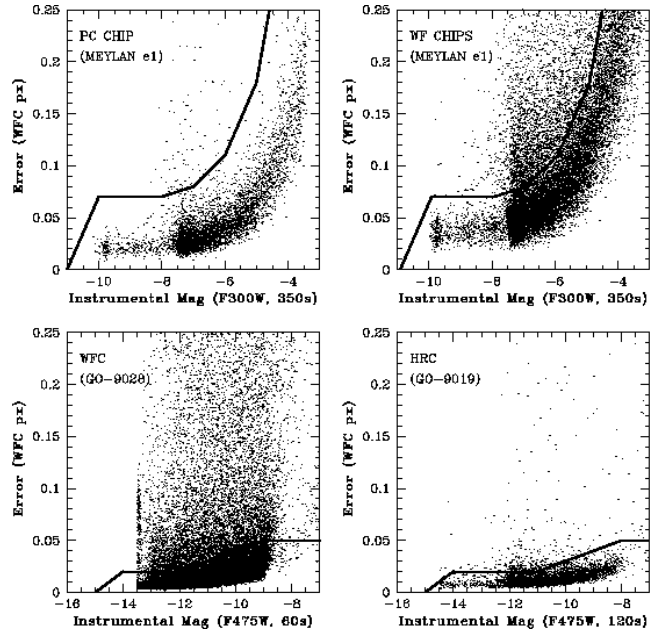}\hfil}}
\caption{
Illustration of the criteria defining stars, on the various chips of
different detectors, with positions that are ``well-measured'' enough to be
used in deriving proper motions. Examples are shown at one epoch for each
of the PC and WF chips on the WFPC2, and the WFC chips and HRC chip of the
ACS. The discriminating lines drawn are applied to the list of
positions and fluxes of all stars found at every epoch on each chip,
with any detections falling above the appropriate line discarded as too
uncertain to contribute meaningfully to the determination of a proper motion.
The horizontal axes are given here in instrumental magnitudes to indicate
the relationship between positional error and raw S/N. The location of the
subgiant branch in 47 Tuc (calibrated $V\simeq17.2$;
cf.~Fig.~\ref{fig:cmdpos}) is clear
from the sharp increase in the density of points in each panel.
\label{fig:magcuts}}
\end{figure}

At this point we examined the errors in the stars' positions as a function of
their flux at each epoch in order to define a criterion to define which stars
were well measured and which were not well measured within each data set.
Figure \ref{fig:magcuts} shows a plot of the position errors as a 
function of the instrumental magnitude $-2.5 {\rm log}_{10}[N_{DN}]$ for
some sample chip-epoch combinations. We drew the fiducial lines shown in
Figure \ref{fig:magcuts} to distinguish well measured from poorly measured
stars; only stars falling below these lines in any of the data sets were
retained for any further analysis. In addition to this, we discarded all stars 
found within the pyramid-affected region of any of the WFPC2 chips 
(ie., any at chip coordinates $x<100$ or $y<100$).  We finally rejected any
saturated bright stars from the star list for every data set.

This left us with an average position, including uncertainties, and a flux for 
every star that we deemed well measured in each chip of each data set. The
positions were still in the
distortion-corrected coordinate systems of the individual epochs, however, and
needed next to be transformed into the system defined by our master frame. 

We should note explicitly here that the GO-9028 WFC data (epoch 2002.26)
were subjected to the above analyses (and to the coordinate transformations
described next) in exactly the same way as were any of the other epochs in
Table \ref{tab:datasets}---even though the master frame in
\S\ref{subsec:master} was itself defined using data from this epoch. This is
because in constructing the master frame we used only 13 images from the
2002.26 WFC set of 20 and made no effort to impose any ``quality control'' on
the stellar positions or magnitudes. However, any given star in the master
list could, in principle, be observed in one or more of the seven other
pointings at
this epoch; or alternately its position might be so poorly constrained
(relative to the criterion illustrated in Figure \ref{fig:magcuts}) that it is
not useful for proper-motion measurements. Thus, to make optimal use of the
GO-9028 WFC data requires that they be re-analyzed in parallel with the other
epochs, with the master frame viewed simply as an externally imposed construct.

\subsubsection{Linking with the Master System}
\label{subsubsec:link}

The transformation into our master coordinate system is non-trivial.
Given the small proper-motion dispersion in 47 Tuc (see Table \ref{tab:basic})
and our short ($\la 7$ year) time baseline, we need to resolve displacements
of order 0\farcs001 and less for the stars in our sample. The pointing of HST
is good to only 0\farcs1 at best, however, and thus we cannot transform our
positions to an absolute frame with anywhere near the accuracy required.

The only information we have about how the coordinate system of one frame 
is related to that of any other comes from the positions of stars that are 
common to both frames.  Each observation typically has thousands of stars
in common with the master frame.  Once these stars are matched up, the two 
sets of positions can be used to define a linear transformation from one 
coordinate system to the other.  Specifically, if we have a list of positions
$\{x_i, y_i; i=1, \dots, {\cal N}\}$ for ${\cal N}$ stars at some epoch, and a
list $\{X_i,
Y_i\}$ for the same stars in the master frame, then we specify
\begin{equation}
\begin{array}{lll}
    X_i & = & A \cdot (x_i-x_0) + B \cdot (y_i-y_0) + X_0 \\
    Y_i & = & C \cdot (x_i-x_0) + D \cdot (y_i-y_0) + Y_0 \ ,
\end{array}
\label{eq:transform}
\end{equation}
for constants $A$, $B$, $C$, $D$, and $(x_0,y_0)$, $(X_0,Y_0)$ to be
determined. It may look like there are eight free parameters in this
transformation, but in fact we have the freedom to pick the zeropoint offset
in one system. In particular, if we choose ($x_0$,$y_0$) to be the centroid of
the ${\cal N}$ matched stars in the non-master frame, then ($X_0$,$Y_0$) 
is necessarily the centroid of the group in the Master system.  With these
set, then, standard least-squares regression can be applied to solve for
the linear terms $A$, $B$, $C$, and $D$ from the $2{\cal N}$ observed pairs of
coordinates. Our last concern is the question of how many stars
should be used to do this.

\subsubsection{Local Transformations}
\label{subsubsec:local}

If we were to use {\it all} the stars in common between the master system
and the data from any other epoch, the solution of equation
(\ref{eq:transform}) would correspond to the best global transformation
between the two coordinate systems.  Unfortunately,
such a chip-wide solution may not give the best transformation for a
given point.  In particular, although all of our frames have been
distortion-corrected, any residual distortions will introduce systematic
errors in a global transformation. Such distortion errors tend to be
cumulative, i.e., they are larger for stars that are farther apart.  As a
result, the distance between two nearby stars can be measured much more
accurately than the distance between stars at different corners
of a chip. We therefore decided to perform more local transformations, using
smaller groups of relatively nearby stars to define different transformations
at different positions in an observed frame. Such a scheme may sound computer
intensive, but it is straightforward enough to implement and is not, in fact,
exceedingly slow. Given that we will never be able to remove distortion
perfectly, this is a useful way to minimize its effect on our results.

Two sources of error influence the choice of the number ${\cal N}_{\rm trans}$
of stars to use in solving the system of equations (\ref{eq:transform}) for
local transformations. First, the positions $\{x_i,y_i\}$ and $\{X_i,Y_i\}$
themselves are obviously subject to some uncertainty. If $\Delta$ denotes a
representative value of this uncertainty, then the average transformation
is fundamentally uncertain at a level
$\sim \Delta/\sqrt{{\cal N}_{\rm trans}}$, which
suggests that we would like ${\cal N}_{\rm trans}$ to be as large as
possible. Second, the uncorrected residual distortion introduces appreciable
systematic error if the ``nearby'' stars used to define the local
transformation at any point come from too far away. This implies that we would
like ${\cal N}_{\rm trans}$ to be as small as possible.

After some experimentation, we found that a reasonable compromise between
these opposing tendencies was reached with ${\cal N}_{\rm trans}=45$ for our
data. In practice, we took a single star from the list of well-measured
positions at one epoch; found this target star's nearest 55 neighbors; and
matched these neighbors to positions in the master frame. We then used these
55 pairs of coordinates ({\it not} including the target star itself) to
solve for the coefficients in equation (\ref{eq:transform}); discarded the 10
stars which deviated most from the solution; and re-solved for the
transformation using the 45 remaining stars. This was then used to transform
the original, target star {\it only} into the reference frame. These steps were
repeated for every well-measured star in the combined frame for each
of our ten epochs, until we had measurements of RA and Dec position (relative
to the cluster center) vs.\ time for all stars in a single, unified coordinate
system.

We also performed this procedure using ${\cal N}_{\rm trans}=20$ and
${\cal N}_{\rm trans}=100$ and did not find significant differences, in
general, from our results with ${\cal N}_{\rm trans}=45$. However, the lower
number is approaching the limit of what we feel comfortable with in terms of
vulnerability to small-number statistics, while the higher number is coming
close to bringing in too-distant neighbors that could well be affected by
uncorrected distortion errors.

There is another, somewhat more subtle source of error in our local
transformations, which comes from the fact that all the stars are physically
moving with respect to each other from one epoch to the next. Thus, even if a
star's position could be measured perfectly in every image, it is impossible to
associate its location $(x,y)$ at some epoch perfectly with its position
$(X,Y)$ in the master frame using a finite network of ${\cal N}_{\rm trans}$
moving neighbors. Put another way, any one network of neighbor stars might in
fact have a real net motion relative to the cluster center, but our approach
assumes that all such network motions are identically zero. This error gets
averaged away over many networks, of course, so that the mean velocity
we estimate for any group of stars will be unbiased.
However, any estimate of the velocity dispersion, $\sigma$, {\it is} affected.
Details of how we correct for
this are given in Appendix \ref{sec:errcor}, which also discusses the
correction of velocity dispersion for unequal measurement errors. Here we
simply note that the net effect of the local-transformation error is an
artificial inflation of the intrinsic stellar $\sigma$ by a factor of
$(1+1/{\cal N}_{\rm trans})^{1/2}$, or about 1.011 for our chosen
${\cal N}_{\rm trans}=45$.

\subsubsection{$V$- and $U$-Band Photometry}
\label{subsubsec:multiphoto}

In addition to calibrated F475W photometry for all stars in our master field
(\S\ref{subsubsec:photo}), we have obtained calibrated $V$-band and
F300W (roughly $U$-band) magnitudes for the subset of stars
falling in the WFPC2 field of the original Meylan exposures (GO-5912,
GO-6467, GO-7503; see Table \ref{tab:datasets}, and note that $V$ exposures
were also taken as part of these programs for photometric purposes only).
As we will describe below, all
stars for which we have derived proper motions were required to be detected in
at least one of these three early epochs, and thus the full proper-motion
sample has multicolor photometry. Although the standard $V$ bandpass is not
substantially different from F475W, it is useful for relating our analyses to
various ground-based observations, and to such things as theoretical stellar
mass-luminosity relations. The shorter $U$-band photometry
is useful for constructing a color-magnitude diagram of our
velocity sample in order to investigate how CMD position might influence
stellar kinematics.

$V$ photometry of the WFPC2 fields was calibrated against an HRC image of
the core of 47 Tuc, using the VEGAMAG zeropoints in \citet{dem04}.
Star-by-star comparison with the calibrated photometry in \citet{bic94}
shows agreement at the $\simeq0.05$-mag level on average. The $U$ magnitudes
were calibrated using the VEGAMAG zeropoint in the WFPC2 Data Handbook
\citep{bag02}.
The main-sequence turn-off in our $V,(U-V)$ CMDs
agrees well with that in \citet{edm03}. We report the $V$ and $U$ photometry
for our proper-motion stars in \S\ref{subsec:pmsample}, where we now define
the velocity sample.

\section{Velocity Samples}
\label{sec:velsamples}

Our multi-epoch astrometry data are most certainly not homogeneous. The 2002
epoch is relatively uniform, thanks to the even ACS/WFC coverage; but the
earlier WFPC2 observations come from a different camera with different
pointings, dithering patterns, fields of view, and resolutions.
Thus, not every star in any one of the ten data sets of Table
\ref{tab:datasets} can be found in all of the other nine.

In \S\ref{subsec:pmsample}, we describe how we go from a heterogeneous sample
of multi-epoch positions to a homogeneous sample of proper motions.
We also examine the quality of the fits of straight lines to the 
position-vs.-time data, and the uncertainties in the final proper motions.
In \S\ref{subsec:rvsample}, we discuss the third component of velocity, that
along the line of sight. Specifically, we have a large sample of ground-based
radial velocities for stars in 47 Tuc, and we are interested in using these in
conjunction with our proper-motion sample to obtain a kinematic estimate of the
distance to the cluster (\S\ref{subsec:distance}).

\subsection{Proper Motions}
\label{subsec:pmsample}

To define a working
sample of plane-of-sky velocities for kinematic analyses, we have chosen to
work only with stars which are observed in at least {\it three} separate
data sets prior to year 2000 (i.e., in three or four of the MEYLANe1, MEYLANe2,
MEYLANe3, and GILLILU1 epochs in Table \ref{tab:datasets}), {\it and} in at
least {\it one} of the ACS data sets (any of the WFC or HRC fields) from 2002.
Thus, we only derive proper motions for stars that have a {\it minimum} of
four separate $(x,y,t)$ measurements spanning a {\it minimum} of about 4.4
years ($2002.26-1997.84=4.42$), and all derived velocities are effectively
``tied down'' by a precise ACS position measurement.
In practice, most of our stars are in fact measured 6 or more
times over the full 6.7-year timespan (1995.82--2002.56) of our epochs,
and a good number do in fact appear in all 10 of the data sets listed in
Table \ref{tab:datasets}.

\begin{figure}[!t]
\centerline{\resizebox{85mm}{!}{\includegraphics{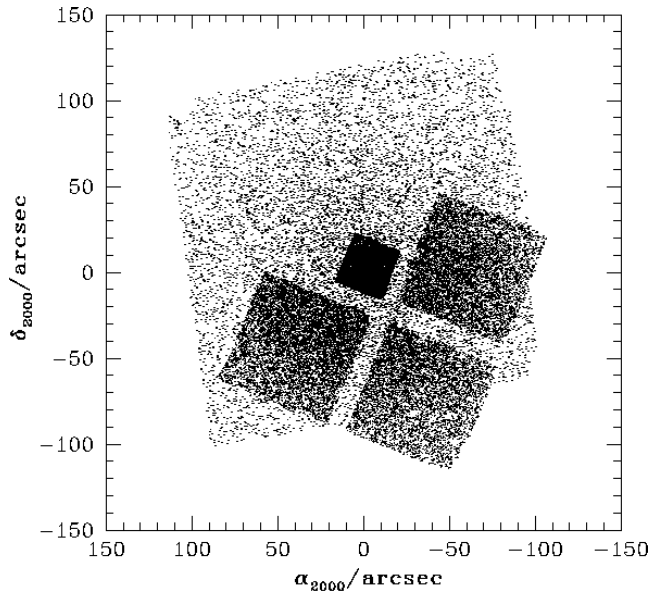}\hfil}}
\caption{
Outline of the fields of view of our master star list (large square area;
Table \ref{tab:master}) derived from ACS program GO-9028, and of the original
WFPC2 observations of GO-5912, 6467, and 7503. Only stars in the intersection
of the two fields are retained for inclusion in the full proper-motion sample
of Table \ref{tab:posdata}.
\label{fig:fields}}
\end{figure}

We further decided to include stars in the proper-motion sample only if
they also appear in the master star list of \S\ref{subsec:master}---even if
another ACS ``epoch'' might contain the star. Ultimately, this
has some effect on the spatial distribution of the velocity sample. Figure
\ref{fig:fields} shows the RA and Dec offsets from our estimated cluster
center for a subset of stars in our master list (the larger, rotated square
field), and the positions of stars in the original MEYLAN epochs
(the four smaller, more densely filled squares). Evidently, the combination of
all our requirements gives us proper motions over a patch on the sky with
roughly the familiar WFPC2 pattern, but with slightly larger gaps than normal
between the chips (cf.~\S\ref{subsec:multiepoch}) and with the outermost
corners shaved off.

We consider a star to be ``observed'' at any pre-2002 (WFPC2/MEYLAN or GILLIL)
epoch only if it is detected in at least {\it ten} of the individual
pointings/ditherings that went into that observation, and if it survives the
culling based on position error vs.\ instrumental flux discussed around Figure
\ref{fig:magcuts} above. For any of the
higher-precision, 2002 ACS observations, we take a star to be detected only
if it is identified in at least {\it four} individual pointings/ditherings
and if it again survives the appropriate culling by position error. 
These criteria are imposed to ensure that, whenever we include the measurement
of a star's position at any one epoch in our proper-motion determination, we
also have enough information to estimate accurately the position
{\it uncertainty} at that epoch.

The estimation of position uncertainties is of critical importance to the
derivation of proper-motion uncertainties, and thus to the ultimate inference
of intrinsic, error-corrected stellar kinematics. As was suggested in
\S\ref{subsec:multiepoch}, we take a star's position in any epoch to be the
mean of the positions measured in the separate ditherings that were combined
for that epoch and transformed into the master reference frame. The
uncertainty is then simply the standard sample deviation of those independent
multiple measurements. Note that we do {\it not} work with the uncertainty in
the mean position (which would involve dividing the standard deviation by
$\sqrt{N_d}$ for $N_d$ ditherings), because the theory of linear
regression---which we use to derive velocities---actually
requires that the square of the errorbars on the data be unbiased estimates
of the variance in the position measurements. 

Thus, given $(x,y)=(\alpha_{2000},\delta_{2000})$ positions and uncertainties
as functions of time for any star that satisfies the minimum criteria just set
out, we
derive a plane-of-sky velocity in each direction by a standard, error-weighted
least-squares fit of a straight line \citep[e.g.,][Section 15.2]{press92},
allowing both the slope and the intercept to vary. This procedure
automatically yields uncertainties in each component of velocity. The
measurement-error distribution in each component of velocity for any single
star will be Gaussian, with a mean of zero and a dispersion equal to the
fitted least-squares errorbar, {\it if} the position measurement errors
are similarly Gaussian distributed. We have assumed that this is the case.

The most important advantage of performing a {\it weighted} least-squares
regression here is that it returns a $\chi^2$ value for each straight-line
fit. Knowing the number of degrees of freedom in the fit ($=N_e-2$, where
$4\le N_e\le 10$ is the number of independent epochs at which the star's
position was measured), it is then possible to calculate the probability
that the $\chi^2$ of the fit could have occurred by chance if
the true motion of the star were really linear \citep[see][]{press92}. We
can then make use of this to exercise some quality control over the proper
motions, by excluding from kinematics analyses any stars with fitted
velocities whose $\chi^2$ probabilities are lower than some specified
threshold.

By fitting straight lines to our position-vs.-time data we are obviously
assuming that the stars are moving at effectively constant velocity. To
justify this in general, note that the characteristic gravitational
acceleration in the core of 47 Tuc is of order (see Table 1, and
\S\S\ref{sec:space} and \ref{sec:veldisp} below)
$a_0\sim \sigma_0^2/r_0
\sim (0.6\,{\rm mas\ yr}^{-1})^2/20\,{\rm arcsec} \sim
2 \times10^{-5}\,{\rm mas\ yr}^{-2}$. Even integrated over 7 years, the
velocity change induced by such an acceleration is only $\sim 10^{-4}$ mas
yr$^{-1}$, which is, as we shall see, a small fraction of the velocity
uncertainties we infer. Of course, this does not exclude the possibility
that some stars could be significantly accelerated by ``nonthermal'' processes
such as stellar or black-hole encounters, or by virtue of being in tight,
face-on binaries, or in some other way. Such stars will simply not be
described well by straight-line motion, and the $\chi^2$ probabilities
inferred from our weighted linear regression will reflect this fact. Thus, a
low $P(\chi^2)$ can reflect legitimately nonlinear data as well as simple
``bad'' measurements.

In all, we have 14,366 stars with RA and Dec positions measured to
satisfactory precision in at least three pre-2000 epochs and at least one
2002 ACS epoch, and with measured $V$ and $U$ magnitudes (as described in
\S\ref{subsubsec:multiphoto}). Table \ref{tab:posdata}, which is published
in its entirety in the electronic version of the {\it Astrophysical Journal},
contains the instantaneous J2000 RA and Dec positions in our
master frame (epoch 2002.26) for all of these stars, expressed both in
arcseconds relative to the cluster center determined in
\S\ref{subsubsec:center} (see also Table 1) and in absolute celestial
coordinates. The F475W, $V$, and $U$ magnitudes of each star are also reported.
The RA and Dec offsets (also in arcsec) from the nominal
master-frame position, and their uncertainties, are then given for every
epoch in which the star was detected. The proper-motion velocities and
uncertainties implied by the weighted straight-line fitting to the offsets
vs.\ time then follow, along with 
the $\chi^2$ values for the fits and the probabilities $P(\chi^2)$ that
these values could occur by chance if the motion is truly linear. We have not
tabulated the intercepts of the linear fits, as these only reflect
the choice of an arbitrary zeropoint in time and are of no physical interest
in the constant-velocity case.
(A sample of Table \ref{tab:posdata} can be found at the end of this preprint.)

Table \ref{tab:posdata} includes all stars
for which we have estimated velocities, regardless of whether or not 
the $\chi^2$ values of the fits have ``good'' probabilities. We stress again
that a reliable sample for kinematics work should exclude stars with
very low $P(\chi^2)$, but that some such stars could still be
of interest for investigations of non--constant-velocity phenomena
(which we do not pursue in this paper).

Two points should be noted regarding the connection between Table
\ref{tab:posdata} and the larger master list of stars in Table
\ref{tab:master}. First, at the 2002.26 WFC-MEUR epoch in Table
\ref{tab:posdata}, the RA and Dec offsets from the absolute master-frame
positions are consistent with 0 within the uncertainties for most stars, but
they do not vanish exactly---even though this is the epoch that was used
to construct the master frame. The reason for this is that the master list in
\S\ref{subsec:master} was defined using a very specific subset of the 20
individual exposures comprising the WFC-MEUR data set, while the relative
positions found in \S\ref{subsec:multiepoch} and used here to compute proper
motions were allowed to come from different combinations of the 20
pointings. As a result, the transformations in the latter case
cannot be expected in general to give positions identical to those in the
master list, and the offsets for this epoch in Table \ref{tab:posdata}
essentially reflect statistical noise. Second, there are some stars in
Table \ref{tab:posdata} for which no offset at all is given at the WFC-MEUR
epoch---an entry of ``n/a'' appears instead---even though all stars in the
proper-motion table are guaranteed by construction to appear in the master
list. In these cases, the uncertainties in
the stars' positions at the master-frame epoch were larger than acceptable
according to our flux-based criterion defined in Figure \ref{fig:magcuts}
above, and thus these data were excluded from our fitting for proper motions.

Having derived proper motions for the best-measured stars in our field, we
now describe the selection of a velocity subsample which is 
best suited for kinematics analyses. In particular, we can be more
specific about how the $\chi^2$ probabilities in Table \ref{tab:posdata} are
used to cull stars with particularly poor proper-motion measurements
(or nonlinear motions), and we can quantify the distribution of the
velocity uncertainties.

\subsubsection{``Good'' and ``Bad'' Proper Motions}
\label{subsubsec:goodbad}

\begin{figure}[!t]
\vspace*{-0.5truecm}
\centerline{\resizebox{90mm}{!}{\includegraphics{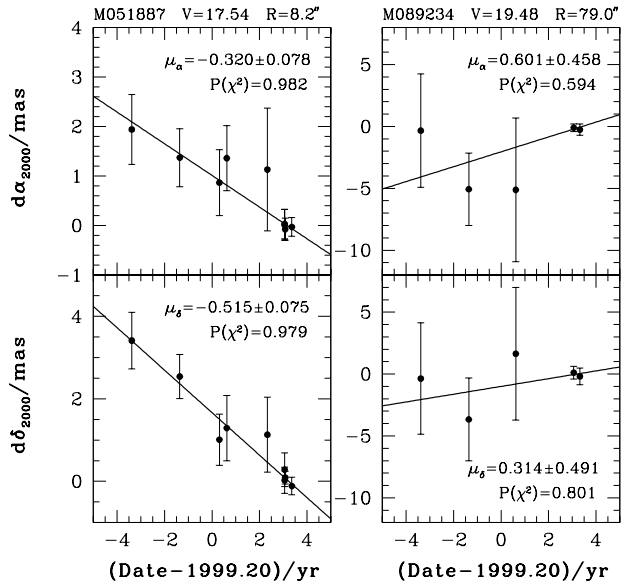}\hfil}}
\caption{
Two examples of stars with good qualities of fit for straight-line
proper motions: $P(\chi^2)\ge 0.001$ in both velocity components. The ID
numbers of the stars correspond to those in Tables \ref{tab:master} and
\ref{tab:posdata}. Plots such as these can be generated for any star in Table
\ref{tab:posdata} using an SM macro available in the electronic edition of the
{\it Astrophysical Journal}.
\label{fig:pmgood}}
\end{figure}

Figure \ref{fig:pmgood} shows plots of position (in milliarcseconds of RA and
Dec offset from the master-frame position) vs.\ time (in years from the
arbitrary zeropoint 1999.20) for two stars with relatively
good proper-motion determinations. In the left panels is a bright star, near
the cluster center, which has position measurements in Table \ref{tab:posdata}
for 8 of our 10 data sets/epochs.
In the right panels is a much fainter star, farther from the cluster center,
which has fewer $(x,y,t)$ datapoints. 
In both cases, the fitted velocities $\mu_\alpha$ and $\mu_\delta$ in the
RA and Dec directions are given in units of mas/yr, with $\mu_\alpha$ defined
to be positive for Eastward motion. The $\chi^2$
probabilities for the linear fits of the two proper-motion components are also
listed, showing that these data are fully consistent, within
measurement error, with the basic assumption of constant velocity.

The much smaller position uncertainties in the most recent, ACS data
relative to the pre-2002 (WFPC2) epochs are noteworthy. The right-hand panels
particularly illustrate how the ACS epochs play a crucial role in defining
the overall motions of our stars throughout the $\sim$7-year baseline of the
observations. Indeed, if only the 4-year span of WFPC2 data had been fit to
find a proper motion for the faint star in Figure \ref{fig:pmgood}, the RA
component would have had the {\it opposite sign} (though with a larger
errorbar) from that obtained when the ACS data are included. Evidently,
relying on only a few epochs of astrometry, even with the HST, can lead to
spurious proper motions for some individual stars.

\begin{figure}[!t]
\vspace*{-0.5truecm}
\centerline{\resizebox{90mm}{!}{\includegraphics{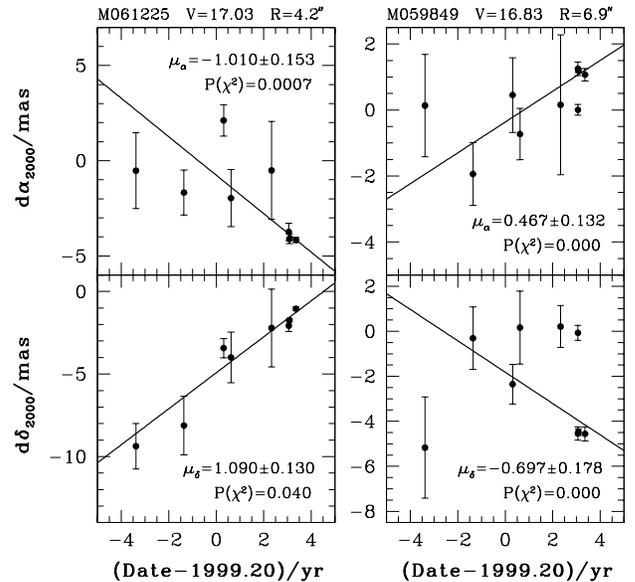}\hfil}}
\caption{
Two examples of stars with poor qualities of fit for straight-line proper
motions. With $P(\chi^2)<0.001$ in one or both velocity components, stars such
as these are not included in any working samples for kinematical analysis in
this paper. They are, however, retained in Table \ref{tab:posdata}.
Plots such as these can be generated for any star in Table
\ref{tab:posdata} using an SM macro available in the electronic edition of the
{\it Astrophysical Journal}.
\label{fig:pmbad}}
\end{figure}

Figure \ref{fig:pmbad} next shows two stars with much less satisfactory
proper-motion fits. Both stars here are fairly bright and near the cluster
center, and both show more scatter about the best-fit linear velocity than is
acceptable. On the left-hand side, the problem is primarily with the RA
displacements in the upper panel, where the early (WFPC2) epochs do not match
well onto the precise later (ACS) position measurements and $P(\chi^2)$ is
uncomfortably small. On the right-hand side, the scatter in
both the RA and Dec motions is so large (relative especially to the small
errorbars on the ACS positions at year $\simeq2002$) that there is effectively
no confidence that the best-fit constant velocity is an accurate
representation of the data.

Plots like those in Figures \ref{fig:pmgood} and \ref{fig:pmbad} can be
generated for any star from the data in Table \ref{tab:posdata}, using an
SM macro ({\tt pmdat.mon}) that we have packaged and made available in the
online edition of the {\it Astrophysical Journal}. After inspecting many such
graphs and comparing the kinematics of samples of stars defined by imposing
various lower limits on the allowed value of $P(\chi^2)$ for the fitted
velocities, we eventually decided to include in detailed analyses only stars
which have
\begin{equation}
\label{eq:pchi}
P(\chi^2) \ge 0.001
\end{equation}
for {\it both} RA and Dec components of proper motion.
This criterion defines a velocity sample which cleanly excludes ``bad''
data such as those illustrated in Fig.~\ref{fig:pmbad} (as well
as---again---potentially perfectly good data that are just
nonlinear) and shows quite robust kinematics, in the sense that samples
defined by revising the
$P(\chi^2)$ threshold in equation (\ref{eq:pchi}) moderately upward---even
by as much as an order of magnitude---do not have significantly different
statistical properties.

\subsubsection{Proper-Motion Uncertainties}
\label{subsubsec:pmerrors}

\begin{figure}[!t]
\vspace*{-0.4truecm}
\centerline{\resizebox{90mm}{!}{\includegraphics{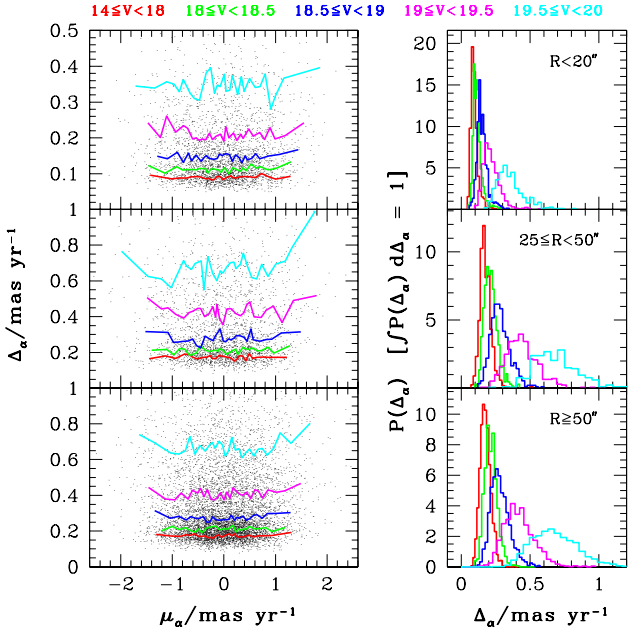}\hfil}}
\caption{
Characterization of velocity uncertainties $\Delta_\alpha$ in the RA component
of proper motion, $\mu_\alpha$. Left-hand panels are scatter plots of
$\Delta_\alpha$, with colored lines connecting median values in bins of
$\mu_\alpha$, for stars of different magnitudes. Right-hand panels are
normalized histograms of the $\Delta_\alpha$ distributions, showing the
tendency for the uncertainties to increase with $V$ magnitude and to be
smallest in general at $R<20\arcsec$, corresponding to the region of the
proper-motion field covered by the PC chip in the early WFPC2 images.
\label{fig:errmua}}
\end{figure}

Figure \ref{fig:errmua} shows the distribution of uncertainties (least-squares
errorbars) in the RA component of proper motion for stars in Table
\ref{tab:posdata} which are brighter
than $V<20$ and have $P(\chi^2)\ge0.001$ for the fitted values of both
$\mu_\alpha$ and $\mu_\delta$. We have divided this subsample into three broad
bins of projected clustercentric radius, $R$. It is immediately apparent that
stars at $R > 20\arcsec$ have systematically higher velocity errorbars,
typically by factors of $\sim$2, than those at $R<20\arcsec$. This is because
this
radius is completely contained in the high-resolution, PC chip of the WFPC2
camera (see Figure \ref{fig:fields}), which affords higher precision in the
measurement of (pre-2002) stellar positions than the WF chips which cover the
larger radii in our field. From this alone it is clear that our velocity
uncertainties correlate with clustercentric position.

Panels on the left-hand side of Fig.~\ref{fig:errmua} plot the proper-motion
uncertainty $\Delta_\alpha$ against the velocity itself, $\mu_\alpha$.
The colored lines, which correspond to different magnitude bins within
each radial bin, connect the median $\Delta_\alpha$ errorbar in each of a
number of discrete $\mu_\alpha$ bins. The right-hand panels then show the
corresponding normalized histogram of velocity uncertainties in each
radius/magnitude bin. These show that, for stars of a given magnitude, the
distribution of $\Delta_\alpha$ not only peaks at smaller values for
$R<20\arcsec$, but it is more {\it sharply} peaked there as well. Put
another way, stars at the larger clustercentric radii have a stronger tail 
toward high velocity errorbars. Conversely---and quite naturally---at any
fixed clustercentric radius, fainter stars always have larger average
uncertainties and broader distributions. Indeed, the stars in the faintest
magnitude bin illustrated here ($19.5\le V<20$) have,
at radii $R\ge 25\arcsec$, ``typical'' velocity errorbars of order 0.6
mas yr$^{-1}$ or more, which is comparable to the intrinsic velocity
dispersion at the center of 47 Tuc (Table \ref{tab:basic}). Thus, 
stars of such faint magnitude will be of limited use for the statistical
characterization of intrinsic cluster kinematics. Stars that are fainter
still have even larger velocity uncertainties and are entirely useless in this
context. Thus, we impose the magnitude limit
\begin{equation}
\label{eq:maglim}
V<20
\end{equation}
when choosing stars for any kinematics analysis. 
This is {\it not} to say that the velocities of individual faint
stars are always unreliable---we shall see evidence to the contrary in
\S\ref{subsec:highpm}---but only that their group properties are poorly
constrained.

The distribution of velocity uncertainties for the Dec component of motion,
$\mu_\delta$, is essentially identical, in all respects, to that shown for
the RA component in Figure \ref{fig:errmua}. The basic message of
these plots is that, due to strong correlations with stellar magnitude and
position, the errorbars on our fitted velocities cover a wide range of values
and cannot be considered even approximately equal. Nor, in general, are they
negligible relative to the intrinsic stellar motions. It is therefore
important to account properly for the effects of measurement error on the
observed proper-motion distribution and its moments (i.e., velocity
dispersion). The details of this are discussed in Appendix \ref{sec:errcor},
where we also consider the statistical implications of the
local-transformation approach to obtaining relative proper motions
(cf.~\S\ref{subsubsec:local}).

\subsubsection{The Sample for Kinematics Analyses}
\label{subsubsec:kinsamp}

\begin{figure}[!t]
\vspace*{+0.1truecm}
\centerline{\resizebox{85mm}{!}{\includegraphics{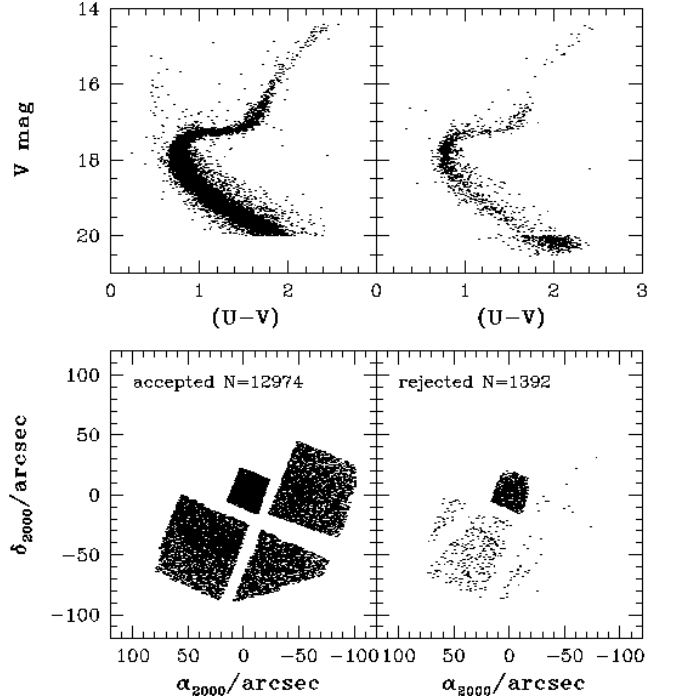}\hfil}}
\caption{
Color-magnitude diagrams and spatial distributions of proper-motion stars
which are used for subsequent kinematics work (left-hand panels), and those
which are not (right-hand side). See text for details of the criteria by
which stars are rejected.
\label{fig:cmdpos}}
\end{figure}

Figure \ref{fig:cmdpos} shows color-magnitude diagrams and spatial
distributions of the 14,366 stars with proper motions listed in Table
\ref{tab:posdata}, split into a
sample in the left-hand panels which we consider useful for kinematics
analyses (i.e., which comprises only stars satisfying both criteria in
eqs.~[\ref{eq:pchi}] and [\ref{eq:maglim}]) and a sample in the right-hand
panels that we exclude from such work (i.e., which consists of stars with
$P(\chi^2)<0.001$ in either component of proper-motion velocity, and/or with
magnitude $V\ge 20$).

The CMD of the upper left-hand panel exhibits a very well defined cluster
sequence, including a red giant branch, a main-sequence turn-off
at about the appropriate $V_{\rm TO}=17.65$ (Table \ref{tab:basic}), and
even a small blue-straggler population. This is, in fact, not unexpected given
that the foreground contamination in our small field (with an area of
$\approx3.5$--4 arcmin$^2$) is expected to amount to perhaps $3\pm2$
brighter than $V=21$ (from the foreground star density estimated by
\citealt{rat85}; see see Table \ref{tab:basic}). Note that we have no stars
brighter than $V=14$ with measured velocities at all. Such stars were
generally saturated in the 2002 ACS frames, and thus did not meet our
selection criteria for the proper-motion sample.

The rejected stars in the right-hand panels of Figure \ref{fig:cmdpos} amount
to only $\sim10\%$ of the full sample in Table \ref{tab:posdata}. 30\% of them
(417/1392) are rejected simply because they are fainter than $V=20$.  
The 975 others are rejected solely because they have a low $\chi^2$
probability for the linear fit to one or both components of their velocity. Of
these, as Figure \ref{fig:cmdpos} suggests, the majority are brighter than
$V=18.5$ and located within $R\le 20\arcsec$ of the cluster center,
where the stellar density is highest and crowding is most problematic.

\begin{figure}[!t]
\vspace*{-0.3truecm}
\hspace*{+0.2truecm}
\centerline{\resizebox{110mm}{!}{\includegraphics{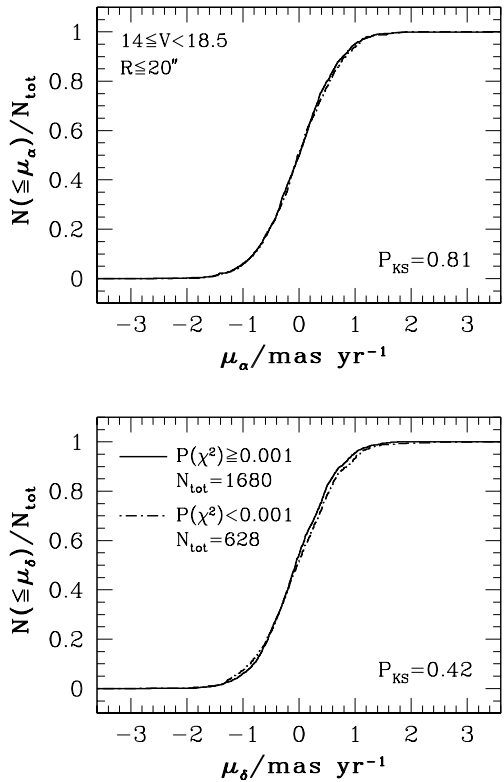}\hfil}}
\caption{
Comparison of the cumulative RA and Dec velocity distributions for bright
stars in the inner 20\arcsec\ of our proper motion field, for the sample of
stars with $P(\chi^2)\ge0.001$ in both proper-motion components,
vs.\ those stars with $P(\chi^2)<0.001$ in at least one component. In each
panel, $P_{\rm KS}$
is the probability (from a Kolmogorov-Smirnov test) that the proper-motion
distributions of the ``good'' and ``bad'' stars are in fact
drawn from the same parent distribution.
The high values of this statistic show that our results are not
biased by excluding the poorest velocity fits from kinematical
analysis.
\label{fig:goodvbad}}
\end{figure}

It is important to check whether the rejection of these stars leads to
significantly different kinematics for the ``good'' proper-motion sample than
would have resulted if all stars were retained. Thus, in Figure
\ref{fig:goodvbad} we show the cumulative distributions of RA and Dec
velocities for stars with $V<18.5$ and $R\le 20$, split into samples with
$P(\chi^2)\ge0.001$ (accepted into a final kinematics sample) and
$P(\chi^2)<0.001$ (rejected). Kolmogorov-Smirnov tests applied to these
indicate that the velocity distributions of the rejected stars are consistent
with their having been drawn from the same parent distribution as the accepted
stars. Thus, excluding the former from
analysis does not give biased kinematics in the end but simply reduces
the net uncertainty in our final results.

Before going on to use our sample of 12,974 good proper motions to
investigate these kinematics in detail, we first briefly describe some
complementary line-of-sight velocity data that we use in
\S\ref{subsec:distance} (only) to derive a kinematic distance to 47 Tuc.

\subsection{Radial Velocities}
\label{subsec:rvsample}

\subsubsection{The Data}
\label{subsubsec:rvdata}

The radial velocities used in this study come from  two different
instruments: the  Geneva Observatory's photoelectric spectrometer CORAVEL
\citep{bmp79} mounted on the 1.5 m Danish  telescope at Cerro La Silla, Chile,
and the Rutgers Fabry-Perot spectrometer \citep[e.g.,][]{geb94} at CTIO.
Details of the CORAVEL data are given by \citet{may83,may84} and
\citet{mdm91}; typically, the measurements
for relatively bright giants and subgiants have uncertainties of
$\simeq0.6$ km s$^{-1}$. The Fabry-Perot velocities include some previously
published \citep{geb95}, and some newly measured (Gebhardt et al.~2006, in
preparation) from two runs at CTIO in 1995: one using the 1.5-meter telescope
on 16--22 June 1995, and one with the 4-meter on 6--7 July 1995.
The velocities normally have precisions of 1 km s$^{-1}$ or better, depending
on the stellar magnitude and crowding.

All together, there are nearly 5,600 stars with CORAVEL and/or Fabry-Perot
radial velocities, with Fabry-Perot contributing a large majority of the
data. However, each of these samples covers a much larger area
of 47 Tuc than does our proper-motion sample. We work here only with
a much smaller subset of stars which lie within a radial distance of
105\arcsec\ of the cluster center and thus overlap the proper-motion field
(cf.~Figure \ref{fig:cmdpos}). In addition, we found it necessary to exclude
from our analysis any Fabry-Perot stars fainter than the horizontal branch in
47 Tuc (which is also roughly the limiting magnitude of the CORAVEL data).
This drastically reduces the sample size and thus requires
some explanation.

The basic strategy of Fabry-Perot observations is described in
\citet{geb94,geb95,geb97}. The idea is to scan an etalon across a strong
absorption line (H$\alpha$ in our case) to build a small spectrum. The
instrumental resolution of the Rutgers system is around 5000, and the
FWHM of the H$\alpha$ line is larger than this. For the 4-meter observations
of 47 Tuc, 21 steps across the H$\alpha$ line were used, while the 1.5-meter
run observed two fields using 17 and 41 steps. The full range in all three
cases was about 5\AA. The exact wavelength
coverage of any particular star depends on its location in the field, due to
a wavelength gradient introduced by the Fabry-Perot. Given the full set of
scans, DAOPHOT \citep{pbs87} and ALLFRAME \citep{pbs98} were used to
determine the brightness of each star in the field. The scans were then
combined to make a small spectrum, which was fitted for the H$\alpha$ velocity
centroid.

There are particular advantages and disadvantages of this technique compared
to the traditional slit \'echelle spectroscopy employed by CORAVEL.
The latter provides
large wavelength coverage, but it is obviously necessary to choose specific
stars on which to place the slits. With this configuration, the observer has
no
control over what light goes down the slit, and neighboring stars can be
be a severe contaminant in crowded regions.  By contrast, Fabry-Perot
spectroscopy provides only a small wavelength coverage but once the scans are
completed, every star in the field produces a spectrum. This provides an
enormous advantage in terms of the number of velocities that can be obtained.
In principle, difficulties due to crowding should also be reduced
since the use of DAOPHOT allows one to apportion the light appropriately in
crowded regions, i.e., accurate photometry can be obtained for faint stars near
bright ones.

Crowding inevitably remains an issue, however, with its main effect being
a progressively stronger bias in the measured velocity dispersion at fainter
magnitudes. If the spectrum of a faint star is affected by either a bright
neighbor or the collective light of the bulk of unresolved cluster stars,
the measured velocity is pulled closer to the cluster mean, and the velocity
dispersion that follows is spuriously low. In the context of our work in
\S\ref{subsec:distance}, where we compare proper-motion to radial-velocity
dispersions to estimate the distance to 47 Tuc, this effect would ultimately
lead to a short value. There are two options for avoiding this. One is to
remove the effect statistically, by using detailed simulations to define
empirical, magnitude-dependent corrections for the stellar velocities. This
approach is pursued by Gebhardt et al.~(2006, in preparation). The other
option is simply to make a magnitude cut in the velocity sample, keeping only
those stars which are bright enough not to be significantly contaminated in
the first place. This is what we have done here.

\begin{figure}[!t]
\vspace*{-0.3truecm}
\hspace*{+0.2truecm}
\centerline{\resizebox{90mm}{!}{\includegraphics{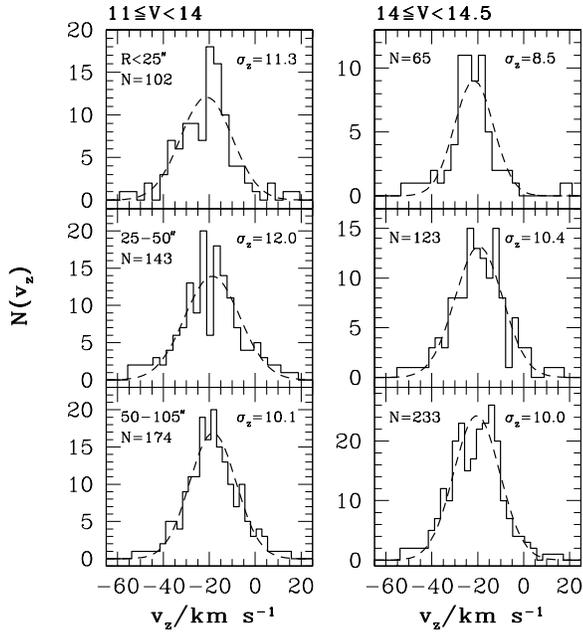}\hfil}}
\caption{
Histograms of CORAVEL and Fabry-Perot radial velocities in the inner
105\arcsec\ of 47 Tuc, comparable to the area covered by our proper-motion
survey. Left-hand panels are for stars brighter than the cluster horizontal
branch, and right-hand panels are for stars just 0.5 mag fainter. The
sharp drop in the (error-corrected) dispersion $\sigma_z$ from the bright to
the faint magnitudes---especially at the smaller clustercentric radii---signals
a crowding-induced bias in the Fabry-Perot velocities.
\label{fig:vzhist}}
\end{figure}

Figure \ref{fig:vzhist} shows the distribution of the combined CORAVEL and
Fabry-Perot radial velocities for stars in three bins of clustercentric
radius. The left-hand panels include only stars with $11\le V<14$, a magnitude
range that has been well explored in previous radial-velocity studies, and
which we have confirmed does not suffer significantly from the contamination
problems just discussed. The limit $V\ge 11$ excludes only a handful of
measured stars in this region of the cluster but is imposed to guard against
radial-velocity ``jitter'' in the brightest cluster giants \citep{gg79}.
The faint limit $V<14$ corresponds to the magnitude of the horizontal branch.
The right-hand panels then include stars just slightly fainter than this, in
the range $14\le V<14.5$.

We have calculated the error-corrected velocity dispersion $\sigma_z$ (see
eq.~[\ref{eq:sigmu}]) for the stars in each of the magnitude ranges and radial
bins defined in Figure \ref{fig:vzhist}. These are listed in the appropriate
panels of the figure, where we have also drawn (strictly for illustrative
purposes) Gaussians with dispersions $\sigma_z$ on top of the velocity
histograms. Comparing these results for
the two magnitude bins at any given radius immediately shows the downward
biasing of $\sigma_z$ caused by crowding. Comparing the different radii
confirms this interpretation of the situation, since (1) the difference
between $\sigma_z$ for the bright and faint samples decreases toward larger
clustercentric radii, where the stellar densities are lower; and (2) the
velocity dispersion at the smallest radii $R < 25\arcsec$ for the fainter
stars especially is dramatically (and unphysically) lower than that at larger
$R$. Although $\sigma_z$ for the stars with $V<14$ is formally lower at
$R<25\arcsec$ than in the range $25\arcsec\le R<50\arcsec$, by some
0.7 km~s$^{-1}$, the uncertainty in the calculated dispersion is also
0.7--0.8 km~s$^{-1}$ in both radial ranges. Thus, the difference in
this case is not highly significant and is likely a reflection of small-number
statistics rather than serious contamination.

Given the results in Figure \ref{fig:vzhist}, we choose to include only stars
brighter than the horizontal branch, $11\le V<14$, in the radial-velocity
sample used in this paper. There are only 419 such stars in the inner
105\arcsec\ of 47 Tuc, and thus any kinematic estimate of the distance 
is fundamentally limited to a precision of no better than
$(2\times419)^{-1/2}\simeq3.5\%$. Clearly, it is desireable to do better than
this, and in principle it is possible to enlarge the radial-velocity sample
by, say, applying position-dependent magnitude cuts to remove contaminated
stars and/or using data from clustercentric radii beyond the proper-motion
field. However, either of these things would require a sophisticated
analysis of the Fabry-Perot data and an in-depth dynamical modeling that are
well beyond the scope of this paper.

\subsubsection{Overlap with the Proper-Motion Sample}
\label{subsubsec:pmrv}

\begin{figure}[!t]
\hspace*{+0.1truecm}
\centerline{\resizebox{75mm}{!}{\includegraphics{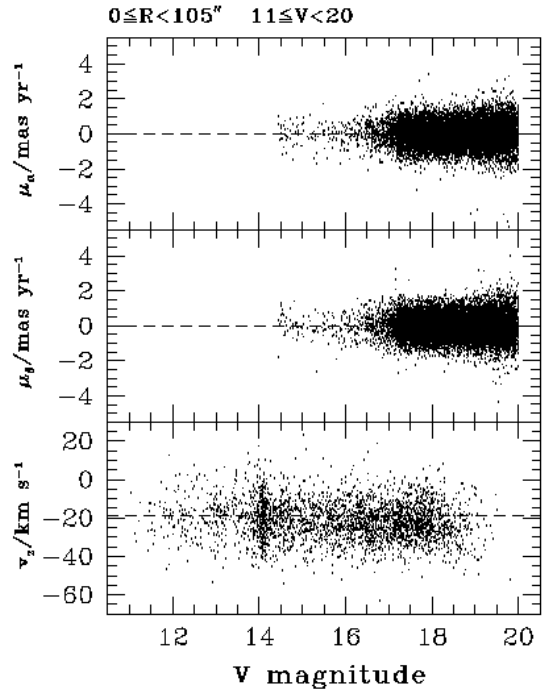}\hfil}}
\caption{
Velocity vs.\ magnitude for stars in our ``good'' proper-motion sample [$V<20$
and $P(\chi^2)\ge 0.001$ for both $\mu_\alpha$ and $\mu_\delta$; from Table
\ref{tab:posdata}], and for all stars at $R<105\arcsec$ in the CORAVEL and
Fabry-Perot
radial-velocity sample. Since the biasing problem illustrated in Figure
\ref{fig:vzhist} precludes the use of any line-of-sight velocities when
$V>14$, we do not currently have a single star for which all three components
of velocity have been reliably measured.
\label{fig:velmags}}
\end{figure}

Figure \ref{fig:velmags} shows the RA and Dec components of proper motion
vs.\ stellar magnitude for our ``good'' proper-motion sample of
\S\ref{subsubsec:kinsamp} [i.e., those stars in Table \ref{tab:posdata} with
$P(\chi^2)\ge 0.001$ and $V<20$], and also for the combined CORAVEL and
Fabry-Perot line-of-sight velocity sample in the inner $R<105\arcsec$ of 47
Tuc. Again, we do not make use here of any data at $V>14$ in the latter case,
but Figure \ref{fig:velmags} emphasizes that radial velocities can indeed be
measured for many such faint stars. The great potential of these data, if
their systematic problems can be solved, is clear.

Figure \ref{fig:velmags} also illustrates the fact that there is currently not
one star in 47 Tuc which
has reliable measurements for all three components of its velocity. As was
also mentioned in \S\ref{subsubsec:kinsamp}, stars brighter than $V=14$ were
not measured in any ACS frames, and thus do not appear in our proper-motion
sample; but fainter than this limit we cannot trust the
radial-velocity measurements in general. As a result, only statistical
comparisons
of motions on the plane of the sky and along the line of sight are feasible at
this point.

\begin{figure}[!t]
\vspace*{-1.2truecm}
\hspace*{+0.2truecm}
\centerline{\resizebox{95mm}{!}{\includegraphics{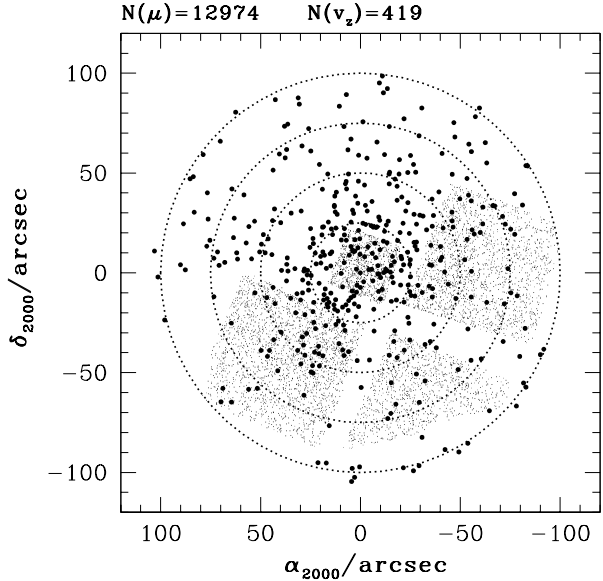}\hfil}}
\caption{
Comparison of the spatial distribution of our good proper-motion sample and
the useable radial-velocity stars. All positions are plotted relative to the
cluster center (\S\ref{subsubsec:center}), and for reference the large, dotted
circles have been plotted with radii of 25\arcsec, 50\arcsec, 75\arcsec, and
100\arcsec.
\label{fig:muvzpos}}
\end{figure}

Finally, Figure \ref{fig:muvzpos} compares the spatial distribution of the
419 stars whose radial velocities can be used for a distance estimate
(large points), to that of our good
proper-motion sample (small dots, plotted for only a fraction of the 12,974
stars to outline the shape of the field). The circles drawn on
this graph have radii of 25\arcsec\ (which completely encloses the PC-chip
area of the proper-motion field), 50\arcsec, 75\arcsec, and 100\arcsec. As we
discuss further in \S\ref{subsec:distance}, the
obvious differences between the non-uniform spatial coverages of the two
kinematics samples mean that some care must be used in properly comparing
the measured velocity dispersions to derive a distance to the cluster.

\section{Spatial Structure and a Model Velocity Distribution}
\label{sec:space}

To optimally explore the implications of the velocity data that we have
assembled, we would like to relate them to a dynamical model
for 47 Tuc. Ideally, this should be constrained by the spatial distribution
of stars as a function of magnitude (mass) and allow for the
possibilities of rotation, velocity anisotropy, axisymmetry (rather than
sphericity), the presence of heavy stellar remnants, and so on. The proper
construction of such a comprehensive, global model is clearly a large task
that would involve bringing in more data than we have collected here
(e.g., the spatial and velocity distributions of stars at radii beyond the
$\simeq100\arcsec$ extent of our proper-motion sample; information on the
stellar luminosity function as a function of position in the cluster; etc.).
Our much more modest goal in this paper is to try and extract as much
model-independent information as possible from our proper motions, and simply
to check whether they are broadly consistent with a very low-level
description of the cluster.

The zeroth-order model we use as a reference point is the
familiar single-mass, isotropic, modified isothermal sphere of \citet{king66}.
Note that \citet{mey88,mey89} has already fitted 47 Tuc with
more realistic models of multimass and anisotropic clusters, but
using only radial velocities and the surface-brightness profile of subgiants
and giants as observational constraints. Thus, applying \citeauthor{mey88}'s
results as they
are to assess fine details of our proper-motion data would not really be any
more appropriate than using a single-mass and isotropic model (and re-fitting
them would, as we've suggested, involve work beyond the scope of this paper).
At any rate, \citeauthor{mey88}'s models imply that velocity anisotropy is
only important at large radii $R\ga500\arcsec$, i.e., in regions of
the cluster not probed by our proper motions.

Similarly, \citet{mey86} and \citet{ak2003a} have quantified the rotation of 47
Tuc, using both line-of-sight velocities and proper motions. However, in the
innermost $R\la 100\arcsec$ covered by our new velocity data, rotation appears
to be present only at the level of $\la3$ km s$^{-1}$, which is only a
fraction of the velocity dispersion there (see \citealt{mey86}, and
\S\ref{subsec:norotation} below). Thus,
comparison of the proper-motion kinematics with a non-rotating and spherical
model can still be profitable in this region.

In \S\ref{subsec:density}, then, we use our master star list from
\S\ref{subsubsec:masterlist} to construct a new density profile for the
brightest stars in the core of 47 Tuc, and combine this with wider-field,
ground-based surface photometry to fit a \citet{king66} model to the cluster's
spatial structure. In \S\ref{subsec:modvel} and Appendix
\ref{sec:kingveldist}, we discuss the proper-motion
velocity distribution predicted by such a model.

\subsection{Density Profile of Turn-off Stars}
\label{subsec:density}

The most important concern for us in constructing a surface-density profile
from our master star list is the question of mass segregation. This involves
massive stars (or heavy remnants) sinking toward the cluster center and
following a more concentrated radial distribution than lower-mass stars, so we
cannot really speak of a single density profile for the cluster. In fact,
the data in Table \ref{tab:master} above could be used to explore
mass segregation in some detail, but once again it is beyond the scope this
paper. Some early work has been done by \citet{par95}, \citet{ak1996}, and
\citet{and97}.

The best constraint on the spatial structure of 47 Tuc outside the core
regions observed by HST is the $V$-band surface photometry
collated from a variety of sources by \citet{mey88}. The cluster light in this
band is dominated by giants and subgiants, all of which have roughly the same
(main-sequence turn-off) mass. Our immediate interest, then, is in defining an
improved profile in the central region for stars of this mass only. The
brightness of the main-sequence turn-off in 47 Tuc is $V=17.65$ (see Table
\ref{tab:basic}), corresponding roughly to magnitude $17.8$ in the F475W
bandpass for which our master star list has photometry. Thus, after first
checking that the radial distribution does not vary significantly as a
function of magnitude for stars brighter than the turn-off, we have used Table
\ref{tab:master} to construct a single density profile for all stars with 
$m_{\rm F475W}\le 17.8$. We emphasize that the full master star list has to be
used for this, and {\it not} any sample of only proper-motion stars from Table
\ref{tab:posdata}, because the latter have a highly irregular distribution on
the sky and a complicated (and unquantified) incompleteness as a function of
magnitude and clustercentric position.

We first define a series of concentric circular annuli $\{R_j\}$, all contained
entirely within the area of the master frame.
The estimates $c_i$ of completeness percentage for individual stars in the
master list (Column 10 of Table \ref{tab:master}; see
\S\ref{subsubsec:compfrac}) are then used to calculate the density in each
annulus: if there are ${\cal N}_{\rm raw}$ stars with $m_{\rm F475W}\le 17.8$
in the range $R_j\le R< R_{j+1}$, then the completeness-corrected number is
${\cal N}_{\rm cor}=\sum_{i=1}^{{\cal N}_{\rm raw}} (100/c_i)$ and the net
density follows as $\Sigma={\cal N}_{\rm cor}/\pi (R_{j+1}^2-R_j^2)$.
The results of this simple exercise are given in Table \ref{tab:density}.
Note, from comparison of the listed ${\cal N}_{\rm raw}$ and
${\cal N}_{\rm cor}$, that the corrections for incompleteness in this
magnitude range are quite modest over our entire field.
Also, no corrections are made for contamination by foreground or background
stars, since the level of such contamination is negligible in this
small area of the sky; see Table \ref{tab:basic} again.

\setcounter{table}{5}
\begin{deluxetable}{ccrrc}
\tabletypesize{\small}
\tablewidth{0pt}
\tablecaption{Density Profile of Cluster Stars With
              $F475W\le17.8$ \label{tab:density}}
\tablecolumns{5}
\tablehead{
\colhead{Annulus}              &
\colhead{$R_{\rm eff}$}        &
\colhead{${\cal N}_{\rm raw}$} &
\colhead{${\cal N}_{\rm cor}$} &
\colhead{$\Sigma(R_{\rm eff})$}              \\
\colhead{[arcsec]}        &
\colhead{[arcsec]}        &
\colhead{}                &
\colhead{}                &
\colhead{[arcmin$^{-2}$]}
}
\startdata
  0.00--1.00  &  0.707 &     11 &   11.601 &   $13294\pm3903$ \\
  1.00--2.00  &  1.581 &     20 &   21.262 &    $8121\pm1761$ \\
  2.00--3.00  &  2.550 &     58 &   62.564 &   $14339\pm1813$ \\
  3.00--4.00  &  3.536 &     55 &   58.602 &    $9593\pm1253$ \\
  4.00--5.00  &  4.528 &     74 &   79.081 &   $10069\pm1132$ \\
  5.00--6.00  &  5.523 &     61 &   65.143 &     $6786\pm841$ \\
  6.00--7.00  &  6.519 &     96 &  102.398 &     $9026\pm892$ \\
  7.00--8.00  &  7.517 &    105 &  111.938 &     $8551\pm808$ \\
  8.00--9.00  &  8.515 &    121 &  128.596 &     $8668\pm764$ \\
  9.00--10.00 &  9.513 &    118 &  124.707 &     $7521\pm674$ \\
 10.00--11.25 & 10.643 &    170 &  179.638 &     $7750\pm578$ \\
 11.25--13.00 & 12.157 &    231 &  243.714 &     $6581\pm422$ \\
 13.00--14.50 & 13.770 &    230 &  241.322 &     $6704\pm432$ \\
 14.50--16.25 & 15.400 &    258 &  270.542 &     $5761\pm350$ \\
 16.25--18.75 & 17.545 &    373 &  389.131 &     $5096\pm258$ \\
 18.75--21.25 & 20.039 &    421 &  437.655 &     $5015\pm240$ \\
 21.25--25.00 & 23.201 &    676 &  699.629 &     $4623\pm175$ \\
 25.00--30.00 & 27.613 &    777 &  797.922 &     $3325\pm118$ \\
 30.00--35.00 & 32.596 &    786 &  802.515 &     $2830\pm100$ \\
 35.00--40.00 & 37.583 &    698 &  709.851 &      $2169\pm81$ \\
 40.00--45.00 & 42.573 &    687 &  696.558 &      $1878\pm71$ \\
 45.00--50.00 & 47.566 &    641 &  648.666 &      $1565\pm61$ \\
 50.00--57.50 & 53.881 &    949 &  959.069 &      $1363\pm44$ \\
 57.50--65.00 & 61.365 &    867 &  874.353 &      $1091\pm37$ \\
 65.00--72.50 & 68.852 &    888 &  894.248 &   $993.7\pm33.2$ \\
 72.50--82.50 & 77.661 &    988 &  993.452 &   $734.5\pm23.3$ \\
\enddata

\tablecomments{Number densities in last column are
$\Sigma(R)\equiv{\cal N}_{\rm cor}/[\pi (R_2^2-R_1^2)]$ with $R_1$ and $R_2$
the inner and outer radii of an annulus (Column 1), in units of arcmin.
Uncertainty in $\Sigma(R)$ is $\sqrt{{\cal N}_{\rm cor}}/[\pi (R_2^2-R_1^2)]$.
To transform into $V$-band surface brightness on the scale of Meylan (1988)
(uncorrected for reddening), set
$\mu_V[{\rm mag\ arcsec^{-2}}]=(24.237\pm0.021) -
 2.5\,\log(\Sigma/{\rm arcmin^{-2}})$.
}
\end{deluxetable}

It remains only to combine these star counts with the wider-field
$V$-band surface photometry taken from \citet{mey88}. We first
define an effective radius for each annulus, $R_{{\rm eff},j}^2 \equiv
(R_j^2+R_{j+1}^2)/2$. The number density at each $R_{\rm eff}$ is then
converted to a surface brightness as
$\mu(R_{{\rm eff},j})=C-2.5\,\log \Sigma(R_{{\rm eff},j})$, where $C$ is
a constant
determined by the overlap of our data with the calibrated
\citeauthor{mey88} photometry. In practice, we only
use the ground-based surface brightnesses $\mu_V$ at clustercentric radii
$R\ge 20\arcsec$ (to guard conservatively against any potential seeing-induced
blurring of the central profile) and obtain $C$ as the
median value of $[\mu_V+2.5\,\log \Sigma(R_{\rm eff})]$ for all of our data
points
with $20\arcsec\le R_{\rm eff}\le 70\arcsec$. We find that the new HST number
counts are brought nicely onto the $V$-band surface-brightness scale of
\citet{mey88} (uncorrected for reddening) with the transformation
\begin{equation}
\label{eq:sb}
\mu_V = (24.237\pm0.021) - 2.5\, \log (\Sigma/{\rm arcmin}^{-2})\ .
\end{equation}

\begin{figure}[!t]
\vspace*{-0.3truecm}
\hspace*{+0.2truecm}
\centerline{\resizebox{90mm}{!}{\includegraphics{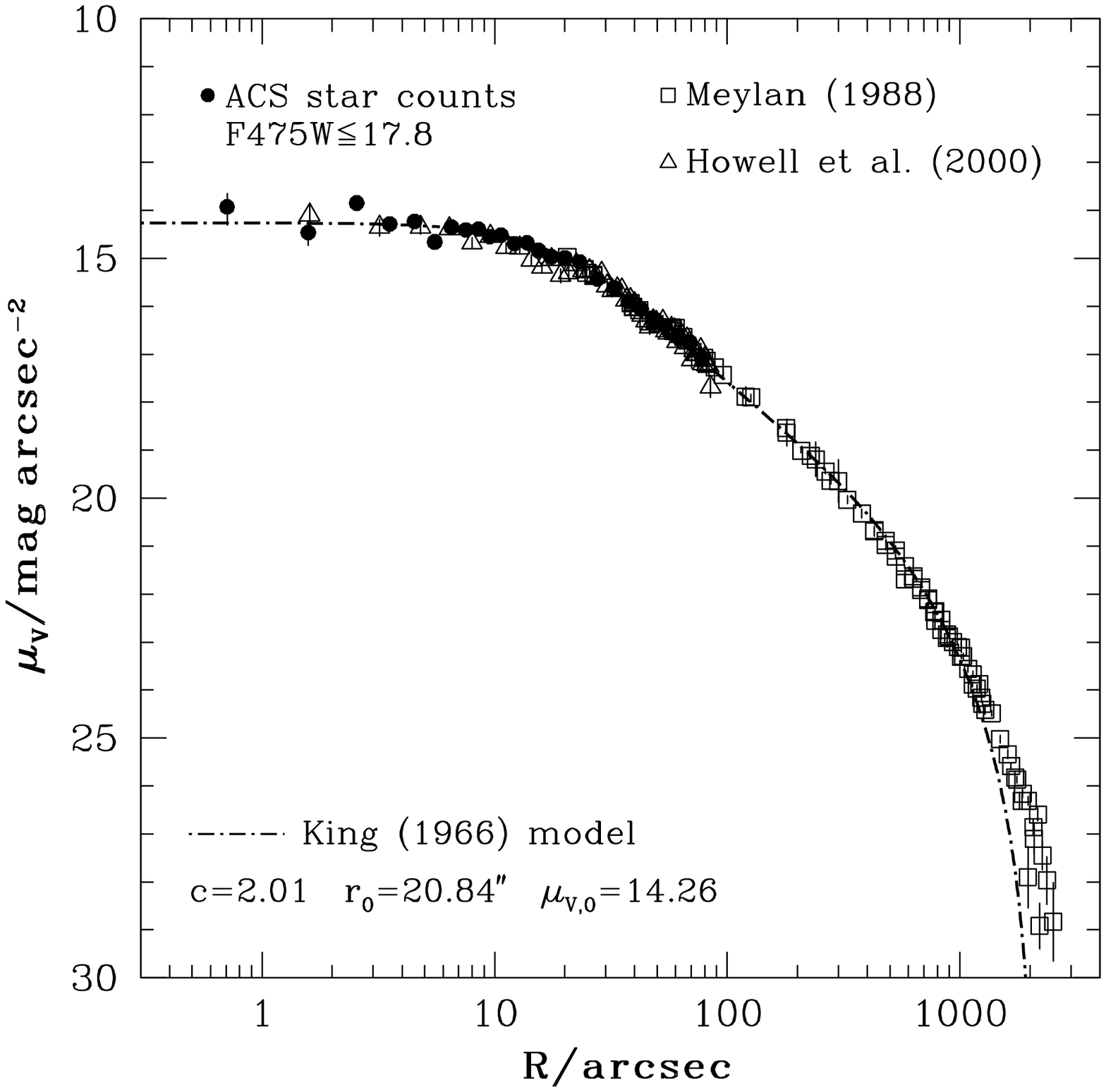}\hfil}}
\caption{
$V$-band surface-brightness profile of 47 Tuc, constructed by combining the
ground-based surface photometry compiled by \citet{mey88} at $R\ge20\arcsec$,
with HST star counts at $R\la 100\arcsec$ from \citet{how00} and our own
master list (Table \ref{tab:master}). Only stars brighter than the
main-sequence turn-off were included in the HST number-density profiles, which
were converted to $\mu_V$ as described in the text. The bold, dash-dot curve
is the fit of a single-mass, isotropic, \citet{king66} model, with the
parameters indicated (also listed in Table \ref{tab:basic}). This model only
attempts to describe the cluster structure at small radii. The poor fit at
$R\ga 1000\arcsec$ is due to neglecting velocity anisotropy and multiple
stellar masses, but the disagreement with $\mu_V$ in the extreme halo does not
impact the core region $R<100\arcsec$ covered by our proper-motion data.
\label{fig:density}}
\end{figure}

Figure~\ref{fig:density} shows the composite surface-brightness profile for 47
Tuc. Open squares represent the data from
\citet{mey88}, while the filled circles correspond to our star counts after
the application of equation (\ref{eq:sb}). We have also plotted, as the open
triangles, earlier HST-based number densities of turn-off stars from
\citet[][their Figure 4]{how00}, transformed by
$\mu_V = (14.927\pm0.033)-2.5\,\log (\Sigma_{\rm HGG}/{\rm arcsec}^{-2})$
to give the best
match to the ground-based surface photometry at $20\arcsec\le R\le 70\arcsec$.
The agreement between the two HST data sets is very good overall, confirming
the conclusion of \citet{how00} that 47 Tuc has a smooth, well-defined,
and reasonably large core that shows no sign of the cluster
having undergone core collapse.

The dot-dash curve in Figure \ref{fig:density} is our fit of a single-mass,
isotropic, \citet{king66} model to the combination of our data points with the
brightness profile of \citet{mey88}. Only data at $R<1100\arcsec$ were
used to constrain the fit since, as we discussed above, we only require
a good description of the central regions covered by our proper-motion
data. Our best estimates of the core radius $r_0$, central surface brightness
$\mu_{V,0}$, and concentration parameter $c\equiv\log(r_t/r_0)$ (for $r_t$ the
tidal radius of the model) are written on Figure \ref{fig:density} and also
listed in Table \ref{tab:basic} above. Note that the fitted $c=2.01$
corresponds to a dimensionless central potential $W_0=8.6$
(see Appendix \ref{sec:kingveldist}). We estimate uncertainties of about
$\pm25\%$ in $r_0$, $\pm0.25$ mag in $\mu_{V,0}$, and $\pm0.1$ dex in $c$
(or $\pm0.4$ in $W_0$). (Note that these uncertainties are correlated, such
that fits with higher $c$ tend to require smaller $r_0$ and brighter
$\mu_{V,0}$.) Thus, while the new values of $r_0$ and $\mu_{V,0}$ here are
formally slightly smaller and brighter than the ``standard'' values in the
literature \citep[e.g.,][]{har96,how00}, the differences are at the
$\sim 1$-$\sigma$ level. Our core radius is in good agreement with the
$r_0=21\arcsec$ recently derived by \citet{map04}, also from HST star counts.

Given our neglect of the data at large $R\ge1100\arcsec$ when fitting the
\citet{king66} model in Figure \ref{fig:density}, its failure in these
outermost parts of the cluster is not surprising. In fact, no
single-mass and isotropic \citeauthor{king66} model can satisfactorily fit the
entire run of surface brightness throughout 47 Tuc. This is already well known
and forms the basis of the multimass and anisotropic modeling of this cluster
by \citet{mey88,mey89}.

\subsubsection{Effect of Mass Segregation}
\label{subsubsec:multir0}

As we suggested above and will show explicitly in \S\ref{sec:veldisp},
a model that assumes velocity isotropy in order to give a description of the
density distribution over radii $R\la 1000\arcsec$ is appropriate enough as a
benchmark in the discussion of our proper-motion observations at $R\la
100\arcsec$. More worrisome in principle is the assumption of a single
stellar mass and the implication that stars in any magnitude range should
have a common spatial structure. This is patently untrue even in the core
of a relaxed cluster; but even so, it turns out that the proper-motion stars
in our case do not, in fact, span a particularly wide range in mass. 

To assess this issue briefly, Table \ref{tab:masslum} summarizes a
theoretical mass--$V$ magnitude relation for stars from 0.5--$0.9\,M_\odot$.
In the second column of the table, the absolute magnitudes associated with
$m_*\le0.8\,M_\odot$ are those calculated by \citet{bar98} for a 12.6-Gyr old
cluster with ${\rm [M/H]}=-0.5$. For $0.8\,M_\odot<m_*<0.9\,M_\odot$
we have relied on the models of \citet{berg01}, as illustrated by
\citet{bril04}, for a 12-Gyr old cluster with
${\rm [Fe/H]}=-0.83$ and ${\rm [\alpha/Fe]}=+0.3$.
The apparent $V$ magnitudes in the third column of Table \ref{tab:masslum}
follow from $M_V$ given a reddening of $E(B-V)=0.04$ mag and an assumed
heliocentric distance $D=4$ kpc to 47 Tuc (see Table \ref{tab:basic}, and
\S\ref{subsec:distance} below). The simple formula on the bottom line,
for $m_*$ as a function of $V$, is our own approximation. As
\S\ref{subsec:pmsample} discussed in detail, our proper-motion sample is
limited by the constraint $V<20$, which corresponds to stellar masses
$0.65\,M_\odot\la m_*\la 0.9\,M_\odot$. The large majority of the
stars for which
we can meaningfully address kinematics questions differ in mass by $< 20\%$
from the main-sequence turn-off and giant-branch stars
($V\la 17.65$, so $m_*\simeq0.85$--$0.9\,M_\odot$) for which we fit
the \citet{king66} model in Figure \ref{fig:density}.

\citeauthor{king66} models with high concentration such as the one in Figure
\ref{fig:density} are not substantially different, in their inner parts, from
bounded (finite central density) isothermal spheres. Crude expectations
for the relative density profiles at small radii for stars of different
masses in 47 Tuc can therefore be obtained by considering models of
two-component clusters in this idealized limit.
\citet{taff75} show that two populations of stars with masses $m_2\ge m_1$
in an equilibrium isothermal sphere have density distributions
related by $\rho_2(r)/\rho_2(0) = [\rho_1(r)/\rho_1(0)]^{m_2/m_1}$.
As expected, then, $\rho_2(r)$
falls off more rapidly than $\rho_1(r)$ and the heavier stars are more
strongly concentrated toward the center. To quantify this, note that the
scale $r_0$ defined by
\citet[][see our equation (\ref{eq:kingr0}) below]{king66}
is, by construction, nearly equal to the radius at which the surface density
of an isothermal sphere falls to half its central value. As a result, it is
also related by a unique numerical factor
to the radius at which the volume density is halved. Thus, we have
calculated $\rho/\rho(0)$ and $[\rho/\rho(0)]^{m_2/m_1}$ vs.\ $(r/r_0)$ for an
isothermal sphere, for a range of fixed mass ratios, and located the
dimensionless
radii at which the dimensionless $\rho$ and $\rho^{m_2/m_1}$ fall (separately)
to $1/2$. For modest values of $m_2/m_1$, the
core radii of the two mass classes turn out to be related roughly by
\begin{equation}
\label{eq:multir0}
r_{0,2}/r_{0,1} \simeq \left(m_2/m_1\right)^{-0.58}\ ,
\end{equation}
accurate to better than 1\% for $1\le m_2/m_1\le 3$ and in good agreement with
the alternative approximation developed by \citet[][their equation 5a]{grin02}.

Returning to 47 Tuc, we do not expect equation (\ref{eq:multir0}) to hold
exactly; but in the absence of detailed modeling it is useful as an
order-of-magnitude guide to the situation {\it in the core region} where all
mass classes are expected to be nearly thermalized (with the velocity
dispersion of each species scaling as $\sigma^2\sim m_*^{-1}$). For example,
the core radius $r_0$ of the faintest stars in our proper-motion sample
should be only slightly larger than that of the turn-off stars with
$m_*\simeq0.85\,M_\odot$: $(0.65/0.85)^{-0.58}=1.17$.
Even the ``average'' cluster stars with $m_*\simeq0.5\,M_\odot$ should be
distributed with a scale radius differing by only $\sim35\%$
[$(0.5/0.85)^{-0.58}=1.36$] from that of the density profile shown in
Figure \ref{fig:density}. We will refer back to
this approximate relation on occasion below, when discussing various aspects
of the velocity distribution of our proper-motion stars.
First, we describe the expected form of this distribution for an
isotropic and single-mass \citeauthor{king66}-model cluster.

\begin{deluxetable}{ccc}
\tablecaption{Stellar Mass-Luminosity Relation \label{tab:masslum}}
\tablewidth{0pt}
\tablecolumns{3}
\tablehead{
\colhead{$m_{*}/M_\odot$} & \colhead{$M_V$} & \colhead{$V=(M_V+13.13)$}
}
\startdata
0.9 & $\la$3.9\phantom{1} &  $\la$17.0\phantom{0}   \\
0.8 & \phantom{$\la$}5.01 & \phantom{$\la$}18.14    \\
0.7 & \phantom{$\la$}6.27 & \phantom{$\la$}19.40    \\
0.6 & \phantom{$\la$}7.58 & \phantom{$\la$}20.71    \\
0.5 & \phantom{$\la$}8.85 & \phantom{$\la$}21.98    \\
 & & \\
\multicolumn{2}{c}{$m_*/M_\odot \simeq 0.9 \phantom{\ -0.081\,(V-17)}$}
                             & $V\la17.0$ \\
\multicolumn{2}{c}{$\phantom{m_*/M_\odot \ }\simeq  0.9-0.081\,(V-17)$}
                             & $V>17.0$ \\
\enddata

\tablecomments{Based on the evolutionary models of \citet{bar98} and
\citet[][see also \citealt{bril04}]{berg01}.
Apparent magnitudes in Column (3) correspond to the
absolute magnitudes in Column (2) for an assumed distance of $D=4.0$ kpc
to 47 Tuc and a reddening of $E(B-V)=0.04$ mag.}
\end{deluxetable}

\subsection{Projected Velocity Distribution}
\label{subsec:modvel}

As was noted above, the inner regions of \citet{king66} models with high
concentrations (or central potentials) are dynamically quite similar to
isothermal spheres. Given that our proper-motion sample covers only (parts of)
the innermost $R\la 5r_0$ in 47 Tuc, we first ask how different from Gaussian
we might expect the observed velocity distribution to be. Very simple
considerations suggest immediately that even for a sample of $\sim$10,000
stars the departure from Gaussianity is somewhat subtle.

First,
\citeauthor{king66} models are defined by a finite escape velocity at every
radius in the cluster. If $W(r)=-\phi(r)/\sigma_0^2$, where $\phi(r)$ is the
gravitational potential and $\sigma_0$ is a parameter closely related (though
not equal) to the central velocity dispersion, then the model escape velocity
at radius $r$ is $v_{\rm max}/\sigma_0=\sqrt{2W(r)}$, which is a
monotonically decreasing function of radius (see Appendix
\ref{sec:kingveldist} for more detail). A star moving outward from radius $r$
with velocity $v_{\rm max}(r)$ would just reach the cluster tidal radius with
zero velocity. Our proper-motion field contains the center of 47 Tuc, where
$W(r)=W_0=8.6\pm0.4$ for the model fit in Figure \ref{fig:density}. Thus, this
model implies that we should see no stars with total proper-motion speed
greater than $\simeq4.15\sigma_0$.

Second, if the velocity distribution were perfectly Gaussian and
isotropic, then the probability of observing a star with relative speed
$\mu_{\rm tot}=(\mu_x^2+\mu_y^2)^{1/2}$ on the plane of the sky would be
\begin{equation}
\label{eq:gausstot}
P_G(\mu_{\rm tot})=\frac{\mu_{\rm tot}}{\sigma_{\rm iso}^2}\,
   \exp\left(-\mu_{\rm tot}^2/2\sigma_{\rm iso}^2\right)\ ,
      \qquad \mu_{\rm tot}\ge 0
\end{equation}
Integrating this, only a tiny fraction $e^{-(4.15)^2/2} \simeq
1.8\times10^{-4}$ of all stars are expected, on average, to have
$\mu_{\rm tot}>4.15\sigma_{\rm iso}$. This shows that simply counting the
numbers of stars with very high velocities in a sample of 10,000 at the
center of a cluster like 47 Tuc cannot easily distinguish between the usual
Gaussian approximation and regular \citeauthor{king66}-model behavior.

Nevertheless, the possibility exists that rather more exotic
physics---such as a central black hole, or a high rate of stellar
collisions---could strongly influence the dynamics, and so we would
like to assess the significance of any fast-moving
stars in the core of 47 Tuc as carefully as we reasonably can.
Appendix \ref{sec:kingveldist} therefore develops
expressions defining the normalized
probability distribution of observable velocities as a function of projected
clustercentric radius in \citet{king66} models. Equation
(\ref{eq:onednorm}) gives $N_1(v_z | R/r_0)$, the distribution of
line-of-sight velocity---or of any {\it one} component of proper motion---at
fixed position $R/r_0$ ($r_0$ is the cluster core radius, determined by
fitting to the surface-density profile as above). Equation
(\ref{eq:twodnorm}) gives $N_2(\mu_x,\mu_y | R/r_0)$, the joint distribution
of any two orthogonal components of velocity on the plane of the sky
[e.g., we could equally well take $(\mu_x,\mu_y)$ to be the RA and Dec proper
motions, or the radial and tangential velocities relative to the cluster
center]. We will work here mainly with the latter function, but note that the
operations we perform on it can also be applied to the one-dimensional
distribution.

As Appendix \ref{sec:kingveldist} discusses, isotropy ensures that the
the proper-motion components $\mu_x$ and $\mu_y$ only appear in the
(position-dependent) velocity distribution $N_2$ in the combination
$(\mu_x^2+\mu_y^2)\equiv \mu_{\rm tot}^2$. We would like to know the
total number of stars, in a sample of size
${\cal N}_{\rm tot}$ distributed between projected radii $R_1\le R\le R_2$,
that lie
within a velocity interval $d\mu_x d\mu_y$ about the speed $\mu_{\rm tot}$. If
$\Sigma(R)$ is the observed projected number density of the velocity
sample, then this spatially-averaged velocity distribution is
\begin{equation}
\label{eq:twodaveideal}
N(\mu_{\rm tot}) = \int_{R_1}^{R_2} N_2(\mu_x,\mu_y | R/r_0)\ 
    \Sigma(R)\, 2\pi R\, dR\ ,
\end{equation}
which clearly requires the appropriate \citeauthor{king66}-model $r_0$ to be
known. Note that $N(\mu_{\rm tot})$ has dimensions of inverse velocity squared
($N_2$ contains a factor $\sigma_0^2$ in its denominator) and satisfies
$\int\!\!\int N(\mu_{\rm tot})\,d\mu_xd\mu_y =
\int N(\mu_{\rm tot})\,2\pi\mu_{\rm tot}d\mu_{\rm tot}={\cal N}_{\rm tot}$.
It is effectively a ``density'' in the $\mu_x$--$\mu_y$ velocity plane.

\begin{figure}[!t]
\vspace*{-0.7truecm}
\hspace*{+0.1truecm}
\centerline{\resizebox{90mm}{!}{\includegraphics{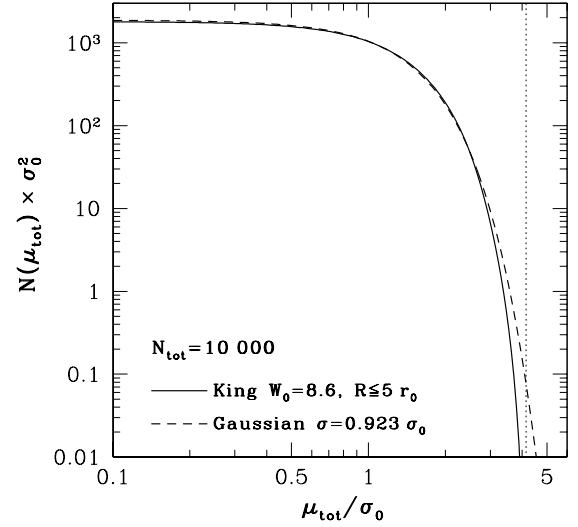}\hfil}}
\vspace*{-0.8truecm}
\caption{
Comparison of the projected, two-dimensional velocity distribution averaged
over the inner 5 core radii of a $W_0=8.6$ \citet{king66} model, vs.\ an
isotropic, bivariate Gaussian distribution with the same average
dispersion. Both theoretical curves are normalized to a total population
of 10,000 stars, similar to our proper-motion sample. Significant differences
between the two appear only at high speeds $\mu_{\rm tot}\ga2\,\sigma_0$,
above which relatively few stars are expected in either case. The vertical,
dotted line is at $\mu_{\rm tot}=4.15\sigma_0$, the maximum velocity allowed
at the center of a $W_0=8.6$ \citeauthor{king66} model. The Gaussian shown
here would predict $<$1 star in 10,000 to be found above such a speed.
\label{fig:modnvel}}
\end{figure}

Figure \ref{fig:modnvel} shows the distribution
$\sigma_0^2 \times N(\mu_{\rm tot})$ for ${\cal N}_{\rm tot}=10,000$ stars
distributed between $0\le R\le 5 r_0$ according to the surface density
profile $\Sigma(R)$ of a \citet{king66} model with $W_0=8.6$---meant to
recall our proper-motion sample. This is the solid curve in the
plot. The vertical dotted line marks
the location of the maximum possible total speed in this model,
$v_{\rm max}/\sigma_0=\sqrt{2W_0}=4.15$. For comparison, the dashed curve
traces the isotropic bivariate Gaussian density
\begin{equation}
\label{equation}
N_G(\mu_{\rm tot}) = \frac{{\cal N}_{\rm tot}}{2\pi\sigma_{\rm iso}^2}
     \exp\left(-\mu_{\rm tot}^2/2\sigma_{\rm iso}^2\right)\ ,
\end{equation}
for $\sigma_{\rm iso}=0.923\sigma_0$ equal to the density-weighted average of
the projected \citeauthor{king66}-model dispersion ($\sigma_z^2$, defined in
eq.~[\ref{eq:modsigz}])
over the inner five core radii. Evidently, the differences between the
two curves are not easily visible---let alone statistically
significant---until $\mu_{\rm tot}\ga3\sigma_0$. The Gaussian distribution
would predict only $e^{-4.5}\times {\cal N}_{\rm tot}\simeq 111$ stars at
such high speeds; the \citeauthor{king66} model, fewer still.

\subsubsection{Accounting for Measurement Errors and Irregular Spatial
  Distributions}
\label{subsubsec:modconvol}

While informative, the idealized calculations just above still lack any
recognition of the facts that observed velocities are subject to measurement
error, and that a kinematical sample of stars need not trace the overall
stellar density profile (as our proper-motion sample clearly does not;
e.g., see Figure \ref{fig:muvzpos} above). Both of these complications will
affect the overall shape of the \citeauthor{king66}-model distribution
$N(\mu_{\rm tot})$ and alter the relative strength of its wings.

To deal with these issues, consider a set of stars with proper-motion
measurements and uncertainties
$\{\mu_{x,i}, \mu_{y,i}, \Delta_{x,i}, \Delta_{y,i}\}$, and clustercentric
radii $\{R_i\}$, where $i=1,\dots,{\cal N}_{\rm tot}$. For each individual
star we know the ideal \citeauthor{king66}-model distribution
$N_2(\mu_x,\mu_y | R_i/r_0)$ from equation \ref{eq:twodnorm}, and the
bivariate Gaussian distribution $P_i(\delta\mu_x,\delta\mu_y)$ of
velocity measurement errors. The predicted observable probability density
at $R_i$ is then the convolution of $N_2$ and $P_i$. Strictly speaking, the
velocity errorbars $\Delta_{x,i}$ and $\Delta_{y,i}$, which appear in $P_i$,
are not exactly equal for all stars, but in general they are nearly so. We
therefore define (for these purposes only) a single errorbar
$\overline{\Delta}_i^2\equiv(\Delta_{x,i}^2+\Delta_{y,i}^2)/2$ for
both velocities of each star, so that the approximate measurement-error
distributions
\begin{equation}
\label{eq:bierrdist}
P_i(\delta\mu_x,\delta\mu_y) = \frac{1}{2\pi \overline{\Delta}_i^2}
     \exp\left[-\frac{(\delta\mu_x)^2 +
     (\delta\mu_y)^2}{2\overline{\Delta}_i^2}\right]
\end{equation}
are isotropic. Then the error-convolved velocity distribution at star $i$,
$$
N_2^* (\mu_x,\mu_y | R_i/r_0, \overline{\Delta}_i) =
 \qquad\qquad \qquad\qquad \qquad\qquad \qquad\qquad
$$
\begin{equation}
\label{eq:twodconv}
 \qquad
\int\!\!\!\int_{-\infty}^{\infty}
               N_2\left(\mu_x^\prime,\mu_y^\prime | R_i/r_0\right)
               P_i\left[(\mu_x-\mu_{x}^\prime),(\mu_y-\mu_{y}^\prime)\right]
       \ d\mu_x^\prime d\mu_y^\prime\ ,
\end{equation}
remains isotropic as well, i.e., it still depends on $\mu_x$ and $\mu_y$ only
through $\mu_{\rm tot}$.

Although cumbersome, equation (\ref{eq:twodconv}) is straightforward to
compute for every star once $r_0$ is known. (Note that the integrals are not
actually infinite, since the original $N_2$ in the integrand vanishes for
proper-motion speeds above the local maximum velocity.)
Then, rather than assume that the proper-motion
stars accurately follow the self-consistent \citeauthor{king66}-model
surface-density profile, we simply find the velocity distribution for stars
distributed on the sky exactly as the observed ones:
\begin{equation}
\label{eq:twodave}
N(\mu_{\rm tot}) = \sum_{i=1}^{{\cal N}_{\rm tot}}
       N_2^*(\mu_x,\mu_y | R_i/r_0, \overline{\Delta}_i)\ .
\end{equation}
It is this last equation (rather than the idealized
eq.~[\ref{eq:twodaveideal}]) that we use to produce models for comparison
against the observed proper-motion distributions in the next Section.

\section{The Observed Proper-Motion Velocity Distribution}
\label{sec:veldist}

\subsection{One-Dimensional Distributions}
\label{subsec:onedvel}

The proper motions in Table \ref{tab:posdata} are given in terms of their
RA (positive Eastward) and Dec (positive Northward) components, which we
write as $(\mu_\alpha, \mu_\delta)$. In a dynamical context, however, it is
more meaningful to refer the velocities to the cluster center. Thus, if the
coordinates of any point are written in terms of the projected clustercentric
radius $R$ and the azimuth $\Theta$ (measured in degrees North of East), it
is a simple matter to make the transformation
\begin{equation}
\label{eq:vels}
\begin{array}{rcl}
\mu_R      & = & \phantom{-} \mu_\alpha\cos\Theta + \mu_\delta\sin\Theta \\
\mu_\Theta & = & -\mu_\alpha\sin\Theta + \mu_\delta\cos\Theta\ .
\end{array}
\end{equation}
The uncertainties in $\mu_R$ and $\mu_\Theta$ then follow from the usual rules:
\begin{equation}
\label{eq:errs}
\begin{array}{rcl}
\Delta_R^2      & = &
               \Delta_\alpha^2 \cos^2\Theta + \Delta_\delta^2 \sin^2\Theta \\
\Delta_\Theta^2 & = &
               \Delta_\alpha^2 \sin^2\Theta + \Delta_\delta^2 \cos^2\Theta\ .
\end{array}
\end{equation}
Note that $\mu_R$ is positive for motion outward from the cluster center
on the plane of the sky; $\mu_\Theta$ is positive for clockwise motion; and
by the right-hand rule, the line-of-sight velocity $v_z$ is positive for
motion away from the observer.

\begin{figure*}[!t]
\vspace*{-0.7truecm}
\hspace*{+0.1truecm}
\centerline{\resizebox{160mm}{!}{\includegraphics{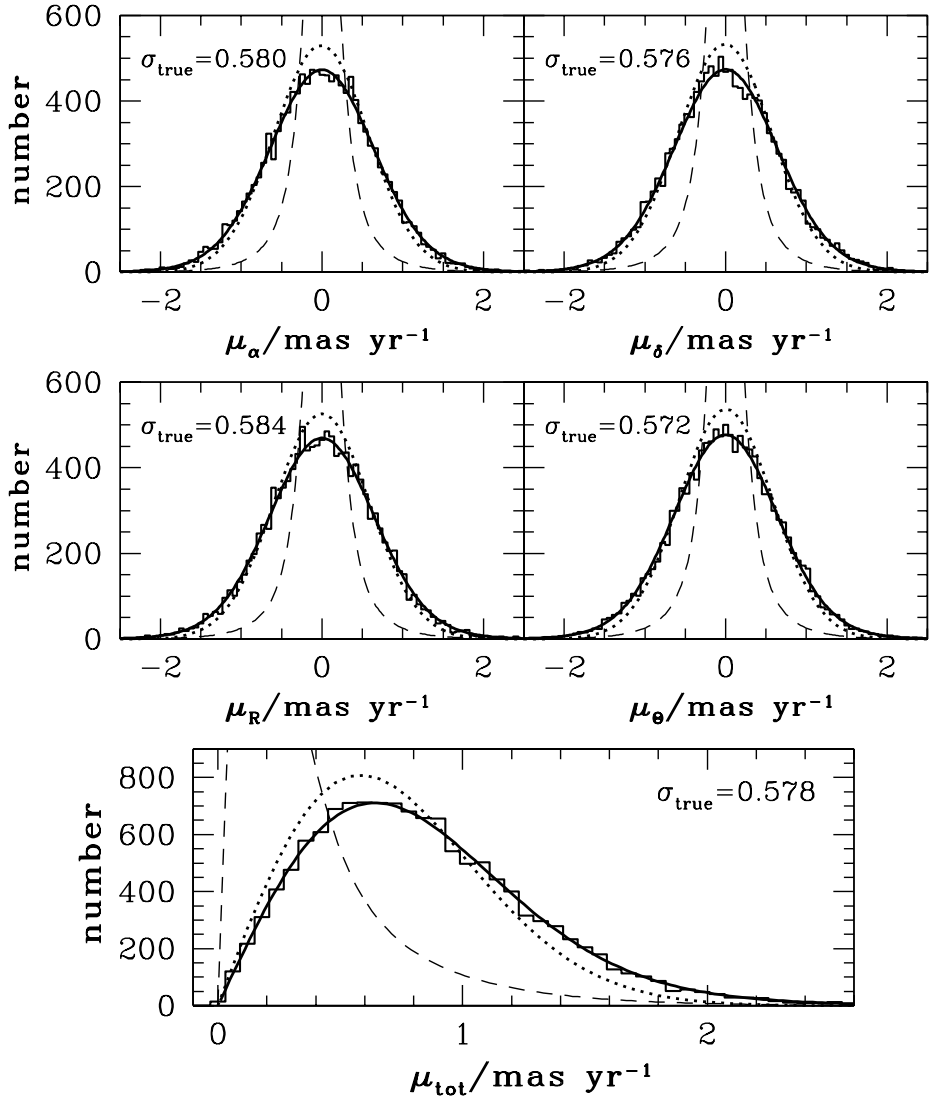}\hfil}}
\caption{
One-dimensional distributions of velocities in the RA and Dec directions and
in the projected radial and azimuthal directions relative to the cluster
center, as well as the distribution of speeds
$\mu_{\rm tot}=(\mu_\alpha^2+\mu_\delta^2)^{1/2}$, all for the sample of
12,974 ``good'' proper motions. Bold, dotted curves in all panels are
Gaussians with the error-corrected dispersions $\sigma_{\rm true}$ indicated
(from eq.~[\ref{eq:sigtrue2}]), while narrower, long-dashed curves are the
velocity measurement-error distributions. Bold, solid curves are the
convolutions of the two. The good agreement with the data in all cases
indicates that the intrinsic velocity distribution in the core of 47 Tuc is
nearly Gaussian, at least for proper-motion speeds $\mu_{\rm tot}\la2$--2.5
mas yr$^{-1}$. Isotropy is also indicated by the similarity of
$\sigma_{\rm true}$ in the four directions shown.
\label{fig:muhists}}
\end{figure*}

To investigate the distribution of these velocities, we first look at the
entire sample of all ``good'' proper motions---the 12,974 stars in Table
\ref{tab:posdata} which have $P(\chi^2)\ge0.001$ for both the fitted
$\mu_\alpha$ and $\mu_\delta$, and magnitudes $V<20$ (see
\S\ref{subsec:pmsample}). Figure \ref{fig:muhists} displays a number of
histograms of the one-dimensional velocities for these stars. The top panels
give the separate $\mu_\alpha$ and $\mu_\delta$ distributions, the middle
panels present those for the $\mu_R$ and $\mu_\Theta$ components, and the
bottom panel shows the histogram of speeds
$\mu_{\rm tot}=(\mu_\alpha^2+\mu_\delta^2)^{1/2} =
(\mu_R^2+\mu_\Theta^2)^{1/2}$. For reference, speeds in mas yr$^{-1}$ are
converted to km s$^{-1}$ by
\begin{equation}
\label{eq:convert}
1\ {\rm mas\ yr}^{-1} =
     4.74\,(D/{\rm kpc})\ {\rm km\ s}^{-1}
\end{equation}
for a distance $D$ to 47 Tuc. With $D\simeq4$ kpc (Table \ref{tab:basic}),
$1\ {\rm mas\ yr}^{-1}\simeq 19\ {\rm km\ s}^{-1}$.

We have calculated the first two moments of each of the first four velocity
distributions in Fig.~\ref{fig:muhists}.
All of the means are consistent with zero---as they should
be, given that we have used each star's nearest neighbors to derive velocities
relative to the net cluster motion on the sky (see \S\ref{subsec:multiepoch}
and Appendix \ref{sec:errcor}).
We have estimated the {\it true} velocity dispersion in
each component $\mu_\alpha$, $\mu_\delta$, $\mu_R$, and $\mu_\Theta$, using
equation (\ref{eq:sigtrue}):
$$
\sigma_{\rm true}^2 = 
    \left(1+\frac{1}{45}\right)^{-1} \times
 \qquad\qquad \qquad\qquad \qquad\qquad \qquad\qquad
$$
\begin{equation}
\label{eq:sigtrue2}
 \quad
    \left\{
    \frac{1}{{\cal N}_{\rm tot}-1} \left[\sum_{i=1}^{{\cal N}_{\rm tot}}
          (\mu_{i,\,{\rm obs}} -
           \langle\mu\rangle_{\rm obs})^2\right]
  - \frac{1}{{\cal N}_{\rm tot}}\sum_{i=1}^{{\cal N}_{\rm tot}} \Delta_i^2
    \right\} \ ,
\end{equation}
where ${\cal N}_{\rm tot}=12,974$; $\Delta_i$ is the uncertainty in the
proper-motion component $\mu_i$; and the leading factor corrects an
unavoidable error introduced by our local transformations
(as discussed in Appendix \ref{sec:errcor}).

The true one-dimensional dispersions computed from this formula are listed
in the appropriate panels of Figure \ref{fig:muhists}. The uncertainty in
each is about $\pm0.004$ mas yr$^{-1}$, and thus the four different
proper-motion components are all consistent with a single, average
$\sigma_{\rm true}=0.578\ {\rm mas\ yr}^{-1} =
11.0\,(D/4\,{\rm kpc})\ {\rm km\ s}^{-1}$. This is the first and simplest
direct evidence that the velocity distribution in the core of 47 Tuc is
(as expected) isotropic. This conclusion is justified more rigorously in
\S\ref{subsec:kinmags} below, where we more carefully examine velocity
dispersion as a function of stellar magnitude and clustercentric radius.

Meanwhile, each of the top four panels in Figure \ref{fig:muhists} has three
curves drawn in it. The dotted curves are Gaussians with dispersion equal to
$(1+1/45)^{1/2}\times \sigma_{\rm true}$, while the much narrower, dashed
curves are the measurement-error distributions given by the sum of Gaussians
in equation (\ref{eq:errdist}), scaled up by the total number of stars.
The thick, solid lines represent the convolution of the ``intrinsic''
dotted Gaussians with the normalized error distributions. These convolutions
are themselves the sum of Gaussians. They closely follow all of the observed
histograms, showing that the intrinsic velocity distribution in the
core of 47 Tuc is in fact very nearly Gaussian. This is {\it not assumed} in
the derivation of eq.~(\ref{eq:sigtrue2}) in Appendix \ref{sec:errcor} but,
from the discussion in \S\ref{subsec:modvel}, it is just what we expect even
if the cluster actually has the structure of a \citet{king66} model.

The curves in the bottom panel of Figure \ref{fig:muhists} assume the
velocity isotropy suggested by the four upper panels. With the
measurement-error distribution given by the sum of Gaussians, if the intrinsic
velocity distribution were also Gaussian then the expected distribution of
observed speeds would be the convolution (cf.~eq.~[\ref{eq:gausstot}])
\begin{equation}
\label{eq:gaussconvol}
{\cal N}_{\rm tot} P^*_G(\mu_{\rm tot}) = \sum_{i=1}^{{\cal N}_{\rm tot}}
       \frac {\mu_{\rm tot}}{\sigma_{\rm iso}^2+\overline{\Delta}_i^2}
       \exp\left(
        -\frac{\mu_{\rm tot}^2}{\sigma_{\rm iso}^2+\overline{\Delta}_i^2}
            \right)  \ ,  \qquad \mu_{\rm tot}\ge 0
\end{equation}
where
$\overline{\Delta}_i^2=(\Delta_{\alpha,i}^2+\Delta_{\delta,i}^2)/2$ is
the isotropized velocity uncertainty of each star. The dotted curve in the
bottom panel of Figure \ref{fig:muhists} is this equation evaluated with
$\sigma_{\rm iso}=(1+1/45)^{1/2}\times0.578$ but $\Delta_i=0$ for all stars,
i.e., it is the Gaussian approximation to the intrinsic
distribution of speeds in our proper-motion sample (with the true dispersion
inflated by transformation error). Conversely, the dashed
curve is equation (\ref{eq:gaussconvol}) with $\sigma_{\rm iso}=0$ but
using the set of $\{\Delta_i\}$ actually measured for the stars in our
sample; this is
the distribution of speeds that would obtain if all apparent motion were due
solely to measurement error. The heavy solid curve running through the
histogram is the full equation (\ref{eq:gaussconvol}), with
$\sigma_{\rm iso}=(1+1/45)^{1/2}\times0.578$ and each $\Delta_i$ as observed.
The main effect of velocity measurement error is perhaps most clearly seen by
comparing the dotted and solid curves in this representation of the data: it
works to take power from the peak of the intrinsic velocity distribution and
move it out to the wings.

\subsection{Two-Dimensional Distributions}
\label{subsec:twodvel}

The distributions in Figure \ref{fig:muhists} are, of course, averages over
the irregular spatial distribution of our velocity sample and a wide range
of stellar magnitudes, $14<V<20$. We expect the velocity distribution to vary
with clustercentric radius and, if mass segregation and energy equipartition
have taken hold to any extent, with stellar mass. It has to be
emphasized in particular that the average dispersion $\sigma_{\rm true}=0.578$
mas yr$^{-1}$ is only an indicative number and is not suitable, e.g., for
direct comparison with the radial-velocity dispersion of giant-branch stars
to estimate a distance to the cluster.

We therefore look now at the velocity distribution for stars in a number of
distinct bins in magnitude and radius. To do this, we take advantage of the
isotropy indicated by Figure \ref{fig:muhists}, and confirmed in
\S\ref{subsec:kinmags} below, to construct the two-dimensional joint
distribution of $(\mu_\alpha,\mu_\delta)$, or equivalently
$(\mu_R,\mu_\Theta)$, rather than the one-dimensional histograms for
individual components. We also switch to compare the data explicitly with
the velocity distributions expected in a \citet{king66} model cluster with
a finite escape velocity.

For these purposes we first restrict the sample of ``good'' proper motions from
Table \ref{tab:posdata} further, by imposing the color
criterion
\begin{equation}
\label{eq:colcut}
\begin{array}{ll}
1.5 \le (U-V) < 2.6\ , & \qquad V  <  16.75 \\
0.6 \le (U-V) < 1.9\ , & \qquad V \ge 16.75
\end{array}
\end{equation}
which is designed to exclude blue-straggler stars and extreme outliers from
the main cluster sequence in the CMD of Figure \ref{fig:cmdpos}. This leaves us
12,791 stars to work with. Given this sample, we define a series of total
proper-motion speeds $\{\mu_{{\rm tot}, j}\}$ and choose a range of stellar
magnitude and clustercentric radius to examine. We calculate the observed
speed for each star in the specified range of $V$ and $R$; count
the number which fall in each
interval $[\mu_{{\rm tot},j}, \mu_{{\rm tot},j+1})$; and divide the population
of each velocity interval by the ``area''
$\pi(\mu_{{\rm tot},j+1}^2-\mu_{{\rm tot},j}^2)$. The result is an isotropized
``density'' profile in projected velocity space, $N(\mu_{\rm tot})$, which can
be compared directly to model distributions of the type discussed in
\S\ref{subsec:modvel}. Uncertainty in the observed $N(\mu_{\rm tot})$ is
estimated by generating 1000 artificial sets of velocity data from the
subsample of real stars in any magnitude/radius bin
\citep[using a standard bootstrap with replacement; e.g.,][Section 15.6]
{press92} and calculating $N(\mu_{\rm tot})$ for each artificial set to find
the 68\% confidence intervals on the actual distribution.

\subsubsection{$N(\mu_{\rm tot})$ vs.~Magnitude for Stars at all Radii
  $R<100\arcsec$} 

\begin{figure}[!t]
\vspace*{-0.3truecm}
\hspace*{+0.1truecm}
\centerline{\resizebox{85mm}{!}{\includegraphics{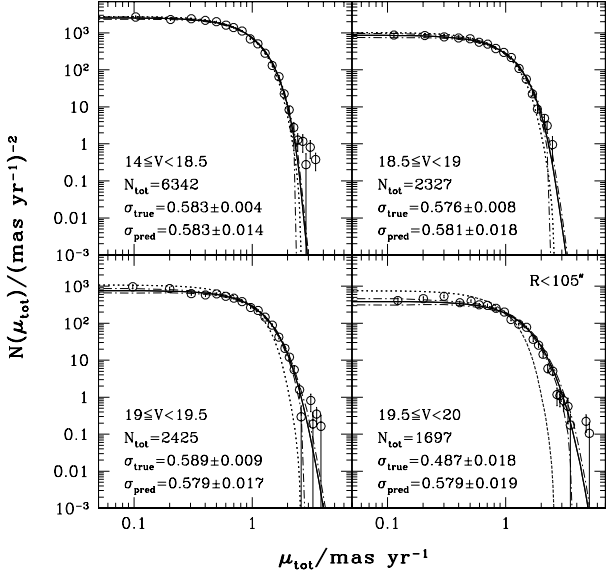}\hfil}}
\vspace*{-0.3truecm}
\caption{
Two-dimensional velocity distributions for stars at all observed
clustercentric radii, divided into four magnitude bins. The total number of
good proper motions and the oberved, error-corrected $\sigma_{\rm true}$ in
each bin are given in each panel, as is the predicted single-mass, isotropic
\citeauthor{king66}-model dispersion accounting for the stars observed
positions. The dotted curve in each case is the intrinsic model velocity
distribution convolved with the observed spatial distribution
(\S\ref{subsec:modvel}). The bold curve is this model convolved with the
velocity measurement-error distribution, and the dash-dot curves are its 68\%
confidence limits. Note the excess stars in the high-velocity tails, and the
fact that the kinematics at $V\ge19.5$ are error-dominated in general.
\label{fig:nmumag}}
\end{figure}

Figure \ref{fig:nmumag} shows the observed
$N(\mu_{\rm tot})$ for the proper-motion sample of 12,791 stars over
our entire field, divided into four broad magnitude bins.
The curves in all panels of the figure are the theoretical distributions for
a \citet{king66} model with $W_0=8.6$ and $r_0=20\farcs84$, computed as
described in Appendix \ref{sec:kingveldist} and \S\ref{subsec:modvel}. In
order to normalize these curves to the data, the model velocity scale
$\sigma_0$ must be known (see eq.~[\ref{eq:twodnorm}]), and we evaluate
this as follows.

Given $W_0$ and $r_0$, we have computed the dimensionless projected
velocity-dispersion profile $\sigma_{\rm mod}^2(R/r_0)/\sigma_0^2$ (given as
$\sigma_z^2$ in 
eqs.~[\ref{eq:modsig}] and [\ref{eq:modsigz}]). Then it is straightforward to
calculate the predicted \citeauthor{king66}-model velocity dispersion for any
collection of stars with a range of observed clustercentric radii:
\begin{equation}
\label{eq:sigpred}
\frac{\sigma_{\rm pred}^2}{\sigma_0^2} =
\frac{1}{{\cal N}} \sum_{i=1}^{{\cal N}}
  \frac{\sigma_{\rm mod}^2(R_i/r_0)}{\sigma_0^2} \ .
\end{equation}
At the same time, we can estimate the true projected velocity
dispersion in any direction on the plane of the sky, corrected for both
measurement and transformation errors, from equation
(\ref{eq:sigtrue2}). Doing this separately for both the RA and Dec (or both
the $R$ and $\Theta$) components of proper motion yields the improved
value $\sigma_{\rm true}^2=(\sigma_{{\rm true}, \alpha}^2 +
\sigma_{{\rm true}, \delta}^2)/2$, which we have calculated for the stars
in each of the magnitude bins in  Figure \ref{fig:nmumag}.\footnotemark
  \footnotetext{To prevent contamination by extreme velocity outliers, we
  estimate $\sigma_{\rm true}$ using only stars with
  total speeds satisfying $\mu_{\rm tot}<3.0$ mas yr$^{-1}$ if $V<19$; 
  $\mu_{\rm tot}<3.2$ mas yr$^{-1}$ if $19\le V<19.5$; and
  $\mu_{\rm tot}<3.6$ mas yr$^{-1}$ if $19.5\le V<20$. The upper limit
  increases with magnitude in recognition of the larger measurement errors
  for fainter stars.}
In particular, for the stars with $V<18.5$ throughout our
entire field (the subsample shown in the top left panel of Figure
\ref{fig:nmumag}), we find $\sigma_{\rm true}=0.583\pm0.004$ mas yr$^{-1}$,
while using their positions in equation (\ref{eq:sigpred}) gives
$\sigma_{\rm pred}^2/\sigma_0^2=0.849\pm0.015$. Requiring that
$\sigma_{\rm pred}=\sigma_{\rm true}$ for these stars
then implies
\begin{equation}
\label{eq:sigma0}
\sigma_0 = 0.633\pm0.010 \ {\rm mas\ yr}^{-1}
         = (12.0\pm0.2)\,(D/4\,{\rm kpc})\ {\rm km\ s}^{-1}\ .
\end{equation}
Note that this quantity really is only a normalization factor---nowhere in a
\citeauthor{king66} model does the velocity dispersion actually equal
$\sigma_0$---and that its value here is specific to the values $W_0=8.6$ and
$r_0=20\farcs84$.

Each panel in Figure \ref{fig:nmumag} lists the error-corrected
$\sigma_{\rm true}$ for the
stars in question, and $\sigma_{\rm pred}$ from equations (\ref{eq:sigpred})
and (\ref{eq:sigma0}). The errorbars on
$\sigma_{\rm true}$ are 68\% confidence intervals obtained numerically
from the same bootstrap experiments used to estimate the uncertainties
in the full velocity distributions. The {\it dotted} curve in each panel is
the intrinsic \citeauthor{king66}-model
$N(\mu_{\rm tot})=\sum_{i=1}^{{\cal N}_{\rm tot}} N_2(\mu_x,\mu_y | R_i/r_0)$,
which is free of velocity-measurement errors but is calculated using an
inflated $\sigma_0=(1+1/45)^{1/2}\times0.633=0.640$ mas yr$^{-1}$ to account
for the error caused by local coordinate transformations
(see \S\ref{subsubsec:local} and Appendix \ref{sec:errcor}). The
{\it solid} line in each case is the model curve after convolution with our
measurement errors (see \S\ref{subsec:modvel}), and the {\it dash-dot} lines
connect its one-sigma uncertainties
(which reflect Poisson noise in the number of stars
predicted in each of the velocity bins that we defined to construct the
observed distribution).

The main conclusion to be taken from these graphs
is that, after convolution with the measurement errors, the
isotropic and single-mass model velocity distributions describe the
proper-motion data as a whole quite well for speeds $\mu_{\rm tot}\la 2$--2.5
mas yr$^{-1}$. This is essentially a re-statement of the good agreement with
Gaussians at these velocities in Figure \ref{fig:muhists}.
The poorest agreement is shown by the faint stars in the lower right-hand
panel of Figure \ref{fig:nmumag}, where the observed $N(\mu_{\rm tot})$ points
tend to fall above the model curve for $\mu_{\rm tot}\la 1$ mas yr$^{-1}$ and
slightly below it at larger speeds. This simply reflects
the discrepancy between the estimate of $\sigma_{\rm true}$ and the expected
$\sigma_{\rm pred}$ for these stars, but this is in turn mostly due to the
fact that---as either Figure \ref{fig:nmumag} or Figure \ref{fig:errmua} above
shows---the observed velocity distribution for $V\ge 19.5$ is strongly
dominated by measurement errors.

From our estimate of $\sigma_0$ in equation (\ref{eq:sigma0}) and the fact
that the maximum total velocity for bound stars at the center of a
\citet{king66} model is $v_{\rm max}/\sigma_0=\sqrt{2W_0}$, if our $W_0=8.6$
model for 47 Tuc were fully correct we would naively expect to see few or no
cluster members with true speeds $\mu_{\rm tot}\ga 2.6$ mas yr$^{-1}$. Indeed,
all the dotted curves in Figure \ref{fig:nmumag} fall to zero at about this
speed. Faster-moving stars are clearly found in our sample, however.
At relatively faint magnitudes $V\ga 19$, most of these are clearly
attributable to measurement error, which inevitably scatters some slow stars
to very high apparent speeds. Thus, although there are a few individual
exceptions, the majority of the
$N(\mu_{\rm tot})$ measurements in the tails of the lower panels of
Figure \ref{fig:nmumag} are statistically consistent with the
error-convolution of the model curves with $\mu_{\rm max}\simeq2.6$ mas
yr$^{-1}$. The same is basically true for the brighter bin
$18.5\le V<19$. It is only for the brightest subsample of stars with
$V<18.5$---for which the \citeauthor{king66} model we have fit should in fact
be most appropriate---that a statistically significant excess of
high-velocity stars is clearly evident.

We discuss these high-velocity stars in more detail in \S\ref{subsec:highpm}
just below but note here that, their specific properties aside, there
are very few of them indeed. Over all magnitudes $V<20$, only 46 of
the 12,974 good proper-motion stars in Table \ref{tab:posdata} have
observed $\mu_{\rm tot}>2.6$ mas yr$^{-1}$, and fewer than 10 of these
can currently be said unequivocally not to be due to velocity-measurement
error. There is, then, little to no evidence for a substantial nonthermal
population of stars in the core of 47 Tuc.

\subsubsection{$N(\mu_{\rm tot})$ vs. Magnitude in the Well-Measured Innermost
  20\arcsec} 

\begin{figure}[!t]
\vspace*{-0.3truecm}
\hspace*{+0.1truecm}
\centerline{\resizebox{85mm}{!}{\includegraphics{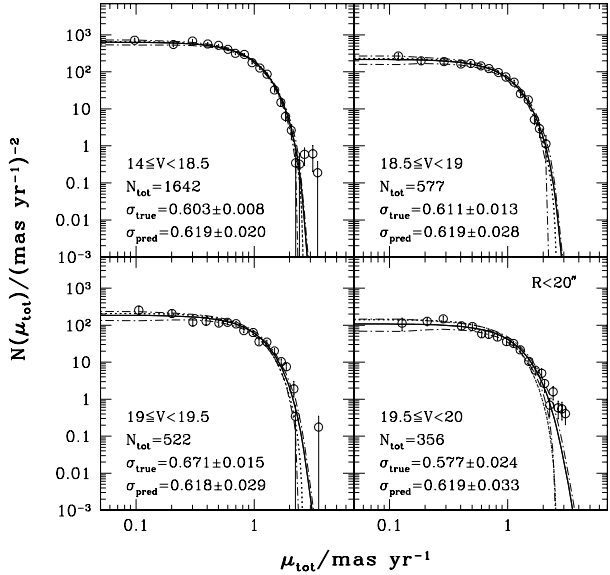}\hfil}}
\vspace*{-0.3truecm}
\caption{
Same as Figure \ref{fig:nmumag}, but only for stars in the innermost
$R<20\arcsec$, where velocity uncertainties are smallest. Note the constancy
in $\sigma_{\rm pred}$ as a function of stellar magnitude, reflecting the
single-mass assumption of the models. This contrasts with the rise in the
observed $\sigma_{\rm true}$ toward fainter $V$, presumably as a result of
mass segregation and equipartition of energy.
\label{fig:nmumagcore}}
\end{figure}

The \citeauthor{king66}-model $\sigma_{\rm pred}$ for the four magnitude bins
of Figure \ref{fig:nmumag} are very similar to each other because the model
assumes that stars of all magnitudes have the same mass and thus the same
spatial and velocity distributions (any slight differences in
$\sigma_{\rm pred}$ are solely due to differences in the spatial distributions
of the observed stars in the subsamples). What is perhaps slightly surprising
is that
the observed $\sigma_{\rm true}$ are also, with the exception of the faintest,
most error-prone stars, very similar both to each other and to the single-mass
model dispersions. From Table \ref{tab:masslum}, the stellar mass
decreases by $\simeq0.04\,M_\odot$ for each 0.5-mag increase in $V$.
Timescales in the dense core of 47 Tuc are short enough that
equipartition of energy has almost certainly been established
\citep{mey88,mey89}, in which case we should expect to
find $\sigma_{\rm true}\sim m_*^{-1/2}$ and thus a
systematic increase of roughly 3\% in velocity dispersion from panel to panel
in Figure \ref{fig:nmumag}. The fact that this is not observed suggests that
the effect may have simply been blurred out by velocity measurement errors.
This is further indicated by the significantly---and
unphysically---{\it lower} value of $\sigma_{\rm true}$ in the faintest
bin, $19.5\le V<20$.

Thus, in Figure \ref{fig:nmumagcore} we show the observed and model
$N(\mu_{\rm tot})$ in the same four magnitude bins as Figure \ref{fig:nmumag},
but now for 3,097 stars in just the innermost region of our field,
$R<20\arcsec$. As was discussed in \S\ref{subsubsec:pmerrors}, this is
where the velocity uncertainties are smallest for any stellar
magnitude (see  Figure \ref{fig:errmua}).
The error-corrected $\sigma_{\rm true}$ in each magnitude bin is again shown
in each panel of Figure \ref{fig:nmumagcore}, along with the model
$\sigma_{\rm pred}$ and $N(\mu_{\rm tot})$ curves for $\sigma_0=0.633$
mas yr$^{-1}$, all as above. The expected
increase of $\sigma_{\rm true}$ toward fainter magnitudes is in better
evidence now, and causes the observed $N(\mu_{\rm tot})$ in the faint bins
to be broader than the single-mass model distributions. Comparison of these
results with those in Figure \ref{fig:nmumag} suggests that it is this
central-most region, with its relatively precise proper motions, that
will provide the most useful kinematical constraints on detailed multimass
dynamical models for 47 Tuc.

\subsubsection{$N(\mu_{\rm tot})$ vs.~Radius for the Brightest Stars}

\begin{figure}[!t]
\vspace*{-0.3truecm}
\hspace*{+0.1truecm}
\centerline{\resizebox{85mm}{!}{\includegraphics{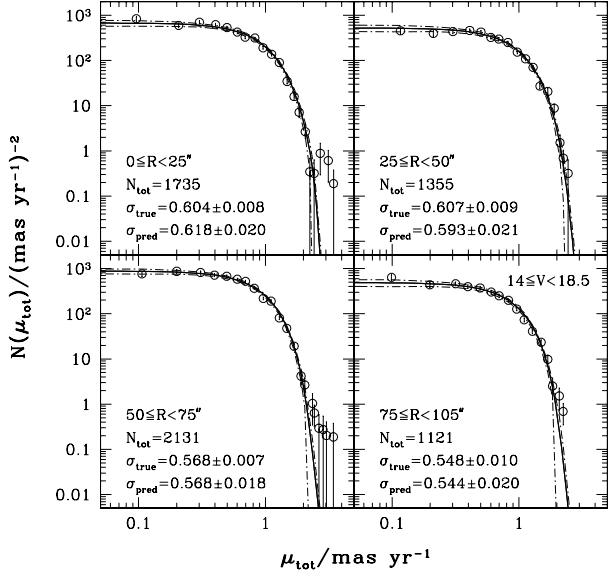}\hfil}}
\vspace*{-0.3truecm}
\caption{
Similar to Figures \ref{fig:nmumag} and \ref{fig:nmumagcore}, but now showing
the velocity distribution for bright stars only, in four intervals of
clustercentric radius. Measurement errors are small at these magnitudes
(Figure \ref{fig:errmua}), so for clarity only the fully error-convolved model
distributions and their 1-$\sigma$ confidence bands are drawn. The somewhat
slower fall-off of observed $\sigma_{\rm true}$ with radius, relative to the
model $\sigma_{\rm pred}$ implies that these main-sequence turn-off mass stars
are moving in a potential that is more extended than the one generated
self-consistently by their spatial distribution. Thus, the core of 47 Tuc
appears to be dominated by average (fainter) main sequence stars, rather than
by more spatially concentrated heavy remnants.
\label{fig:nmurad}}
\end{figure}

Finally, Figure \ref{fig:nmurad} presents the velocity distribution of the
6,342 brightest proper-motion stars in four concentric circular annuli
covering our observed field. From Table \ref{tab:masslum}, the magnitude
range $V<18.5$ chosen here corresponds to stellar masses
$m_*\simeq0.77$--$0.9\,M_\odot$. The mean magnitude is
$\langle V\rangle\simeq17.7$, and thus
$\langle m_*\rangle\simeq0.84\,M_\odot$---essentially the main-sequence
turn-off mass, and thus very similar
to the stars used to constrain the \citeauthor{king66} model fit in
\S\ref{subsec:density}. These are also the stars for which the velocity
errorbars are lowest (Fig.~\ref{fig:errmua}). In fact, for clarity in Figure
\ref{fig:nmurad} we only show the error-convolved
\citeauthor{king66}-model distributions and their one-sigma confidence bands,
as these distributions are not greatly different from the
intrinsic ones (dotted curves in Fig.~\ref{fig:nmumag} and
Fig.~\ref{fig:nmumagcore}).

Perhaps the most striking aspect of Figure \ref{fig:nmurad} 
is that the estimates of $\sigma_{\rm true}$ come within about 2\%
of the single-mass \citeauthor{king66}-model $\sigma_{\rm pred}$ at all radii.
That is, the kinematics of stars with the main-sequence turn-off mass are
described to rather high accuracy by a dynamical model which assumes that
their spatial distribution traces the distribution of total mass in the
cluster---even though this includes heavier neutron stars and white dwarfs
and a large number of much less massive main-sequence stars, all stratified by
mass segregation into different relative density profiles. At the same
time, however, the slight disagreement between the observed and predicted
velocity dispersions is informative. Looking at the top two panels in
particular, $\sigma_{\rm true}$ stays essentially constant as the
clustercentric radius increases out to 50\arcsec, while $\sigma_{\rm pred}$
falls by just over $4\%$, an amount which exceeds the observational errorbars
by a factor of $\sim$3. Thus, the stars in 47 Tuc are moving in a
net gravitational potential which is {\it less} centrally concentrated than
that generated by the turn-off and giant-branch stars alone.

This implies that the core of 47 Tuc {\it cannot} be dominated by heavy
stellar remnants such as neutron stars or white dwarfs, which would have
a significantly smaller core radius than the bright stars and lead to a steeper
drop-off in velocity dispersion with clustercentric radius (for neutron stars
with $m_*=1.4\,M_\odot$, eq.~[\ref{eq:multir0}] gives $r_0\simeq16\arcsec$
vs.\ the $20\farcs84$ measured for $m_*\simeq0.85$--$0.9\,M_\odot$).
This conclusion is
consistent with other lines of argument. The multimass and anisotropic
models of \citet{mey88,mey89} require, essentially to obtain a good fit to
the detailed shape of the surface-brightness profile in Figure
\ref{fig:density} above, that heavy stellar remnants make up only $\sim$0.1\%
or less of the current total mass of 47 Tuc. The most recent theoretical and
observational estimates of the total neutron-star population in 47 Tuc lie
in the range ${\cal N}_{\rm rem}\sim300$--1500, easily compatible with
\citeauthor{mey89}'s indirect
inference \citep{ivan05,hein05}. But in this case, even allowing for mass
segregation to place most of the remnants in the core leads to a structure
in which normal stars with $m_*\la0.85\,M_\odot$ dominate the gravity in
the inner few core radii by at least an order of magnitude
\citep[see Figure 5 of][]{mey88}. For an average stellar mass
around $m_*\simeq0.5\,M_\odot$, we expect the total mass distribution to
have a large core with $r_0\simeq27\farcs4$ (from eq.~[\ref{eq:multir0}]
again), which is then consistent with the shallow radial dependence of the
proper-motion $\sigma_{\rm true}$ in Figure \ref{fig:nmurad}. This will also
be seen, more quantitatively, in \S\ref{subsec:bhmods} below.

A second point to be taken from Figure \ref{fig:nmurad} is its
demonstration that the ``high-velocity'' stars---those which
contribute to $N(\mu_{\rm tot})$ measurements which lie significantly above
the tails of the \citeauthor{king66}-model curves---are not confined
exclusively to the very center of 47 Tuc but can be found out to 
five core radii away. This is, in fact, expected for proper-motion
high-velocity stars, whereas such stars observed in radial velocities are all
expected to be observed close to the cluster center in projection.
We now present more details on all of the very fast-moving stars in our
full sample.

\subsection{High-Velocity Stars}
\label{subsec:highpm}

As was discussed above, the nominal central escape velocity for our
\citet{king66} model of 47 Tuc is
$v_{\rm max}=\sqrt{2W_0}\times \sigma_0\simeq2.63\ {\rm mas\ yr}^{-1}$, or
$49.8\,(D/4\,{\rm kpc})\ {\rm km\ s}^{-1}$. The presence
of stars in our sample with observed proper-motion speeds above this value
is due in part to unavoidable measurement error (although Figure
\ref{fig:nmumag} shows that this cannot account for all of them), and at some
level it may also reflect the inevitable limitations of single-mass modeling
(as in Figure \ref{fig:nmumagcore}, fainter stars can attain broader velocity
distributions and still remain bound to the
cluster). Nevertheless, such stars are also of potential physical
interest as the possible byproducts of close stellar encounters and strong
scattering in the very dense core environment.

\begin{figure*}[!t]
\hspace*{+0.1truecm}
\centerline{\resizebox{160mm}{!}{\includegraphics{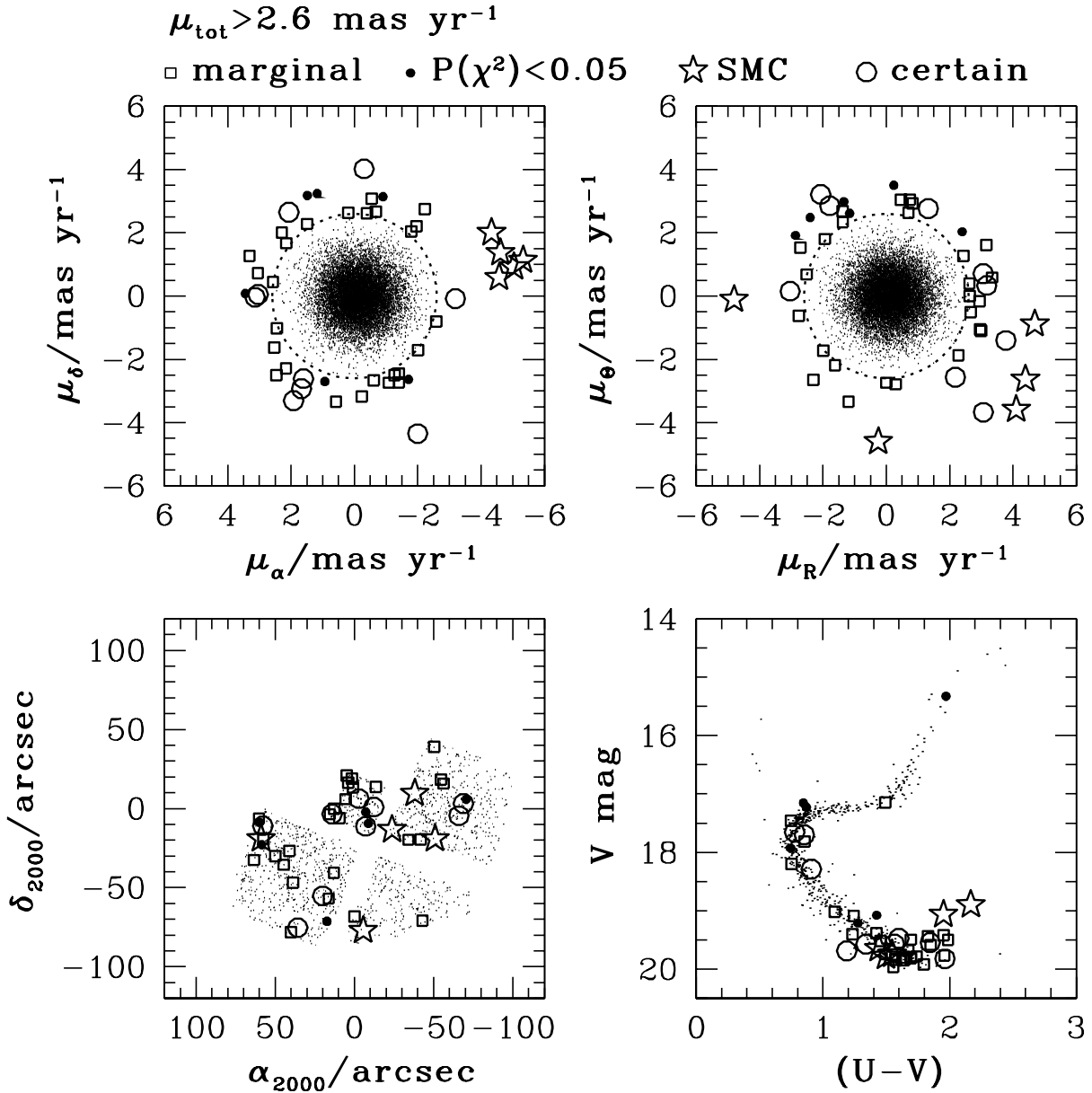}\hfil}}
\caption{
Distribution of ``high-velocity'' proper-motion stars ($\mu_{\rm tot}>2.6$ mas
yr$^{-1}$, indicated by the dotted circles in the upper panels and
corresponding roughly to the nominal escape velocity at the center of 47 Tuc),
in velocity space, in position relative to the cluster center, and on a
color-magnitude diagram. See the text and Table \ref{tab:highpm} for details,
but note that these fast-moving stars are not confined to the very center of
the cluster and show no preference for one particular direction of motion.
None is a blue straggler.
\label{fig:highpm}}
\end{figure*}

Thus, Table \ref{tab:highpm} lists the basic properties of all the stars in
Table \ref{tab:posdata} with $P(\chi^2)\ge0.001$ in both the RA and Dec
components of proper motion, $V<20$, $0.2\le (U-V) < 2.6$, and
$\mu_{\rm tot}>2.6$ mas yr$^{-1}$. There are only 46 such stars in our 
full sample of 12,974 ``good'' proper motions. Figure
\ref{fig:highpm} identifies these stars relative to the rest of the velocity
sample in plots of their distributions in the projected velocity planes
$(\mu_\alpha,\mu_\delta)$ and $(\mu_R,\mu_\Theta)$, their
$(\alpha_{2000}, \delta_{2000})$ positions relative to the cluster center,
and a $(V,\,U-V)$ color-magnitude diagram.

\begin{deluxetable*}{lcccrrrrrrr}
\tabletypesize{\scriptsize}
\tablewidth{0pt}
\tablecaption{Stars with Proper-Motion Speed $\mu_{\rm tot}>2.6$ mas yr$^{-1}$
\label{tab:highpm}}
\tablecolumns{11}
\tablehead{
\colhead{} & \colhead{} & \colhead{} & \colhead{$\mu_{\rm tot}$} &
\colhead{$\alpha_{2000}$} & \colhead{$\delta_{2000}$} &
\colhead{$\mu_{\alpha}$}  & \colhead{$\mu_{\delta}$}  &
\colhead{$\mu_{R}$} & \colhead{$\mu_{\Theta}$} & \colhead{}\\
\colhead{ID} & \colhead{$V$} & \colhead{$(U-V)$} &
\colhead{[mas yr$^{-1}$]} & \multicolumn{2}{c}{[arcsec]} &
\multicolumn{2}{c}{[mas yr$^{-1}$]} & \multicolumn{2}{c}{[mas yr$^{-1}$]} &
\colhead{comment} \\
 \colhead{(1)} &  \colhead{(2)} &  \colhead{(3)} &  \colhead{(4)} &
 \colhead{(5)} &  \colhead{(6)} &  \colhead{(7)} &  \colhead{(8)} &
 \colhead{(9)} & \colhead{(10)} & \colhead{(11)}
}
\startdata
M055901 & 15.33 & 1.97 & 3.15 & $ -7.25$ & $ -2.19$ & $-1.71\pm0.12$ &
         $-2.64\pm0.10$ & $ 2.40\pm0.12$ & $ 2.03\pm0.10$ & low $P(\chi^2)$ \\
M077141 & 17.15 & 1.49 & 2.75 & $  4.86$ & $ 20.91$ & $-0.68\pm0.20$ &
          $ 2.66\pm0.20$ & $ 2.44\pm0.20$ & $ 1.27\pm0.20$ & marginal \\
M005828 & 17.15 & 0.85 & 3.45 & $ 17.39$ & $-71.22$ & $ 1.18\pm0.27$ &
         $ 3.24\pm0.38$ & $-2.87\pm0.38$ & $ 1.91\pm0.28$ & low $P(\chi^2)$ \\
M063627 & 17.23 & 0.87 & 2.86 & $-70.65$ & $  5.85$ & $ 0.93\pm0.19$ &
         $-2.70\pm0.21$ & $-1.15\pm0.19$ & $ 2.61\pm0.21$ & low $P(\chi^2)$ \\
M027453 & 17.46 & 0.75 & 2.65 & $ 44.63$ & $-35.47$ & $-0.40\pm0.19$ &
          $ 2.62\pm0.37$ & $-1.94\pm0.27$ & $ 1.80\pm0.31$ & marginal \\
M058939 & 17.66 & 0.78 & 3.19 & $-12.25$ & $  0.96$ & $-3.19\pm0.14$ &
          $-0.08\pm0.10$ & $ 3.17\pm0.14$ & $ 0.33\pm0.10$ & certain \\
M053721 & 17.70 & 0.85 & 3.04 & $-66.05$ & $ -4.50$ & $ 3.04\pm0.18$ &
          $ 0.05\pm0.15$ & $-3.04\pm0.18$ & $ 0.16\pm0.15$ & certain \\
M057847 & 17.81 & 0.85 & 2.64 & $ 12.79$ & $ -0.15$ & $-2.01\pm0.25$ &
          $-1.71\pm0.23$ & $-1.99\pm0.24$ & $-1.73\pm0.23$ & marginal \\
M049017 & 17.93 & 0.75 & 3.45 & $ -8.75$ & $ -9.45$ & $ 3.45\pm0.25$ &
         $ 0.08\pm0.16$ & $-2.40\pm0.21$ & $ 2.48\pm0.22$ & low $P(\chi^2)$ \\
M073236 & 18.20 & 0.75 & 2.71 & $  3.87$ & $ 16.45$ & $-2.59\pm0.29$ &
          $-0.80\pm0.22$ & $-1.37\pm0.23$ & $ 2.34\pm0.29$ & marginal \\
M047472 & 18.28 & 0.91 & 3.07 & $ -7.18$ & $-11.18$ & $ 1.61\pm0.20$ &
          $-2.61\pm0.19$ & $ 1.33\pm0.20$ & $ 2.77\pm0.20$ & certain \\
M003793 & 18.90 & 2.17 & 4.61 & $ -5.60$ & $-76.99$ & $-4.57\pm0.48$ &
          $ 0.59\pm0.54$ & $-0.25\pm0.54$ & $-4.60\pm0.48$ & SMC \\
M029657 & 19.02 & 1.09 & 3.15 & $ 63.47$ & $-32.54$ & $ 2.16\pm0.45$ &
          $-2.29\pm0.60$ & $ 2.97\pm0.48$ & $-1.05\pm0.57$ & marginal \\
M067102 & 19.06 & 1.95 & 4.77 & $-38.13$ & $  9.66$ & $-4.33\pm0.71$ &
          $ 2.01\pm0.81$ & $ 4.69\pm0.72$ & $-0.89\pm0.80$ & SMC \\
M037308 & 19.08 & 1.42 & 3.51 & $ 58.57$ & $-22.83$ & $ 1.49\pm0.57$ &
         $ 3.18\pm0.46$ & $ 0.23\pm0.55$ & $ 3.50\pm0.47$ & low $P(\chi^2)$ \\
M039992 & 19.09 & 1.24 & 3.01 & $-34.14$ & $-19.69$ & $ 2.53\pm0.44$ &
          $-1.63\pm0.53$ & $-1.38\pm0.47$ & $ 2.68\pm0.51$ & marginal \\
M049738 & 19.21 & 1.27 & 3.27 & $ 60.06$ & $ -8.68$ & $-0.90\pm0.55$ &
         $ 3.14\pm0.58$ & $-1.34\pm0.55$ & $ 2.98\pm0.58$ & low $P(\chi^2)$ \\
M040209 & 19.38 & 1.42 & 2.61 & $-41.94$ & $-19.46$ & $ 2.57\pm0.71$ &
          $ 0.44\pm0.52$ & $-2.52\pm0.68$ & $ 0.68\pm0.56$ & marginal \\
M013200 & 19.40 & 1.23 & 3.19 & $ 16.25$ & $-56.91$ & $-0.23\pm0.54$ &
          $-3.18\pm0.69$ & $ 2.99\pm0.68$ & $-1.10\pm0.55$ & marginal \\
M031525 & 19.42 & 1.95 & 2.65 & $ 49.89$ & $-30.12$ & $ 2.45\pm0.83$ &
          $-1.02\pm0.90$ & $ 2.62\pm0.85$ & $ 0.39\pm0.88$ & marginal \\
M075395 & 19.44 & 1.83 & 3.51 & $  1.45$ & $ 18.89$ & $ 2.47\pm0.73$ &
          $-2.50\pm1.05$ & $-2.30\pm1.05$ & $-2.65\pm0.73$ & marginal \\
M061148 & 19.47 & 1.60 & 3.83 & $-68.89$ & $  3.26$ & $ 1.92\pm0.69$ &
          $-3.31\pm1.12$ & $-2.07\pm0.69$ & $ 3.22\pm1.12$ & certain \\
M003409 & 19.49 & 1.69 & 3.14 & $ 40.08$ & $-78.17$ & $ 3.05\pm0.97$ &
          $ 0.73\pm0.74$ & $ 0.74\pm0.79$ & $ 3.05\pm0.93$ & marginal \\
M072700 & 19.50 & 1.99 & 2.72 & $-56.36$ & $ 15.83$ & $ 2.15\pm0.82$ &
          $ 1.67\pm0.70$ & $-1.62\pm0.81$ & $-2.19\pm0.71$ & marginal \\
M034129 & 19.52 & 1.45 & 2.80 & $ 40.97$ & $-26.76$ & $-1.27\pm0.78$ &
          $-2.50\pm1.01$ & $ 0.30\pm0.85$ & $-2.79\pm0.95$ & marginal \\
M004332 & 19.55 & 1.85 & 4.79 & $ 35.85$ & $-75.30$ & $-2.00\pm0.68$ &
          $-4.35\pm0.91$ & $ 3.07\pm0.87$ & $-3.68\pm0.73$ & certain \\
M047722 & 19.57 & 1.34 & 3.37 & $ 57.87$ & $-10.89$ & $ 1.67\pm0.61$ &
          $-2.93\pm0.53$ & $ 2.18\pm0.61$ & $-2.57\pm0.54$ & certain \\
M014212 & 19.57 & 1.56 & 3.36 & $ 20.07$ & $-55.21$ & $ 2.07\pm0.63$ &
          $ 2.65\pm0.65$ & $-1.78\pm0.65$ & $ 2.85\pm0.63$ & certain \\
M007099 & 19.58 & 1.85 & 3.39 & $ -0.08$ & $-68.26$ & $ 0.58\pm0.77$ &
          $-3.34\pm0.84$ & $ 3.34\pm0.84$ & $ 0.58\pm0.77$ & marginal \\
M040651 & 19.65 & 1.44 & 4.81 & $ 58.89$ & $-18.96$ & $-4.61\pm0.40$ &
          $ 1.37\pm0.39$ & $-4.81\pm0.40$ & $-0.11\pm0.39$ & SMC \\
M023785 & 19.67 & 1.67 & 2.95 & $ 12.94$ & $-40.61$ & $-1.09\pm0.52$ &
          $-2.74\pm0.61$ & $ 2.28\pm0.60$ & $-1.87\pm0.53$ & marginal \\
M054717 & 19.68 & 1.19 & 3.14 & $ 14.08$ & $ -3.44$ & $ 3.14\pm0.26$ &
          $-0.04\pm0.36$ & $ 3.06\pm0.27$ & $ 0.71\pm0.36$ & certain \\
M040527 & 19.74 & 1.55 & 5.12 & $-50.95$ & $-19.10$ & $-5.04\pm0.52$ &
          $ 0.92\pm0.63$ & $ 4.40\pm0.53$ & $-2.63\pm0.61$ & SMC \\
M070760 & 19.74 & 1.49 & 2.95 & $-13.65$ & $ 13.66$ & $-1.96\pm0.50$ &
          $ 2.20\pm0.59$ & $ 2.94\pm0.55$ & $-0.17\pm0.55$ & marginal \\
M054596 & 19.75 & 1.55 & 2.74 & $ 15.58$ & $ -3.56$ & $-0.61\pm0.54$ &
          $-2.67\pm0.47$ & $-0.00\pm0.53$ & $-2.74\pm0.48$ & marginal \\
M051847 & 19.77 & 1.96 & 3.55 & $ 60.31$ & $ -6.46$ & $ 3.31\pm0.99$ &
          $ 1.27\pm0.64$ & $ 3.16\pm0.99$ & $ 1.62\pm0.65$ & marginal \\
M045617 & 19.78 & 1.50 & 5.45 & $-23.80$ & $-13.26$ & $-5.33\pm0.82$ &
          $ 1.13\pm0.86$ & $ 4.11\pm0.83$ & $-3.58\pm0.85$ & SMC \\
M074956 & 19.78 & 1.74 & 3.07 & $-54.67$ & $ 18.42$ & $-1.41\pm0.96$ &
          $-2.73\pm1.38$ & $ 0.46\pm1.01$ & $ 3.04\pm1.34$ & marginal \\
M019505 & 19.80 & 1.63 & 3.12 & $ 38.84$ & $-46.89$ & $-0.55\pm0.63$ &
          $ 3.07\pm0.78$ & $-2.72\pm0.72$ & $ 1.53\pm0.70$ & marginal \\
M052255 & 19.80 & 1.69 & 3.05 & $  9.34$ & $ -6.03$ & $ 2.29\pm0.62$ &
          $ 2.01\pm0.52$ & $ 0.84\pm0.59$ & $ 2.93\pm0.55$ & marginal \\
M091341 & 19.81 & 1.70 & 2.72 & $-50.32$ & $ 39.02$ & $-1.80\pm0.77$ &
          $ 2.04\pm0.93$ & $ 2.67\pm0.83$ & $-0.51\pm0.87$ & marginal \\
M064031 & 19.82 & 1.96 & 4.03 & $ -2.89$ & $  6.28$ & $-0.31\pm0.49$ &
          $ 4.02\pm0.96$ & $ 3.78\pm0.89$ & $-1.40\pm0.60$ & certain \\
M006034 & 19.84 & 1.57 & 3.54 & $-43.05$ & $-70.74$ & $-2.23\pm0.87$ &
          $ 2.75\pm1.06$ & $-1.19\pm1.01$ & $-3.33\pm0.93$ & marginal \\
M040685 & 19.85 & 1.64 & 2.72 & $ 57.48$ & $-18.91$ & $ 1.49\pm0.67$ &
          $ 2.28\pm0.70$ & $ 0.70\pm0.67$ & $ 2.63\pm0.69$ & marginal \\
M070501 & 19.91 & 1.80 & 2.64 & $  1.00$ & $ 13.37$ & $ 0.19\pm0.58$ &
          $ 2.63\pm0.84$ & $ 2.64\pm0.84$ & $ 0.00\pm0.58$ & marginal \\
M063762 & 19.96 & 1.56 & 2.83 & $  5.52$ & $  5.99$ & $-1.41\pm0.34$ &
          $-2.45\pm0.41$ & $-2.76\pm0.38$ & $-0.62\pm0.37$ & marginal \\
\enddata
\end{deluxetable*}

We have inspected the position vs.\ time data for each of the high-velocity
stars in Table \ref{tab:highpm} and Figure \ref{fig:highpm} individually, and
separated them into four categories:

(1) {\it Marginal}: 26 stars have total speeds just barely above 2.6 mas
    yr$^{-1}$, and/or velocity uncertainties that could bring them below
    $\mu_{\rm tot}<2.65$ mas yr$^{-1}$. Table \ref{tab:highpm} and the lower
    right-hand panel of Figure \ref{fig:highpm} show that the large majority
    of these stars are fairly faint, with $V > 19$, suggesting that their
    (modestly) high apparent speeds exceed the $v_{\rm max}$ limit simply
    because of measurement error.

(2) {\it Low $P(\chi^2)$}: Six stars with $P(\chi^2)\ge 0.001$ in both
    velocity components nevertheless have $P(\chi^2)<0.05$ in at least one
    component. Thus, it can't be said with better than 95\% confidence that
    these stars follow genuinely linear motion, let alone whether their fitted
    velocities are completely reliable. Most of these stars are fairly bright,
    and additional data from future epochs could either confirm or remove
    them as high-velocity candidates.

(3) {\it SMC}: Five stars are grouped together at $\mu_\alpha=4.3$--5.3 mas
    yr$^{-1}$ West and $\mu_\delta=0.6$--2.0 mas yr$^{-1}$ North in the upper
    left-hand panel of Figure \ref{fig:highpm} but are not so closely
    associated with each other in the $(\mu_R,\mu_\Theta)$ plane which refers
    their motion to the center of 47 Tuc. The Small Magellanic Cloud is in the
    background of our proper-motion field, and its average motion relative to
    47 Tuc is $(\mu_\alpha, \mu_\delta) \simeq (-4.7, +1.3)$ mas yr$^{-1}$
    \citep{ak2003a}.
    Thus, we identify these five stars as red giants in the SMC. Their
    faint $V$ magnitudes and relatively red $(U-V)$ colors (see
    the lower right-hand panel of Figure \ref{fig:highpm}) are consistent with
    this interpretation.

(4)  {\it Certain}: Nine stars remain which pass all of the first three tests,
     and which we therefore take to be genuinely high-velocity stars that could
     well be associated in some way with 47 Tuc.

In fact, all of the ``certain'' high-velocity stars have
$\mu_{\rm tot} > 3.0\ {\rm mas\ yr}^{-1} = 56.9\,(D/4\,{\rm kpc})$ 
km s$^{-1}$, which is obviously a lower limit to their total speed
including motion along the line of sight. This makes it quite unlikely that
any dynamical model could allow for them to be bound to the cluster currently.
If they were ever members of 47 Tuc, they must have been produced by extreme
events inside the cluster core. On the other hand, it may appear
suspicious that six of the nine stars are gathered at the bottom of the main
sequence in the lower right of Figure \ref{fig:highpm}. By construction of our
categories, the relative velocity uncertainties on these particular faint
stars are not extreme, and their speeds seem secure enough; but this
positioning on the CMD suggests the possibility that at least some of them
might be foreground field stars (of which we do expect to see a handful
throughout our field; cf.~the contamination estimate in Table
\ref{tab:basic}).

Of the three brighter ``certain'' high-velocity stars, two (M058939 and
M053721, both at the main-sequence turn-off) were previously identified
by \citet{mcl03}, who also pointed out that
they have $|\mu_R|\gg |\mu_\Theta|$ and thus appear to be moving almost purely
radially, with respect to the cluster center, on the plane of the sky. This is
true of only one other ``certain'' star (M054717), although it is much fainter
and falls blueward of the main cluster sequence in the CMD of Figure
\ref{fig:highpm}. Even among the marginal and low-$P(\chi^2)$ stars in Table
\ref{tab:highpm}, only a handful show such a preference in their direction
of motion. Since we do not have information on the
line-of-sight velocity component for any of these stars, it is impossible
to say exactly how their total space motions relate to the cluster center.
But with this caveat, even if all of the non-SMC stars with apparent
$\mu_{\rm tot}>2.6$ mas yr$^{-1}$ are bona-fide high-velocity stars and
genuinely associated with 47 Tuc, only a small minority show any obvious hint
that they might have been ejected directly outward from the very center of the
cluster (as could be expected to happen, for example, as a result of
interactions with centrally concentrated binaries or a large black hole).

Our fastest stars are rather more extreme than the two ``high-velocity'' stars
found in the radial-velocity survey by \citet{mdm91} that prompted our current
study. Unfortunately, these two likely cluster members are too bright
($V<14$) to have been included in our proper-motion sample. However, it is
worth noting that their line-of-sight velocities are $|v_z|=32.4$ and
36.7 km s$^{-1}$ relative to the cluster mean, which would correspond here
to about 1.7 and 1.9 mas yr$^{-1}$ (for $D=4$ kpc) and would not even
be considered as high velocities by our definition. Meylan et al.~deemed
these stars potentially interesting based in part on a lower estimate of
the central escape velocity than we have now, and in part on the fact that
they appeared in a much smaller sample of velocities. Now, however, it seems
that they may simply be part of the normal population of the velocity
distribution at $v\sim 3\sigma_0$ \citep[see also][]{geb95}.

It must be emphasized that there are an additional 1,392 stars in Table
\ref{tab:posdata} with measured proper motions that have $\chi^2$
probabilities less than 0.001 in at least one component. As we have already
discussed, although many of these measurements are effectively noise, some
could be associated with perfectly reliable data which sample motions that
are nonlinear due to strong accelerations.
Any such stars should probably be considered alongside those in Table
\ref{tab:highpm} in any future attempts to understand the extreme tail of the
47 Tuc velocity distribution in detail.

But even ignoring any nonlinear motions, it is still significant that the 
fraction of stars with anomalously high straight-line proper motions
is so small: the number is at most
$41/12,974\simeq0.3\%$ within $R\la 5$ core radii of the cluster center,
and it could easily be lower than this by a factor of 4 or more.
Together with the scattered distribution of these stars in the
$(\mu_R,\mu_\Theta)$ plane and the highly regular appearance of
the velocity distribution of all other stars below $v_{\rm max}\simeq2.6$ mas
yr$^{-1}$, this should serve as a useful constraint on theoretical analyses
of any processes that could produce very fast-moving stars in dense globular
cluster cores
(such as the mere presence of a central black hole, or close encounters
between various combinations of single stars, binaries,
stellar- or intermediate-mass black holes, and black-hole binaries;
see, e.g., \citealt{dav95,druk03,map05}).
In this context, note that for a \citet{king66} model with $W_0=8.6$,
roughly 75--80\% of all stars projected to within
$R\le 5 r_0$ on the plane of the sky are in fact expected to lie in a
sphere of radius $5 r_0$ around the cluster center.

\section{Velocity Dispersions}
\label{sec:veldisp}

Having seen that there is nothing particularly untoward in the full
proper-motion velocity distribution at the center of 47 Tuc, the last analyses
we undertake involve examining its second moment (velocity
dispersion) in particular,
as a function of stellar magnitude and clustercentric radius in various
subsamples of our data. We do not attempt to model the effects of the
cluster rotation in any detail---even
though in principle this affects both the radial and the
azimuthal components of proper-motion dispersion---both because the patchy
azimuthal coverage of our velocity sample (e.g., Figure \ref{fig:muvzpos})
makes it difficult to constrain this well using only our data, and because the
rotation in the inner few core radii is known to be small anyway
\citep{mey86}.

In \S\ref{subsec:strag}, we consider the velocity dispersion of blue stragglers
in our sample. Section \ref{subsec:kinmags} next looks at the dispersion of
normal main-sequence stars as a function of magnitude and radius, and obtains
quantitative measures of the velocity anisotropy. Section
\ref{subsec:distance} compares the proper-motion and line-of-sight dispersions
for turn-off mass stars to derive a kinematic distance to 47 Tuc. Having
finally estimated the line-of-sight velocity dispersion in the cluster core,
we draw on it in \S\ref{subsec:norotation} to argue that the presence of
ordered stellar motions
(rotation) will affect our estimates of the kinematics on these small
radial scales at the level of $\sim2\%$ or even less---a source of
error that we do not correct for. Last, \S\ref{subsec:bhmods}
fits the proper-motion $\sigma(R)$ profile with
very simple modifications of \citet{king66} models to allow for the
possibility of a central point mass.

Throughout this section, we report all velocity dispersions after correction
for both velocity-measurement errors and local transformation errors, as
calculated from equation (\ref{eq:sigtrue2}). For convenience, however, we
drop the subscript ``true'' on $\sigma$ from now on. In addition, we prevent
the highest-velocity stars of \S\ref{subsec:highpm} from inflating the
estimated dispersions by only using stars with
\begin{equation}
\label{eq:velcut}
\mu_{\rm tot}< \left\{
\begin{array}{ll}
  3.0\,{\rm mas\ yr}^{-1}\ , & \phantom{19.5\le\ }V<19    \\
  3.2\,{\rm mas\ yr}^{-1}\ , & \phantom{.5}19\le V<19.5   \\
  3.6\,{\rm mas\ yr}^{-1}\ , & 19.5\le V<20               \\
\end{array}
\right.
\end{equation}
in the calculations---as we did earlier, in \S\ref{subsec:twodvel}. All
uncertainties are estimated through numerical bootstrap experiments, which
were also described in \S\ref{subsec:twodvel}.

\subsection{Blue Stragglers}
\label{subsec:strag}

Figure \ref{fig:cmdpos} above clearly shows the presence of a number of blue
stragglers in our ``good'' proper-motion sample, and it is worth noting that
none of these appear as high-velocity stars (cf.~Figure
\ref{fig:highpm}). In fact, \citet{ka2001} used an earlier, more restricted
version of the current data set to show qualitatively that the blue stragglers
in 47 Tuc have a lower velocity dispersion than the giants. We can now
quantify this result.

For our purposes here, we define blue stragglers as stars with $15.25\le
V<17.25$ (recall that the main-sequence turn-off is at $V=17.65$; Table
\ref{tab:basic}) and $(U-V) < 0.7$. Throughout our entire field, there are 23
proper-motion stars which satisfy these criteria, and 18 of these have
$R<16\farcs5$. By comparison, there are 441 proper-motion
stars in the same magnitude range but on the red-giant branch of the cluster
[which we take to be $1.5\le (U-V)<2.5$], and of these 191 lie within
$R\le 20\arcsec$. Although our velocity sample does not
provide an unbiased sampling of the true radial distribution of any stellar
population, this already indicates that the blue stragglers are more centrally
concentrated than normal cluster members of the same magnitude.

\begin{figure}[!b]
\vspace*{-0.3truecm}
\hspace*{+0.2truecm}
\centerline{\resizebox{90mm}{!}{\includegraphics{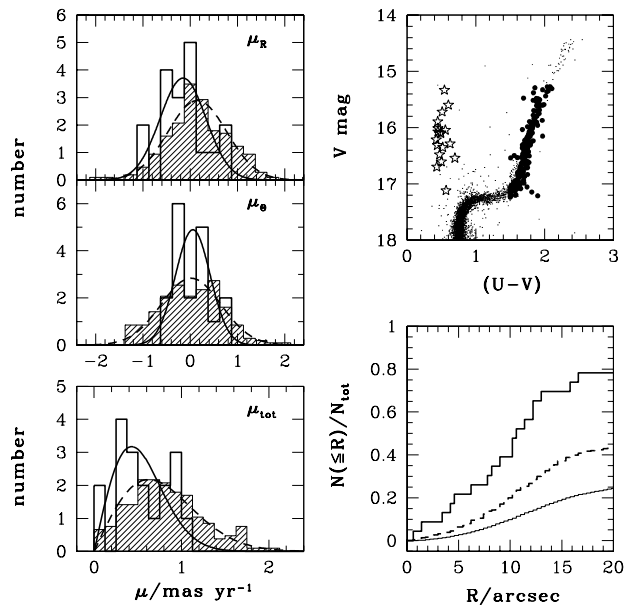}\hfil}}
\caption{
Kinematics of blue stragglers vs.\ giant-branch stars of the same magnitude,
in the inner $R<20\arcsec$ (roughly one core radius) of 47 Tuc. Only stars
plotted as open stars (the blue stragglers) or large filled circles (red
giants) are used to
construct the velocity and speed distributions in the left-hand panels; see
Table \ref{tab:strag} for the definition of the two subsamples.
In these panels, the open histograms pertain to the blue stragglers, and the
bold solid curves are Gaussians with error-corrected velocity dispersions
given in Table \ref{tab:strag}. The shaded histograms are for the red-giant
stars, and the bold dashed curves are Gaussians with intrinsic dispersions
also given in Table \ref{tab:strag}.
The lower $\sigma_\mu$ for the blue stragglers is
consistent with them being on average twice as massive as main-sequence
turn-off stars. The cumulative radial distributions of the blue stragglers,
red giants, and all stars in our proper-motion sample (the
curves from top to bottom in the lower right-hand panel) support this
qualitatively, although quantitatively the spatial distribution of stars in
the velocity sample does not accurately reflect the true density profile.
\label{fig:strag}}
\end{figure}

Figure \ref{fig:strag} and Table \ref{tab:strag} compare in more detail the
relative spatial distributions and kinematics of the blue-straggler and
red-giant branch stars in the inner $R<20\arcsec$ of 47 Tuc. The upper
right-hand panel of Figure \ref{fig:strag} illustrates the CMD-based
definitions of these two subsamples, as explained just above and also written
in Table \ref{tab:strag}. The left-hand panels of the figure then show their
one-dimensional distributions of $\mu_R$, $\mu_\Theta$, and speed
$\mu_{\rm tot}=(\mu_R^2+\mu_\Theta^2)^{1/2}$, very similarly to Figure
\ref{fig:muhists} above. The curves in each panel are Gaussians with
dispersions equal to the measurement- and transformation-error
corrected values given by equation (\ref{eq:sigtrue2}), which are listed in
Table \ref{tab:strag}; for the $\mu_{\rm tot}$ distribution, we use the average
value $\sigma_\mu^2 \equiv (\sigma_R^2+\sigma_\Theta^2)/2$ in equation
(\ref{eq:gausstot}). The lower right-hand panel then shows the cumulative
radial distributions, out to $R=20\arcsec$, of the blue stragglers
(upper curve), the
red-giant branch stars (middle curve), and all stars in the full sample of
good proper motions (lower curve). The differences between the three are
clearly significant to a high level of confidence (although they are also
strongly affected by complex selection biases in these velocity samples).

The interesting point here is the ratio of the blue-straggler and
red-giant velocity dispersions:
$\sigma_\mu({\rm BS})/\sigma_\mu({\rm RGB}) = 0.69\,(+0.07, -0.14)$,
which is, to well within the errors, indistinguishable from $1/\sqrt{2}$. This
is just what we would expect if blue stragglers are on average about twice as
massive as the normal main-sequence turn-off stars in 47 Tuc and the two
populations are in energy equipartition. This is the first evidence of its
kind for 
a relatively high mass for blue stragglers, and it is consistent with
the idea that these peculiar stars represent some combination of
tight (contact) or even coalesced equal-mass binaries, and
the remnants of collisional mergers between main-sequence turn-off stars
\citep[see, e.g.,][]{dav04,map04}.

A typical mass ratio of 2 would also imply that thermalized blue stragglers
should be spatially distributed with a core radius roughly $2^{-0.58}=0.67$
times that of the
turn-off and giant-branch stars (see eq.~[\ref{eq:multir0}]). For 47 Tuc,
then, $r_0({\rm BS})\simeq 0.67\times20\farcs84\approx14\arcsec$, which is
indeed consistent with an observed $r_0({\rm BS})=10\arcsec$--18\arcsec, as
estimated by \citet{guh93}.

\begin{deluxetable}{ccc}[!b]
\tablecaption{Kinematics of Stars with $15.25\leq V<17.25$ and
$R\leq20\arcsec$ \label{tab:strag}}
\tablewidth{0pt}
\tablecolumns{3}
\tablehead{
\colhead{} & \colhead{Blue Stragglers} & \colhead{Red Giant Branch} \\
\colhead{} & \colhead{$0\leq(U-V)<0.7$} &
\colhead{$1.5\leq(U-V)<2.5$}
}
\startdata
                         ${\cal N}$ & 18 & 190 \\
       $\sigma_{R}$ (mas yr$^{-1}$) & $0.485^{+0.051}_{-0.099}$ &
                                      $0.621^{+0.031}_{-0.036}$ \\
  $\sigma_{\Theta}$ (mas yr$^{-1}$) & $0.367^{+0.037}_{-0.069}$ &
                                      $0.631^{+0.026}_{-0.030}$ \\
 $\sigma_{\rm tot}$ (mas yr$^{-1}$) & $0.430^{+0.030}_{-0.065}$ &
                                      $0.626^{+0.020}_{-0.025}$ \\
              $\langle\beta\rangle$ & $0.527^{+0.186}_{-0.427}$ &
                                      $-0.046^{+0.177}_{-0.275}$ \\
\enddata
\end{deluxetable}

As well as they mesh with each other, these arguments based on kinematics and
spatial distributions are still indirect.
Spectroscopic mass measurements have been made for five blue
stragglers in 47 Tuc by
\citet[][$m=0.29$--$1.88\,M_\odot$]{dem05},
and for one other in the cluster by
\citet[][$m=1.7\pm0.4\,M_\odot$]{shara97}.
The weighted average mass of these 6 stars (using weights based on the
relative errors as described by \citeauthor{dem05}) is
$\langle m \rangle\simeq 1.4\pm0.5\,M_\odot$,
which is to be compared with the main-sequence turn-off
mass of about $0.85\,M_\odot$ (Table \ref{tab:masslum}). While perhaps
roughly consistent with the indirect suggestions that $m({\rm BS})=2\times
m({\rm TO})$ on average, the number of direct mass measurements is still
very small, and the spread and uncertainties in the individual values are
quite large. More work is needed in this area.

The last line of Table \ref{tab:strag} gives rough estimates of the velocity
anisotropy $\langle\beta\rangle$ obtained by comparing the radial and azimuthal
proper-motion dispersions of both the blue stragglers and the red giants
(see eq.~[\ref{eq:aniso}] below). While the giant-branch
stars are clearly consistent with the expected isotropy
($\langle\beta\rangle=0$), it is perhaps noteworthy that the formal
$\langle\beta\rangle > 0$ for the blue stragglers implies an apparent radial
bias in their orbits. However, given the small number of stars, this
result is not significant at even the one-sigma level.

\subsection{Dispersion as a Function of Stellar Magnitude and Projected Radius}
\label{subsec:kinmags}

In order to define the run of proper-motion dispersion vs.\ clustercentric
radius, we first break our good velocity sample of 12,974 stars into four
broad magnitude/mass bins as in \S\ref{subsec:twodvel}: $V<18.5$,
corresponding to $0.77\la m_*\la 0.9\,M_\odot$; $18.5\le V<19$, meaning
$0.73\la m_*\la 0.77\,M_\odot$; $19\le V<19.5$, or $0.69\la m_*\la
0.73\,M_\odot$; and $19.5\le V<20$, for $0.65 \la m_*\la 0.69\,M_\odot$.
We also apply the color selection specified in equation (\ref{eq:colcut}),
to exclude blue stragglers in particular from the analysis.
In each of these ranges we use stars surviving the velocity cuts in equation
(\ref{eq:velcut}) to compute the mean velocities in the RA and Dec
directions, and the corresponding error-corrected dispersions, in a series of
narrow, concentric annuli.
From these we confirm that $\langle\mu_\alpha\rangle =
\langle\mu_\delta\rangle = 0$ to within the uncertainties at all radii in
general---as expected from the relative nature of our proper motions---and
then set
$\langle\mu_R\rangle = \langle\mu_\Theta\rangle \equiv 0$ to compute the true
radial and azimuthal dispersions $\sigma_R$ and $\sigma_\Theta$, and the
average $\sigma_\mu=[(\sigma_R^2+\sigma_\Theta^2)/2]^{1/2}$, in
the same annuli (with ${\cal N}_{\rm tot}-1$ replaced by ${\cal N}_{\rm tot}$
in eq.~[\ref{eq:sigtrue2}]). Finally, we find the relative difference
$(\sigma_R^2-\sigma_\Theta^2)/\sigma_\mu^2$.

Table \ref{tab:kinmags} presents the results of all these calculations
performed in a series of narrow annuli in each magnitude bins.
(Table \ref{tab:kinmags} can be found at the end of this preprint).
Note that we increase the annulus size as the stellar magnitudes grow fainter,
in order to avoid drastic declines in the numbers of stars per annulus. As a
result, in the bins with $V>18.5$,  we actually use two sets of overlapping
annuli to aid in evaluating the extent to which the choice of bin positioning
affects the measurements.
All of the data required to repeat this exercise
for any other choice of magnitude or radial binning (or any alternate
definition of the velocity sample) are contained in Table
\ref{tab:posdata}, which is available in full in the online edition of the
{\it Astrophysical Journal}. 

\begin{figure}[!b]
\vspace*{-0.3truecm}
\hspace*{+0.1truecm}
\centerline{\resizebox{90mm}{!}{\includegraphics{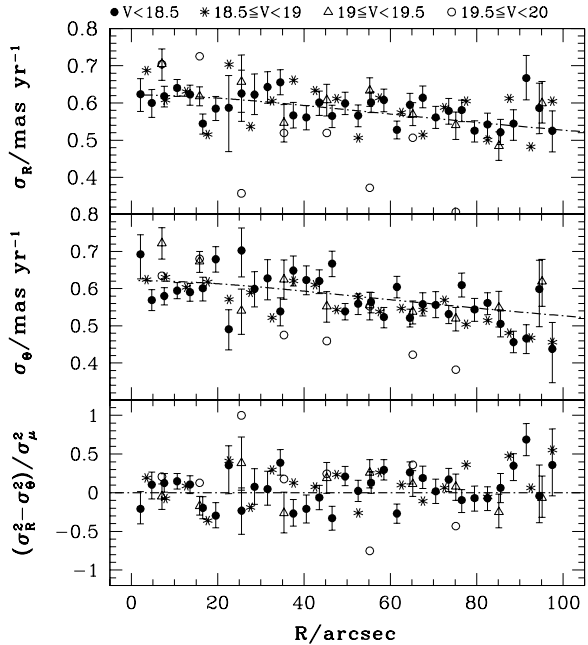}\hfil}}
\caption{
Proper-motion dispersions in the projected radial and azimuthal directions
relative to the cluster center, and their relative difference, as functions of
radius in four magnitude bins. Curves in the upper two panels are
one-dimensional, projected $\sigma(R)$ profiles for a $W_0=8.6$ \citet{king66}
model with $r_0=20\farcs84$ and $\sigma_0=(1+1/45)^{1/2}\times0.633$ mas
yr$^{-1}$ (see text). Note that the stellar kinematics at $V\ge 19.5$ and
outside of $R>20\arcsec$ in particular are error-dominated. The relative
difference of variances in the bottom panel is related to the velocity
anisotropy; see Table \ref{tab:aniso} for average values over the entire field.
\label{fig:kinmags}}
\end{figure}

Figure \ref{fig:kinmags} shows $\sigma_R$, $\sigma_\Theta$, and
$(\sigma_R^2-\sigma_\Theta^2)/\sigma_\mu^2$ as functions of $R$ from
Table \ref{tab:kinmags}. Note that we have only plotted every second number
from the table for the fainter magnitudes $V>18.5$, so that every point shown
is statistically independent of the others. For clarity, we have attached
errorbars only to the data in the first and third magnitude bins.

\setcounter{table}{10}
\begin{deluxetable*}{lccccrr}
\tabletypesize{\scriptsize}
\tablewidth{0pt}
\tablecaption{Aperture Dispersions and Average Anisotropy
               \label{tab:aniso}}
\tablecolumns{7}
\tablehead{
\colhead{Stellar Magnitude} & \colhead{${\cal N}$}            &
\colhead{$\langle\sigma_R^2\rangle^{1/2}$}                    &
\colhead{$\langle\sigma_\Theta^2\rangle^{1/2}$}               &
\colhead{$\langle\sigma_\mu^2\rangle^{1/2}$}                  &
\colhead{$\frac{\langle\sigma_R^2\rangle-\langle\sigma_\Theta^2\rangle}
               {\langle\sigma_\mu^2\rangle}$}           &
\colhead{$\langle\beta\rangle$}                         \\
\colhead{} & \colhead{} &
\colhead{[mas yr$^{-1}$]} & \colhead{[mas yr$^{-1}$]} &
\colhead{[mas yr$^{-1}$]} & \colhead{} & \colhead{}
}
\startdata
$\phantom{19.5\le\ }V<18.5$  &
   6336 & $0.589\pm0.006$ & $0.577\pm0.006$ &
   $0.583\pm0.004$ & $0.04\pm0.03$ &
   $0.06\pm0.04$ \\
 & & & & & & \\
$18.5\le V < 19$             &
   2327 & $0.587\pm0.010$ & $0.565\pm0.011$ &
   $0.576\pm0.008$ & $0.08\pm0.05$ &
   $0.11\pm0.07$ \\
 & & & & & & \\
$\phantom{.5}19\le V < 19.5$ &
   2421 & $0.591\pm0.013$ & $0.586\pm0.013$ &
   $0.589\pm0.009$ & $0.02\pm0.06$ &
   $0.03\pm0.09$ \\
 & & & & & & \\
$19.5\le V < 20$             &
   1693 & $0.479\pm0.026$ & $0.495\pm0.024$ &
   $0.487\pm0.018$ & $-0.07\pm0.14$ &
   $-0.11^{+0.20}_{-0.31}$   \\
\enddata
\end{deluxetable*}

In each of the top two panels of Figure \ref{fig:kinmags} we draw
the predicted one-dimensional, projected velocity-dispersion profile for a
\citet{king66} model with $W_0=8.6$ and $r_0=20\farcs84$
(Figure \ref{fig:density} in \S\ref{subsec:density}), and with $\sigma_0
=0.633\ {\rm mas\ yr}^{-1} = 12.0\,(D/4\,{\rm kpc})\ {\rm km\ s}^{-1}$
as in equation (\ref{eq:sigma0}) of \S\ref{subsec:twodvel}. These curves were
calculated as described in Appendix \ref{sec:kingveldist}
(see eqs.~[\ref{eq:modsig}] and [\ref{eq:modsigz}]), and then smoothed by
averaging $\sigma_{\rm mod}^2(R)$ over a 3\arcsec-wide annulus at each point,
to match the binning of the brightest data. In this specific model, the
observable central dispersion in any component of velocity is
$\sigma(R=0)=0.982\,\sigma_0=0.622$ mas yr$^{-1}$, or 11.8 km s$^{-1}$ for
$D=4$ kpc. Note that the model predicts a decrease of about 15\%
in $\sigma_R=\sigma_\Theta$ from the center to $R=100\arcsec$. Although this
amounts to just under 0.1 mas yr$^{-1}$, it is nicely resolved by the
measured dispersions for the brighter (more numerous and less uncertain) stars.
The clear tendency for a decreasing $\sigma$ as a function of $R$
suggests---rather more strongly than do the weakly populated tails of the full,
two-dimensional velocity distribution at any radius---that a
\citeauthor{king66}-model distribution function (with a finite escape velocity
that decreases monotonically towards larger radii) is indeed a closer
approximation to the cluster dynamics than is a truly isothermal sphere
with a perfectly Gaussian velocity distribution. 

Also in the top two panels of Fig.~\ref{fig:kinmags}, note the slightly higher
dispersion of the fainter stars inside $R\la20\arcsec$, which has already been
discussed in connection with Figure \ref{fig:nmumagcore} above. Apparent too
are the low values and large scatter in the error-corrected $\sigma_R$ and
$\sigma_\Theta$ at larger radii, for stars in the range $19.5\le V < 20$. As
was also discussed above, this is due to the generally large velocity
uncertainties at these magnitudes in these areas of the field. Our faint stars
might provide useful kinematics constraints on better dynamical models for 47
Tuc if the definition of the velocity sample is revised to exclude stars with
uncertainties close to or larger than the true velocity dispersion; but, as
Figure \ref{fig:errmua} makes clear, this will drastically decrease the sample
size.

The bottom panel of Figure \ref{fig:kinmags} shows the relative difference of
variances, $(\sigma_R^2-\sigma_\Theta^2)/\sigma_\mu^2$, as a function of $R$.
This is naturally the most uncertain kinematical parameter that we calculate,
and its noisy profile reflects this. The fact that it rarely differs
significantly from 0, and the absence of any sustained trend in clustercentric
radius, suggest that $\sigma_R^2=\sigma_\Theta^2$ on the whole. Again, then,
in keeping with the earlier modeling of \citet{mey88,mey89} and our
discussion in \S\ref{sec:veldist}, the velocity distribution in the inner
several core radii of 47 Tuc appears to be essentially isotropic.

This impression can be better quantified by considering how the observable
velocity dispersions in the $R$ and $\Theta$ directions and along the line of
sight, $z$, are related to the intrinsic,
unprojected $\sigma_r$, $\sigma_\theta$, and $\sigma_\phi$. For a spherically
symmetric star cluster with an ellipsoidal velocity-dispersion tensor
($\sigma_\theta=\sigma_\phi$), \citet{leon89} and \citet{gen00} show that
\begin{equation}
\label{eq:velcomps}
\begin{array}{rcl}
\sigma_z^2(r) & = &
    \sigma_r^2(r)\,\cos^2\theta + \sigma_\theta^2(r)\,\sin^2\theta \\
\sigma_R^2(r) & = &
    \sigma_r^2(r)\,\sin^2\theta + \sigma_\theta^2(r)\,\cos^2\theta \\
\sigma_\Theta^2(r) & = & \sigma_\theta^2(r)\ ,
\end{array}
\end{equation}
where $r$ is the unprojected clustercentric radius and
$\theta$ is the angle between the unit vectors $\widehat{{\bf r}}$ and
$\widehat{{\bf z}}$. With velocity anisotropy parametrized by the usual
function $\beta(r)\equiv 1-\sigma_\theta^2/\sigma_r^2$ \citep[e.g.,][]{bt87},
averaging the last two of equations (\ref{eq:velcomps}) over all $\theta$
and all $r$ leads to \citep{leon89}
\begin{equation}
\label{eq:aniso}
\langle\beta\rangle \ \equiv\ 
 1-\frac{\langle\sigma_\theta^2\rangle}
        {\langle\sigma_r^2\rangle}
 \ = \ \frac{3\langle\sigma_R^2\rangle-3\langle\sigma_\Theta^2\rangle}
            {3\langle\sigma_R^2\rangle-\langle\sigma_\Theta^2\rangle}\ 
\end{equation}
Thus, for an isotropic velocity ellipsoid, $\langle\beta\rangle = 0$ and we
have $\langle\sigma_R^2\rangle = \langle\sigma_\Theta^2\rangle$. For
purely radial orbits, $\langle\beta\rangle = 1$ and
$\langle\sigma_\Theta^2\rangle=0$, the same as the unprojected
$\sigma_\theta$. For purely circular orbits, the unprojected
$\sigma_r=0$, so $\langle\beta\rangle=-\infty$ and the averaged proper-motion
dispersions satisfy $\langle\sigma_\Theta^2\rangle=3\langle\sigma_R^2\rangle$. 

Equation (\ref{eq:aniso}) is strictly only valid as an estimator of the
{\it globally} averaged anisotropy, and stars spread over an entire cluster
should be used to compute it; it is not a particularly meaningful
quantity in narrow annuli such as those considered in Table \ref{tab:kinmags},
for example. However, as was also mentioned at the end of
\S\ref{subsec:highpm}, the majority of stars with projected radius
$R< 5\,r_0$ in a \citeauthor{king66}-model cluster similar to 47 Tuc do
actually lie in a sphere of the same radius. Thus, calculating a single
$\langle\beta\rangle$ for all the stars in our proper-motion sample will still
yield a useful approximation to the volume-averaged $\langle\beta\rangle$
over the inner $100\arcsec\simeq4.8$ core radii of this cluster (with a small
contamination from stars at larger distances along the line of sight).

Table \ref{tab:aniso} reports the $R$ and $\Theta$ aperture dispersions for
our entire field, as well as their relative difference and the corresponding
averages $\langle\sigma_\mu^2\rangle^{1/2}$ and $\langle\beta\rangle$, with
the sample broken into the same magnitude bins as before. Note that the
dispersions in the fifth column of the table are just those used in Figure
\ref{fig:nmumag} above and that, as was also discussed in
\S\ref{subsec:twodvel}, the results at the faintest magnitudes are dominated
by measurement errors and are thus unreliable. Apart from this, there is an
apparent preference for a slight radial anisotropy
($\langle\sigma_R^2\rangle > \langle\sigma_\Theta^2\rangle$ and
$\langle\beta\rangle>0$) in the average. Although the departure from isotropy
is not highly significant statistically, the sense of it is
just what we expect for a dynamical structure in which the stellar
orbits are isotropic at the center of the cluster and become gradually more
radially biased toward very large radii \citep[see][]{mey88}.

\subsection{A Kinematic Distance}
\label{subsec:distance}

Section \ref{subsec:rvsample} presented details of a set of radial velocities
we have obtained for stars in the inner $R< 105\arcsec$ of 47 Tuc. Here we
compare the velocity dispersion of these stars to that of our proper-motion
stars to estimate a distance to the cluster. Recall that we have useful
radial velocities only for 419 bright giants with magnitudes $11\le V<14$,
(corresponding to masses $m_*\simeq0.9\,M_\odot$), whereas all of our
proper-motion stars are fainter than this and can have masses as low as
$m_*\simeq0.65\,M_\odot$. In
order not to bias the derived distance by comparing the kinematics of stars
with too widely different masses, we work only with rather bright
proper-motion stars:
$V<18.5$, and thus $0.78\,M_\odot\la m_*\la 0.9\,M_\odot$ according to the
formula in Table \ref{tab:masslum}.
Applying the color cuts of equation (\ref{eq:colcut}) to exclude blue
stragglers, and the velocity cut in equation (\ref{eq:velcut}), we are left
with 6,336 proper-motion stars having an average magnitude $\langle
V\rangle=17.74$ and hence an average mass of $0.84\,M_\odot$.

Since the stars in the core of 47 Tuc are expected to be in energy
equipartition,
we might naturally worry that a proper-motion sample with $V<18.5$ could
have a dispersion some $(0.9/0.84)^{1/2}=1.035$ times larger than a sample of
stars with masses strictly equal to those in our radial-velocity sample---in
which case, the true distance to the cluster might be {\it longer} than
what we derive in this Section, by about 3.5\% (roughly 0.15 kpc).
While this is a valid concern in principle, in practice we have found that
carrying out the analysis below
using a smaller proper-motion sample of 1,895 stars with $V<17.5$ (for a mean
mass of $0.88\,M_\odot$)
formally changes our inferred distance in the opposite way to what we would
expect, but by a statistically insignificant amount:
$D(V\!\!<\!\!17.5)/D(V\!\!<\!\!18.5)\simeq 0.99\pm0.05$.
Given our inability to resolve directly the signature of energy equipartition
in the
kinematics of the brightest proper-motion stars, we simply present the
results of our work with the larger, $V<18.5$ sample and do not attempt to
compensate for any anticipated effect.

Another potential complication is that in using these data to estimate the
distance to 47 Tuc, we assume that the
cluster is not rotating. As we have mentioned already, it is known
that 47 Tuc {\it does} rotate \citep{mey86,ak2003a}, but
we are saved from having to apply much more sophisticated dynamical modeling
(with more free parameters, constrained by additional data with further
sources of uncertainty) by the fact that the
level of rotation in the central-most regions of the cluster presently under
consideration is low relative to the random motions. In
\S\ref{subsec:norotation}, we give a brief but more quantitative
a posteriori justification for our neglect of this issue.

\begin{figure}[!b]
\vspace*{-0.3truecm}
\hspace*{+0.1truecm}
\centerline{\resizebox{90mm}{!}{\includegraphics{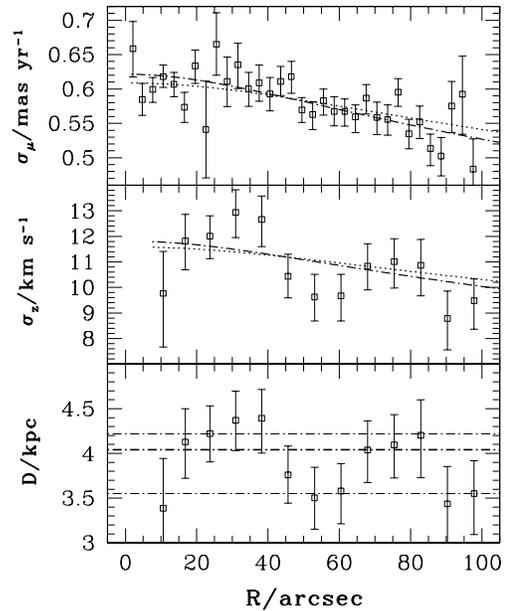}\hfil}}
\caption{
Comparison of proper-motion dispersion profile $\sigma_\mu(R)$ for stars with
$V<18.5$ (first section of Table \ref{tab:kinmags}) against the line-of-sight
dispersion profile $\sigma_z(R)$ for stars with $V<14$, and the resulting
kinematic distance estimate in a number of circular annuli. Only every second
point in the lower two panels is statistically independent, as they refer to
overlapping annuli; see Table \ref{tab:distance}. Curves in the upper two
panels are \citet{king66} model profiles for $W_0=8.6$, with two different core
radii $r_0$ and two different $\sigma_0$ parameters; see text. The bold
horizontal line in the bottom panel is at the median of the independent
annulus-by-annulus distance estimates, and the lighter lines denote its 68\%
confidence interval (see eq.~[\ref{eq:dlocal}]). Taking the ratio of either
pair of model curves in the upper panels yields essentially the same distance
(eq.~[\ref{eq:dglobal}]).
\label{fig:distance}}
\end{figure}

Given equations (\ref{eq:velcomps}) and the definition
$\beta(r)\equiv 1-\sigma_\theta^2/\sigma_r^2$, it is straightforward to show
that the velocity dispersions projected along the line of sight and into the
radial and azimuthal directions on the plane of the sky are connected by
\begin{equation}
\label{eq:sigzrt}
\sigma_z^2(r) = \sigma_\Theta^2\,\left(\frac{2-\beta}{1-\beta}\right)
                - \sigma_R^2
\end{equation}
at any radius $r$ in a spherical and non-rotating cluster, for all three
dispersions measured in the
same units. If we then perform the same spatial averaging as in
\S\ref{subsec:kinmags}, substitute equation (\ref{eq:aniso}) for $\beta$, and
use equation (\ref{eq:convert}) to convert between proper motions measured in
mas yr$^{-1}$ and radial velocities in km s$^{-1}$, the distance follows as
\begin{equation}
\label{eq:distance}
\frac{D}{{\rm kpc}}\ = \ \frac{1}{4.74}\,
   \sqrt{\frac{2\langle\sigma_z^2\rangle}
              {\langle\sigma_R^2\rangle+\langle\sigma_\Theta^2\rangle}}
   \ \equiv\ \frac{\langle\sigma_z^2\rangle^{1/2}}
                  {4.74\,\langle\sigma_\mu^2\rangle^{1/2}}\ .
\end{equation}
For non-zero anisotropy, this formula is really applicable only to a global
sample of stars covering the entire cluster, although---as with the expression
above for $\langle\beta\rangle$---its use for our $R\la5\,r_0$ field is still
justified. And in any case, if $\beta\equiv0$ it is
more generally valid for any local sample; $\langle\sigma_\mu^2\rangle$ is
then simply an improved estimate of a single proper-motion variance
given two independent estimates from the $R$ and $\Theta$ components. Thus,
we have estimated the distance to 47 Tuc in two different (though
interdependent) ways, both of which are illustrated in
Figure \ref{fig:distance}.

\begin{deluxetable*}{lcrrcrccc}
\tabletypesize{\scriptsize}
\tablewidth{0pt}
\tablecaption{Line-of-Sight vs.\ Proper-Motion Kinematics
               \label{tab:distance}}
\tablecolumns{9}
\tablehead{
\colhead{} & \colhead{} &
\multicolumn{2}{c}{LOS: $11\leq V<14$}      &  \colhead{} &
\multicolumn{2}{c}{PM:  $V<18.5$} &  \colhead{} &
\colhead{} \\
\cline{3-4} \cline{6-7}
\colhead{Radii} & \colhead{} &
\colhead{${\cal N}$} & \colhead{$\sigma_{\rm z}$}   & \colhead{} &
\colhead{${\cal N}$} & \colhead{$\sigma_{\mu}$}     & \colhead{} &
\colhead{$D$} \\
\colhead{[arcsec]} & \colhead{} &
\colhead{} & \colhead{[km s$^{-1}$]}   & \colhead{} &
\colhead{} & \colhead{[mas yr$^{-1}$]} & \colhead{} &
\colhead{[kpc]} \\
\colhead{(1)} & \colhead{}    &
\colhead{(2)} & \colhead{(3)} & \colhead{}    & 
\colhead{(4)} & \colhead{(5)} & \colhead{}    & 
\colhead{(6)}
}
\startdata
\phantom{1}0.0--15.0 &   & 35 & $9.76^{+1.64}_{-2.09}$ &   &
     1283 & $0.608^{+0.010}_{-0.009}$ &   & $3.39^{+0.56}_{-0.73}$ \\
\phantom{1}7.5--22.5 &   & 76 & $11.82^{+1.05}_{-1.13}$ &   &
     1314 & $0.604^{+0.009}_{-0.009}$ &   & $4.13^{+0.37}_{-0.41}$ \\
15.0--30.0 &   & 104 & $12.01^{+0.80}_{-0.88}$ &   &
     569 & $0.600^{+0.013}_{-0.013}$ &   & $4.22^{+0.31}_{-0.32}$ \\
22.5--37.5 &   & 96 & $12.94^{+0.89}_{-0.98}$ &   &
     498 & $0.624^{+0.016}_{-0.015}$ &   & $4.37^{+0.33}_{-0.34}$ \\
30.0--45.0 &   & 79 & $12.66^{+0.91}_{-1.07}$ &   &
     824 & $0.608^{+0.011}_{-0.011}$ &   & $4.39^{+0.32}_{-0.39}$ \\
37.5--52.5 &   & 77 & $10.43^{+0.86}_{-0.84}$ &   &
     1086 & $0.585^{+0.010}_{-0.009}$ &   & $3.76^{+0.32}_{-0.32}$ \\
45.0--60.0 &   & 68 & $9.63^{+0.88}_{-0.94}$ &   &
     1279 & $0.580^{+0.008}_{-0.010}$ &   & $3.50^{+0.34}_{-0.35}$ \\
52.5--67.5 &   & 53 & $9.67^{+0.84}_{-0.99}$ &   &
     1349 & $0.570^{+0.009}_{-0.009}$ &   & $3.58^{+0.31}_{-0.37}$ \\
60.0--75.0 &   & 54 & $10.84^{+0.87}_{-0.94}$ &   &
     1260 & $0.566^{+0.009}_{-0.009}$ &   & $4.04^{+0.33}_{-0.36}$ \\
67.5--82.5 &   & 49 & $11.01^{+0.90}_{-1.03}$ &   &
     1125 & $0.567^{+0.010}_{-0.009}$ &   & $4.10^{+0.34}_{-0.37}$ \\
75.0--90.0 &   & 37 & $10.87^{+1.01}_{-1.20}$ &   &
     952 & $0.546^{+0.010}_{-0.010}$ &   & $4.20^{+0.40}_{-0.47}$ \\
82.5--97.5 &   & 37 & $8.79^{+1.07}_{-1.23}$ &   &
     550 & $0.540^{+0.014}_{-0.015}$ &   & $3.44^{+0.42}_{-0.49}$ \\
90.0--105.0 &   & 42 & $9.48^{+0.86}_{-1.12}$ &   &
     169 & $0.563^{+0.025}_{-0.026}$ &   & $3.55^{+0.37}_{-0.46}$ \\
\enddata
\end{deluxetable*}

The top panel of Figure \ref{fig:distance} plots $\sigma_\mu$ against $R$ from
the first section of Table \ref{tab:kinmags}. The curves running
through the points are \citet{king66} models with two different core radii,
which we will discuss below. The middle panel of the figure then shows
the error-corrected line-of-sight dispersion $\sigma_z$ as a function of
radius (and \citeauthor{king66}-model profile curves again), for stars binned
in 15\arcsec-wide annuli. Unlike the proper motions, we have plotted these
results for both of two slightly offset, overlapping
sets of annuli, so that only every second $\sigma_z$ point is statistically
independent. Table \ref{tab:distance} gives the radii of the annuli defined
for the radial velocities, followed by the error-corrected $\sigma_z$ and the
true proper-motion $\sigma_\mu$  in each, and then the distance inferred by
assuming velocity isotropy to apply equation (\ref{eq:distance}) in each
annulus. The bottom panel of Figure \ref{fig:distance} plots
these distance estimates, again for all of the overlapping annuli. Only seven
of them are independent, and their median value is
\begin{equation}
\label{eq:dlocal}
D=4.04\,(+0.18, -0.49)\ {\rm kpc}\ ,
\end{equation}
as indicated by the horizontal lines in the graph. The uncertainties here
delimit the 68\% confidence interval on the median for 1000 artificial data
sets generated by bootstrapping with replacement from the measured distances.

Clearly, it is the relatively small number of radial velocities at
our disposal which limits the precision of any distance estimate here, and the
noisiness of the profile $\sigma_z(R)$ which limits its accuracy in any single
radial bin. It would thus be preferable to use all of the velocities over our
entire field at once to obtain a single $D$. This is nontrivial, however,
since neither component of the dispersion is spatially constant and the
radial-velocity and proper-motion stars have significantly different (and
highly irregular) distributions on the sky; see Figure \ref{fig:muvzpos}
above. The total $\sigma_\mu$ and $\sigma_z$ for our full samples therefore
represent differently weighted averages of a single underlying
velocity-dispersion profile, and if we were to put them directly into equation
(\ref{eq:distance}) a biased distance could result in principle. It is
possible, though, first to correct for the different spatial samplings using
our \citet{king66} model description of 47 Tuc.

The bold, dash-dot curve in the top panel of Figure \ref{fig:distance}
is the projected and smoothed one-dimensional velocity-dispersion profile
plotted in Figure \ref{fig:kinmags}. As was also discussed in
\S\ref{subsec:twodvel}, to normalize this model to the data we first computed
the dimensionless profile
$\sigma_{\rm mod}^2/\sigma_0^2$ vs.\ $(R/r_0)$, as described in Appendix
\ref{sec:kingveldist}, for $W_0=8.6$. Then, given any fixed value of the 
\citeauthor{king66} core radius $r_0$, the average model dispersion for a set
of stars with measured clustercentric radii $\{R_i\}$ is given by equation
(\ref{eq:sigpred}). To repeat:
\[
\frac{\langle\sigma_{\rm mod}^2\rangle}{\sigma_0^2} \ = \ 
   \frac{1}{{\cal N}} \sum_{i=1}^{{\cal N}}
   \frac{\sigma_{\rm mod}^2(R_i/r_0)}{\sigma_0^2}\ .
\nonumber
\]
With $r_0=20\farcs84$ from our fit to the surface-brightness profile in
\S\ref{subsec:density}, the measured positions of the 6,336 proper-motion
stars being used here yield
$\langle\sigma_{\rm mod}^2\rangle/\sigma_0^2=0.849\pm0.015$. Equating
$\langle\sigma_{\rm mod}^2\rangle^{1/2}$ to the true
$\langle\sigma_\mu^2\rangle^{1/2}=0.583\pm0.004$ mas yr$^{-1}$ from
Table \ref{tab:aniso} then implies $\sigma_0=0.633\pm0.010$ mas yr$^{-1}$,
just as in equation (\ref{eq:sigma0}) above. Turning now to the
radial-velocity sample, the 419 observed stellar positions correspond to
$\langle\sigma_{\rm mod}^2\rangle/\sigma_0^2=0.851\pm0.059$ (only
{\it fortuitously} close to the result for the proper-motion stars) if
$r_0=20\farcs84$ still, while the observed
$\langle\sigma_z^2\rangle^{1/2} = 11.1\pm0.4$ km s$^{-1}$ after correction for
measurement errors. Thus, $\sigma_0=12.05\pm0.85$ km s$^{-1}$. The model
$\sigma_z(R)$ profile for $W_0=8.6$ and $r_0=20\farcs84$, normalized by this
value of $\sigma_0$ and smoothed with an annular filter of width 15\arcsec\ to
mimic the binning of the data, is drawn as the heavy dash-dot curve in the
middle panel of Figure \ref{fig:distance}. An estimate of the distance to 47
Tuc, which effectively smooths over the fluctuations of $\sigma_z$ and
$\sigma_\mu$ in narrow annuli and does not require an assumption of strict
velocity isotropy, then follows from using the proper-motion and
radial-velocity $\sigma_0$ values in equation (\ref{eq:distance}): 
\begin{equation}
\label{eq:dglobal}
D = 4.02\pm0.35\ {\rm kpc}\ ,
\end{equation}
in good agreement with our first result. Note that the
$\simeq$9\% relative uncertainty in $D$ is twice the amount caused by number
statistics alone, due to the necessity of correcting for the different spatial
distributions of the two velocity samples.

As we discussed in connection with Figure \ref{fig:nmurad} above, and can also
be seen from the top panel of Figure \ref{fig:distance}, the observed
$\sigma_\mu(R)$ profile in the inner $\approx40\arcsec$ ($\simeq2$ core radii)
appears rather flatter than expected for a $W_0=8.6$ \citet{king66} model with
the core radius appropriate to stars brighter than the main-sequence turn-off.
This point will also appear again in \S\ref{subsec:bhmods}. Thus, we have
repeated the exercise just described using a $W_0=8.6$ \citeauthor{king66}
model with $r_0=27\farcs5$ rather than $20\farcs84$. In this case, we find
$\sigma_0=0.620\pm0.010$ mas yr$^{-1}$ from the proper-motion sample, and
$\sigma_0=11.8\pm0.8$ km s$^{-1}$ from the radial velocities. The dotted
curves in the top two panels of Figure \ref{fig:distance} show the model
profiles with this larger $r_0$ and the lower $\sigma_0$ values, again
smoothed by amounts matching the binnings of the data. Taking the
ratio of the normalizations gives a distance identical to that in equation
(\ref{eq:dglobal}), demonstrating that the result is insensitive to fine
details of the model used to smooth over the spatial distributions.

Recent determinations of the distance to 47 Tuc by standard CMD fitting
range from $D=4.45\pm0.15$ kpc \citep{perc02} to $D=4.85\pm0.18$ kpc
\citep{grat03}, while a fit to
the white-dwarf cooling sequence by \citet{zoc01} returned $D=4.15\pm0.27$
kpc. Our result clearly supports the shorter white-dwarf value
(which in turn implies a cluster age of 13 Gyr; see \citeauthor{zoc01}),
although the lowest of the CMD-based distance estimates may still be
consistent within the uncertainties.

\subsection{Neglecting Rotation}
\label{subsec:norotation}

It would be surprising if more sophisticated modeling did not uncover slight
biases in our distance estimate due to our simplifying assumptions
of spherical symmetry and zero rotation. But it seems very doubtful that
the value of $D$ could change
by more than the 9\% margin of error already allowed by the formal
uncertainties. In fact, we expect that any rotation-related biases in
the kinematics we have derived will be present only at the level of
$\sim$2\%, which is comparable to the minimal, $1/\sqrt{{\cal N}}$ statistical
errorbars that are unavoidable even in the brightest and lowest-uncertainty
subsamples of proper-motion stars.

The signature of rotation in the radial velocities appears as a
sinusoidal dependence of $v_z$ on the projected azimuth $\Theta$, and simply 
fitting a sine curve to the 419 reliable line-of-sight velocities available
at $R<105\arcsec$ yields a rotation amplitude $V_{\rm rot}\simeq3$ km
s$^{-1}$. The azimuthally averaged velocity {\it variance} due to rotation
alone is therefore
$\langle V_{\rm rot}^2 \sin^2\Theta\rangle=4.5\ ({\rm km\ s}^{-1})^2$.
Meanwhile, the observed, error-corrected variance for all 419 radial-velocity
stars together is $\langle\sigma_z^2\rangle=(11.1\ {\rm km\ s}^{-1})^2$
(see \S\ref{subsec:distance}). The velocity dispersion over our field that is
due to truly random stellar motions, rather than rotation, is then
$(11.1^2-4.5)^{1/2}=10.9\ {\rm km\ s}^{-1}$---just over 98\% of the 
observed value.

The velocity distribution on these spatial scales in 47 Tuc is essentially
isotropic (\S\ref{subsec:kinmags}), and there is evidence that the inclination
angle of the rotation axis is $\sim$45$^\circ$ \citep{ak2003a}, so any
signature of rotation in our proper-motion dispersions can also be expected to
appear at roughly $\sim$2\% levels, which is too small to discern accurately
once all sources of statistical uncertainty (number statistics; measurement
and transformation errors; irregular spatial distributions) are taken into
account. In any event, our distance estimate particularly should be rather
robust, since it relies on the {\it ratio} $\sigma_z/\sigma_\mu$.

\subsection{Does 47 Tuc Harbor a Central Black Hole?}
\label{subsec:bhmods}

Our final analysis looks at the kinematics of the innermost
$\sim$3\arcsec\ of 47 Tuc, the area in which the effects of a compact
central mass concentration (if one is present) should be most clearly
observable.

Returning for a moment to the results of \S\ref{subsec:distance}, it is a
simple matter to calculate the observable velocity dispersion
(for stars at the turn-off mass) at the exact center of 47 Tuc from our
estimates of the \citeauthor{king66}-model scale parameter,
$\sigma_0$. With $W_0=8.6$, the projected one-dimensional
$\sigma_{\rm mod}(R=0)/\sigma_0$ is 0.982. For the $\sigma_0$ values
associated with $r_0=27\farcs5$ (arguably a better description of the inner
one core radius of the proper-motion velocity-dispersion profile in
Fig.~\ref{fig:distance}), we thus have
\begin{equation}
\label{eq:sigcenter}
\left.
\begin{array}{rcl}
\sigma_\mu(R=0) & = & 0.609\pm0.010\ {\rm mas\ yr}^{-1} \\
\sigma_z(R=0)   & = & 11.6\pm0.8\    {\rm km\ s}^{-1}\ .
\end{array}
\right. \qquad (m_*\simeq0.85\,M_\odot)
\end{equation}
Obviously this assumes that velocity dispersion is nearly constant
as $R$ approaches 0. On the other hand, the innermost point
in the top panel of Figure \ref{fig:distance} appears to fall above this
extrapolation. Even though
the discrepancy is hardly significant at even the one-sigma level, it is
nevertheless of potential interest in light of recent claims for evidence of
massive black holes at the centers of globular clusters.

It is now well established that the masses of supermassive black holes at
the centers of galaxies correlate tightly with the central velocity
dispersions of the stellar bulges \citep{fer00,geb00}. Recently, claims have
been made, on the basis of radial-velocity studies, that the Galactic globular
cluster M15 \citep{gers02} and the very massive cluster G1 in Andromeda
\citep{geb02,geb05} may harbor large central black holes with masses
$M_\bullet\sim 10^3$--$10^4\,M_\odot$. If so, such black holes
could lie on a simple extension of the $M_\bullet$--$\sigma$ relation for
galaxies. For M15, the data show less than a 1-$\sigma$ significance for a
black hole \citep{gers03,mcn03}, and theoretical models show there is no need
to invoke one \citep{baum03a}. Given the data of \citet{geb02},
\citet{baum03b} concluded there is no need for a black hole in G1 either,
although
\citet{geb05} use more recent data to argue for an improved significance of
detection. It is thus unclear whether globular clusters do in fact contain
``intermediate-mass'' central black holes, and it is worthwhile asking whether
our proper motions in 47 Tuc can add anything to this discussion.

The galactic $M_\bullet$--$\sigma$ relation given by \citet{trem02}
uses the bulge velocity dispersion averaged over an effective radius, which in
a $W_0=8.6$ \citet{king66} model is $\sigma(R\le R_e)=0.91\,\sigma(R=0)$.
Given equation (\ref{eq:sigcenter}), then,
$\sigma(R\le R_e)\simeq10.6$ km~s$^{-1}$ in 47 Tuc, and extrapolating
the \citeauthor{trem02}
relation gives a ``predicted'' black-hole mass of roughly 
$1000\,M_\odot$. This corresponds to a sphere of influence
$GM_\bullet/\sigma(0)^2\approx 0.032\ {\rm pc}=1\farcs6\,(D/4\,{\rm kpc})$
and suggests a potential observational signature in our
proper-motion data. On the other hand,
there is no a priori justification for extrapolating the empirical scaling for
galaxies down to the globular cluster range, and the results of doing so are
highly dependent on the adopted parametrization of the relation. Thus, for
example, the steeper dependence of black-hole mass on (central) galactic
velocity
dispersion advocated by \citet{fer02} leads to $M_\bullet\approx360\,M_\odot$
in 47 Tuc, which would have a much smaller effect on the observable
kinematics. Conversely, the log-quadratic relation of \citet{wy06}
implies an untenable (though highly uncertain)
$M_\bullet\sim2$--$3\times10^{5\pm2}\,M_\odot$.

Aside from these rough expectations, it is also worth noting that
\citet{grin01} use the positions and
dispersion measures of millisecond pulsars in the core of 47 Tuc, together with
a {\it Chandra} upper limit on the X-ray luminosity of a central source, to
estimate an upper limit of $M_\bullet\la 470\,M_\odot$ for the mass of any
central black hole. This corresponds to a sphere of
influence, $r\la 0.015\ {\rm pc} \simeq 0\farcs8\,(D/4\,{\rm kpc})$,
which would be very difficult to probe with the current kinematical data.
On the other hand, an X-ray upper mass limit is fairly sensitive to the
assumed efficiency and spectral temperature for radiation from gas accreting
onto a black hole
$(M_\bullet\sim\varepsilon^{-1/2}\,T$)---uncertainties that are, of course,
avoided with kinematical mass estimates.

\subsubsection{Simple Models for the Velocity Dispersion Profile}
\label{subsubsec:bhvdp}

To address this issue, we have constructed a series of modified
\citet{king66} models allowing for the presence of a central point mass,
and fitted these to the 
proper-motion dispersion profile $\sigma_\mu(R)$. We first find the
dimensionless density profile $\rho_K(\widetilde{r})/\rho_0$ and enclosed mass
profile $GM_K(\widetilde{r})/\sigma_0^2 r_0$, both as functions of
$\widetilde{r}\equiv r/r_0$, for a regular \citeauthor{king66} model
with given $W_0$. Then we add a black hole with dimensionless mass
$GM_\bullet/\sigma_0^2 r_0$, assume that the stellar density profile is
unchanged, and solve the isotropic Jeans equation,
\begin{equation}
\label{eq:kingbh}
\frac{d}{d\widetilde{r}}
\left(\frac{\rho_K}{\rho_0}\frac{\sigma_r^2}{\sigma_0^2}\right) =
  - \frac{1}{\widetilde{r}^2} \frac{\rho_K(\widetilde{r})}{\rho_0}
    \left[\frac{GM_K(\widetilde{r})}{\sigma_0^2 r_0} +
          \frac{GM_\bullet}{\sigma_0^2 r_0}\right]\ ,
\end{equation}
for a new $\sigma_r^2(\widetilde{r})/\sigma_0^2$ profile (by integrating
inward from the original tidal radius, where $\rho\sigma_r^2=0$). The
projected, one-dimensional velocity-dispersion profile to be compared with the
observed $\sigma_\mu(R)$ follows from the usual averaging of $\sigma_r^2$
along the line of sight (e.g., eq.~[\ref{eq:modsigz}]).

Note that these models do not have the density cusp that is usually associated
intuitively with the presence of a central black hole. This is because they
do not attempt to address the origin of the black hole or the dynamical
evolution of the stellar distribution function in its presence. Rather, they
simply describe the self-consistent kinematics of stars around a dark point
mass {\it given} that the density structure is well matched by
a normal \citeauthor{king66} model. \citet{baum05} use detailed N-body
simulations to show that the visible stars in globular clusters with
intermediate-mass central black holes can indeed be reasonably well
described by \citeauthor{king66}-type density profiles (the expected density
cusp is essentially confined to heavy stellar remnants), and that the stellar
kinematics are accurately reproduced by isotropic Jeans modeling. The greatest
limitation to our simple approach is the assumption of a single stellar mass,
since in principle neutron stars or heavy white dwarfs that are strongly
concentrated to the center by mass segregation can affect the kinematics of
observable stars in ways similar to a true point mass. Although we have
already argued that the total mass of such remnants in 47 Tuc is quite
small (\S\ref{subsec:twodvel}; see the discussion around Figure
\ref{fig:nmurad}), ignoring the issue altogether means that fitting the
projected solutions of equation (\ref{eq:kingbh}) to an observed
$\sigma_\mu(R)$ profile is likely to overestimate the mass of any black hole
in general.

To fit the black-hole \citeauthor{king66} models, we first calculated
the $\sigma_\mu(R)$ profile again for our ``good'' proper-motion stars
with $V<18.5$ and restricted colors as in equation (\ref{eq:colcut}), but
binned into 1\arcsec-wide annuli in order to resolve the small-radius
behavior better than we did in \S\ref{subsec:kinmags} and
\S\ref{subsec:distance}. (However, we did check that our results are
insensitive to the binning.) 
We then solved equation (\ref{eq:kingbh}) for the dimensionless profile
$\sigma_r/\sigma_0$~vs.~$r/r_0$,
for a number of models with fixed $W_0$ in the range $W_0=8.6\pm0.5$
and dimensionless black-hole masses in the range
$0\le GM_\bullet/\sigma_0^2 r_0 \le 0.50$. Every model profile was
projected along the line of sight, smoothed with an annular filter of width
$\Delta R=1\arcsec$ to match our treatment of the data, and fit to the
observed $\sigma_\mu(R)$ profile by finding the normalizations $r_0$ and
$\sigma_0$ that minimize the usual error-weighted $\chi^2$ statistic. We
used only points with $R<90\arcsec$ when computing $\chi^2$, since the number
statistics at larger radii are very poor.

For any value of $W_0$, the best-fit black-hole mass is, of course, that which
yields the lowest $\chi^2$ in the model grid of
$GM_{\bullet}/\sigma_0^2 r_0$ values that we defined. The 68\% (1-$\sigma$)
confidence interval is defined by the values of $GM_{\bullet}/\sigma_0^2 r_0$
for which $\chi^2\le\chi_{\rm min}^2+1$; the 95\% confidence interval is set
by $\chi^2\le \chi_{\rm min}^2+3.84$; and the 99\% confidence interval,
by $\chi^2\le \chi_{\rm min}^2+6.63$.

The first part of Table \ref{tab:bhfit18.5} presents the details of several
of the fits to $\sigma_\mu(R)$ obtained for $W_0=8.6$, which best matches the
spatial structure of 47 Tuc for $R\la 1000\arcsec$
(see \S\ref{sec:space}). The columns in this table are the dimensionless
black-hole mass specified for each fit, followed by the $\chi^2$, the fitted
model scale velocity and radius (both allowed to vary freely in the fitting),
the physical value of the black-hole mass, and comments identifying the formal
best fit and approximate 68\%, 95\%, and 99\% confidence intervals. Note that
the dimensional black hole mass is obtained as
\begin{equation}
\label{eq:bhmass}
\frac{M_\bullet}{M_\odot} = 25.40 \, \frac{G M_\bullet}{\sigma_0^2 r_0} \,
   \left(\frac{D}{{\rm kpc}}\right)^3
   \left(\frac{\sigma_0}{{\rm mas\ yr}^{-1}}\right)^2
   \left(\frac{r_0}{{\rm arcsec}}\right) \ ,
\end{equation}
which clearly is quite sensitive to the assumed distance to the
cluster. Naturally, we take $D=4.0$ kpc, from \S\ref{subsec:distance}.

\begin{deluxetable}{crcccr}
\tabletypesize{\scriptsize}
\tablewidth{0pt}
\tablecaption{Central Black-Hole Model Fits to $V\leq 18.5$ Kinematics
              \tablenotemark{a} \label{tab:bhfit18.5}}
\tablecolumns{6}
\tablehead{
\colhead{$GM_{\bullet}/\sigma_0^2r_0$}   &
\colhead{$\chi^{2}$}                     &
\colhead{$\sigma_0$}                     &
\colhead{$r_0$}                          &
\colhead{$M_{\bullet}$\tablenotemark{b}}  \\
\colhead{}                &
\colhead{}                &
\colhead{[mas yr$^{-1}$]} &
\colhead{[arcsec]}        &
\colhead{[$M_\odot$]}     &
\colhead{}
}
\startdata
\cutinhead{$W_0=8.6$}
0.00 & 98.47 & 0.614 & 26.34 &     0 &   best fit \\
0.01 & 98.57 & 0.612 & 26.94 &   164 &            \\
0.02 & 98.75 & 0.610 & 27.56 &   334 &            \\
0.03 & 99.03 & 0.608 & 28.19 &   509 &            \\
0.04 & 99.40 & 0.606 & 28.85 &   690 & $<\!68\%$ C.I. \\
0.05 & 99.89 & 0.605 & 29.51 &   877 &            \\
0.06 & 100.49 & 0.603 & 30.17 &  1070 &            \\
0.07 & 101.21 & 0.601 & 30.86 &  1270 &            \\
0.08 & 102.04 & 0.599 & 31.56 &  1470 & $<\!95\%$ C.I. \\
0.09 & 103.01 & 0.597 & 32.25 &  1680 &            \\
0.10 & 104.11 & 0.595 & 32.86 &  1890 &            \\
0.11 & 105.34 & 0.593 & 33.48 &  2110 & $>\!99\%$ C.I. \\
0.12 & 106.72 & 0.592 & 33.97 &  2320 &            \\
0.13 & 108.24 & 0.590 & 34.37 &  2530 &            \\
0.14 & 109.91 & 0.588 & 34.68 &  2730 &            \\
0.15 & 111.72 & 0.587 & 34.92 &  2930 &            \\
0.16 & 113.66 & 0.586 & 35.08 &  3130 &            \\
%
\cutinhead{$W_0=8.1$}
0.00 & 98.25 & 0.615 & 30.72 &     0 &   best fit \\
0.04 & 99.49 & 0.608 & 33.44 &   805 & $>\!68\%$ C.I. \\
0.08 & 102.68 & 0.600 & 36.80 &  1720 & $>\!95\%$ C.I. \\
0.10 & 105.15 & 0.596 & 38.47 &  2220 & $>\!99\%$ C.I. \\
\cutinhead{$W_0=9.1$}
0.00 & 98.76 & 0.613 & 22.39 &     0 &   best fit \\
0.05 & 99.77 & 0.603 & 25.29 &   749 & 68\% C.I. \\
0.09 & 102.30 & 0.597 & 27.50 &  1430 & $<\!95\%$ C.I. \\
0.12 & 105.37 & 0.592 & 28.78 &  1970 & 99\% C.I. \\
\enddata

\tablenotetext{a}{Fits are to the proper-motion velocity-dispersion profile
of stars with $R<90\arcsec$, $V<18.5$, and $P(\chi^2)\geq0.001$ in both RA
and Dec components of proper motion. Stars are binned in $1\arcsec$-wide
annuli, resulting in 87 points fitted.}

\tablenotetext{b}{Dimensional black-hole masses assume
a distance of $D=4.0$ kpc to 47 Tuc. For any other distance, multiply by
$(D/4.0\,{\rm kpc})^3$.}
\end{deluxetable}

Assuming $W_0=8.6$, the best fit to $\sigma_\mu(R)$ for this sample of stars
with $V<18.5$ is achieved with $M_{\bullet}=0$, although
$M_{\bullet}\la 700\,M_\odot$ is allowed at the 1-$\sigma$ level, and
$M_{\bullet}\la 1500\,M_\odot$ at the 95\% ($\la2$-$\sigma$) confidence level.
This conclusion is essentially independent of the assumed value of $W_0$, as
Table \ref{tab:bhfit18.5} also shows: for $W_0=8.1$ and $W_0=9.1$, the
best-fit black-hole mass is also zero, and the 68\%, 95\%, and 99\% upper
limits are very similar to those found with $W_0=8.6$.
All in all, the proper-motion dispersions cannot be used to argue for the
presence of an intermediate-mass black hole at the center of 47 Tuc---although
neither do they strongly disallow such a possibility.

What is perhaps more convincingly seen in Table
\ref{tab:bhfit18.5} is the fact that the velocity-dispersion profile of
these stars (with $\langle m_*\rangle\simeq 0.84\,M_\odot$) always
prefers a ``core''
radius, $r_0$, which is rather larger than the $\approx21\arcsec$ implied by
the density profile of the same population. This is yet another
demonstration that (as was discussed in
\S\ref{subsec:twodvel} and \S\ref{subsec:distance}) the bright stars in
47 Tuc appear to be moving in response to a more extended mass
distribution that is
reminiscent of what we naively expect for the dominant, but unseen (in this
cluster), population of stars with $m_*\simeq0.5\,M_\odot$.

\begin{figure}[!t]
\hspace*{+0.1truecm}
\centerline{\resizebox{90mm}{!}{\includegraphics{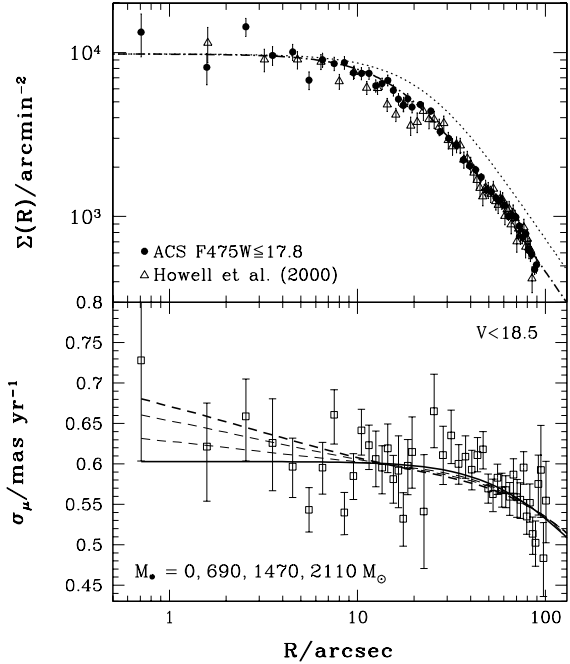}\hfil}}
\caption{
{\it Upper panel}: Number-density profile of stars brighter than the
main-sequence turn-off ($V_{\rm TO}=17.65$, or ${\rm F475W}\simeq17.8$),
as derived from our master list in Table \ref{tab:master} and compared to
similar results from \citet{how00}. To obtain our numbers (filled circles),
stars were binned into 1\arcsec-wide circular annuli at $R<20\arcsec$ and
into 3\arcsec-wide annuli at $20\arcsec\le R<90\arcsec$. Bold, dash-dot curve
is a $W_0=8.6$ \citeauthor{king66}-model curve with $r_0=20\farcs84$; the
dotted curve has a larger core radius of $r_0=27\farcs5$.
{\it Bottom panel}: Proper-motion
velocity-dispersion profile for stars with $V<18.5$, binned in radius as for
the density
profile. Curves represent fits of $W_0=8.6$ \citeauthor{king66} models with
central black holes of the masses indicated, which correspond to the formal
best fit (lower, solid curve) and the 68\%, 95\%, and 99\% (uppermost dashed
curve) upper limits. Normalizations $r_0$ and $\sigma_0$ for each case are
given in Table \ref{tab:bhfit18.5}.
\label{fig:muvdens1}}
\end{figure}

Figure \ref{fig:muvdens1} illustrates these results. In the upper panel,
we show the number density of turn-off mass stars (F475W magnitudes
$\le 17.8$), constructed from our master
star list (Table \ref{tab:master}) as described in \S\ref{subsec:density}, but
now binned in annuli with $\Delta R=1\arcsec$ for $R<20\arcsec$ and
$\Delta R=3\arcsec$ for $R>20\arcsec$. The bold, dash-dot curve is the
$W_0=8.6$ \citeauthor{king66}-model fit from Figure \ref{fig:density}, while
the dotted curve has the larger $r_0=27\farcs5$ from Figure \ref{fig:distance}
above. The bottom panel of Figure \ref{fig:muvdens1} then shows $\sigma_\mu$
vs.\ $R$ for our proper-motion stars with $V<18.5$, calculated in annuli
with the same $\Delta R$ inside and outside $R=20\arcsec$ as in the upper
panel. The model curves drawn in the bottom panel of Figure \ref{fig:muvdens1}
correspond to the best-fit $M_\bullet=0$ and the rough 68\%, 95\%, and 99\%
upper limits on $M_\bullet$ for the $W_0=8.6$ models detailed in Table
\ref{tab:bhfit18.5}.
As was mentioned above, these curves were obtained by fitting the $\sigma_\mu$
profile calculated in 1\arcsec-wide annuli at all radii $R<90\arcsec$; the
coarser binning at large $R$ in this figure is applied simply to make the
plot clearer.

We have also defined an alternative velocity sample by selecting all stars
from Table \ref{tab:posdata} which have $V<20$, $P(\chi^2)\ge0.001$ in both
components of proper motion, colors obeying equation (\ref{eq:colcut}), 
{\it and} velocity errorbars
$\Delta_\alpha$ and $\Delta_\delta$ both less than 0.25 mas yr$^{-1}$. The
last of these selection criteria allows for more robust determination of the
error-corrected velocity dispersions in this sample. We then fit our modified
\citeauthor{king66} models with central black holes to the true
$\sigma_\mu(R)$ in 1\arcsec-wide annuli, exactly as we did before.

\begin{deluxetable}{crcccr}
\tabletypesize{\scriptsize}
\tablewidth{0pt}
\tablecaption{Central Black-Hole Model Fits to $V\leq 20$ Kinematics
              \tablenotemark{a} \label{tab:bhfit20}}
\tablecolumns{6}
\tablehead{
\colhead{$GM_{\bullet}/\sigma_0^2r_0$}   &
\colhead{$\chi^{2}$}                     &
\colhead{$\sigma_0$}                     &
\colhead{$r_0$}                          &
\colhead{$M_{\bullet}$\tablenotemark{b}}  \\
\colhead{}                &
\colhead{}                &
\colhead{[mas yr$^{-1}$]} &
\colhead{[arcsec]}        &
\colhead{[$M_\odot$]}     &
\colhead{}
}
\startdata
\cutinhead{$W_0=8.6$}
0.00 & 97.74 & 0.613 & 23.06 &     0 &            \\
0.01 & 97.24 & 0.612 & 23.03 &   140 & $<\!68\%$ C.I. \\
0.02 & 96.89 & 0.611 & 23.21 &   282 &            \\
0.03 & 96.65 & 0.610 & 23.53 &   427 &            \\
0.04 & 96.49 & 0.609 & 23.92 &   576 &            \\
0.05 & 96.42 & 0.607 & 24.36 &   729 &   best fit \\
0.06 & 96.44 & 0.605 & 24.83 &   887 &            \\
0.07 & 96.56 & 0.603 & 25.32 &  1050 &            \\
0.08 & 96.78 & 0.602 & 25.85 &  1220 &            \\
0.09 & 97.12 & 0.600 & 26.39 &  1390 &            \\
0.10 & 97.55 & 0.598 & 26.95 &  1570 & $>\!68\%$ C.I. \\
0.11 & 98.11 & 0.596 & 27.52 &  1750 &            \\
0.12 & 98.80 & 0.594 & 28.08 &  1930 &            \\
0.13 & 99.63 & 0.592 & 28.61 &  2120 &            \\
0.14 & 100.61 & 0.590 & 29.12 &  2310 & $>\!95\%$ C.I. \\
0.15 & 101.73 & 0.588 & 29.57 &  2490 &            \\
0.16 & 103.00 & 0.586 & 30.01 &  2680 & $>\!99\%$ C.I. \\
0.17 & 104.41 & 0.584 & 30.37 &  2870 &            \\
0.18 & 106.00 & 0.583 & 30.68 &  3050 &            \\
%
\cutinhead{$W_0=8.1$}
0.00 & 97.79 & 0.613 & 21.61 &     0 &            \\
0.01 & 97.32 & 0.612 & 21.59 &   131 & 68\% C.I. \\
0.05 & 96.33 & 0.609 & 28.45 &   856 &   best fit \\
0.09 & 97.44 & 0.600 & 31.06 &  1640 & $>\!68\%$ C.I. \\
0.13 & 100.80 & 0.591 & 33.95 &  2510 & $>\!95\%$ C.I. \\
0.15 & 103.49 & 0.587 & 35.16 &  2950 & $>\!99\%$ C.I. \\
\cutinhead{$W_0=9.1$}
0.00 & 97.87 & 0.612 & 19.60 &     0 &            \\
0.01 & 97.44 & 0.611 & 19.63 &   119 & $<\!68\%$ C.I. \\
0.06 & 96.53 & 0.604 & 21.09 &   752 &   best fit \\
0.11 & 97.60 & 0.596 & 23.20 &  1470 & 68\% C.I. \\
0.15 & 100.29 & 0.590 & 24.75 &  2100 & 95\% C.I. \\
0.18 & 103.58 & 0.585 & 25.61 &  2560 & $>\!99\%$ C.I. \\
\enddata

\tablenotetext{a}{Fits are to the proper-motion velocity-dispersion profile
of stars with $R<90\arcsec$, $V<20$, $P(\chi^2)\geq0.001$ in both RA and
Dec components of proper motion, {\it and} uncertainties of $<0.25$ mas
yr$^{-1}$ in both components. Stars are binned in $1\arcsec$-wide
annuli, resulting in 87 points fitted.}

\tablenotetext{b}{Dimensional black-hole masses assume
a distance of $D=4.0$ kpc to 47 Tuc. For any other distance, multiply by
$(D/4.0\,{\rm kpc})^3$.}
\end{deluxetable}

\begin{figure}[!t]
\hspace*{+0.1truecm}
\centerline{\resizebox{90mm}{!}{\includegraphics{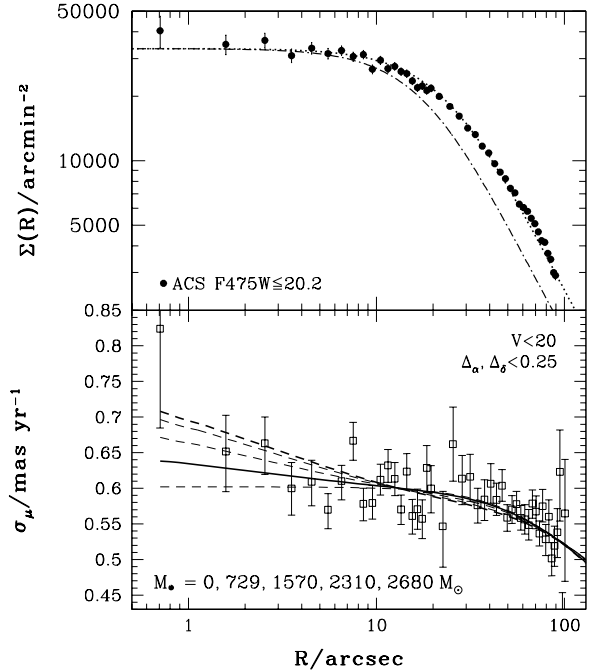}\hfil}}
\caption{
As for Figure \ref{fig:muvdens1}, but now for stars with $V<20$. Note the
better description provided for the density profile of this sample (which
includes stellar masses down to $m_*\simeq0.65\,M_\odot$; Table
\ref{tab:masslum}) by the $r_0=27\farcs5$ \citeauthor{king66}-model curve in
the upper panel. The kinematics sample in the lower panel is further
restricted by placing upper limits of 0.25 mas yr$^{-1}$ on the uncertainty in
both velocity components [as well as the usual limits on $P(\chi^2)$ and the
$(U-V)$ colors]. Curves in the bottom panel are again fits of $W_0=8.6$
black-hole \citeauthor{king66} models to the observed $\sigma_\mu(R)$. The
bold, solid curve is the formal best fit, the
lowermost curve has $M_\bullet=0$, and the other dashed curves represent the
68\%, 95\%, and 99\% upper limits on $M_\bullet$. Details of all these fits
are given in Table \ref{tab:bhfit20}.
\label{fig:muvdens2}}
\end{figure}

Table \ref{tab:bhfit20}, which has the same format as Table
\ref{tab:bhfit18.5}, shows that the best-fit black-hole mass inferred from
fitting this different sample of proper-motion stars is now
$M_{\bullet}\sim 700$--$800\,M_\odot$ (depending slightly on the assumed value
of $W_0$). The 68\% confidence interval is about
$(100\,M_\odot,1500\,M_\odot)$ and the 95\% confidence interval is roughly
$(0,2300\,M_\odot)$. At best this might suggest a $\sim$1-$\sigma$
consistency with the hypothesis of an intermediate-mass black hole, but
the fact that different velocity-sample definitions yield
formal best-fit masses that differ by nearly $1000\,M_\odot$ clearly
emphasizes the weak nature of any
black-hole signal in our data. It also has to be stressed that the sample
of stars with $V<20$ covers a non-negligible range of masses ($m_*\simeq
0.65$--$0.9\,M_\odot$; Table \ref{tab:masslum}), and moreover that the mixture
of mass classes varies significantly with radius as a result of selection
on the basis of errors (see Fig.~\ref{fig:errmua})---effects that are
ignored in our simple models.

Figure \ref{fig:muvdens2} shows, in its upper panel, the
completeness-corrected density profile of stars with estimated $V\la 20$ in
our master star list, Table \ref{tab:master}. As in Figure \ref{fig:muvdens1},
\citet{king66} models with $W_0=8.6$ and
$r_0=20\farcs84$ or $r_0=27\farcs5$ are drawn here.
The latter now provides a better description of the
spatial distribution in this closer-to-average mass range. The lower panel of
Figure \ref{fig:muvdens2} displays the true $\sigma_{\mu}(R)$ profile
and the best-fit
black-hole \citeauthor{king66} model with $W_0=8.6$ from Table
\ref{tab:bhfit20}, along with the 68\%, 95\%, and 99\% limits on $M_{\bullet}.$

Whatever small signature of a central mass concentration might appear in
Figures \ref{fig:muvdens1} and \ref{fig:muvdens2} obviously comes from the
innermost $R<3\arcsec$---or 0.06 pc, which would contain the sphere of
influence of any $<2000$-$M_\odot$ black hole---and in particular from the
high apparent $\sigma_\mu$ in the central 1\arcsec. However, there are only
8 stars at $R<1\arcsec$ in the $V<18.5$ velocity sample and 11 stars in the
error-selected $V<20$ sample, which is why the significance of the rise is
unavoidably low.

Sample size aside, it is also clear that
much more sophisticated modeling is required if these proper-motion data
are to be used either to prove or to disprove the presence of an
intermediate-mass black hole in 47 Tuc. Allowances must be made for a
nontrivial stellar mass spectrum; careful consideration must be given to
sample selection effects; and, most likely, higher-order moments of the
velocity distribution will need to be examined in addition to the dispersion.
On this last note, we now construct the full proper-motion distributions
in the central 3\arcsec\ of the cluster.

\subsubsection{Velocity Distributions in the Innermost ${\mathit 3\arcsec}$} 
\label{subsubsec:bhnmu}

\begin{figure}[!t]
\vspace*{-0.6truecm}
\centerline{\resizebox{100mm}{!}{\includegraphics{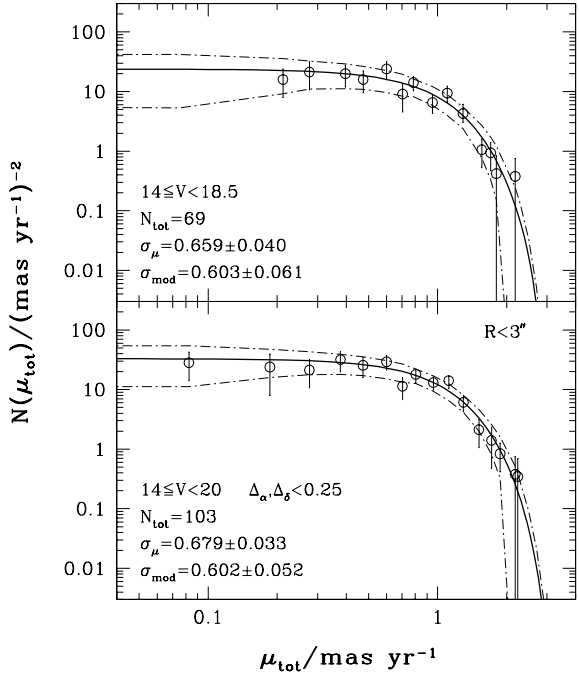}\hfil}}
\caption{
Two-dimensional velocity distributions within 3\arcsec\ of the cluster center
for the proper-motion samples to which black-hole models are fit in Figures
\ref{fig:muvdens1} and \ref{fig:muvdens2}. Solid curves are error-convolved
\citet{king66} model $N(\mu_{\rm tot})$ distributions for the $W_0=8.6$ fits
with {\it no} black hole in the two previous figures (top lines of Tables
\ref{tab:bhfit18.5} and \ref{tab:bhfit20}). The dash-dot curves
are the $\pm1$-$\sigma$ uncertainties on the model. The observed
(error-corrected) $\sigma_\mu$ indicated overlap with the model expectations
for no black hole at the one-sigma level, and more generally the observed
full distributions appear consistent with the model curves at the same level
(except for a paucity of slow-moving stars in the upper panel).
\label{fig:nmuinner}}
\end{figure}

Figure \ref{fig:nmuinner}
shows the two-dimensional velocity distribution $N(\mu_{\rm tot})$ for
stars with $R<3\arcsec$ in the two subsamples. These are obtained
from the data in the same way as described in \S\ref{subsec:twodvel}. The
solid curves in both panels are the error-convolved distributions for
regular $W_0=8.6$ \citeauthor{king66} models with no central black hole, also
computed as in \S\ref{subsec:twodvel} but with $r_0$ and $\sigma_0$ now set
by our fits to the velocity-dispersion profile alone (i.e., taken from the
first line of Table \ref{tab:bhfit18.5} for the upper panel, and the first
line of Table \ref{tab:bhfit20} for the lower panel). The dotted lines are
the one-sigma uncertainties on the predicted $N(\mu_{\rm tot})$. By and
large, the data and the model distributions with no black hole agree
reasonably well. The absence of a handful of expected stars with
$\mu_{\rm tot}\la0.2$ mas yr$^{-1}$ in the upper panel may be of some interest,
but it necessarily has low significance given the small sample size.
In any event, note that none of the high-velocity stars
discussed in \S\ref{subsec:highpm} fall in the zone $R<3\arcsec$.

\begin{figure}[!t]
\vspace*{-0.5truecm}
\centerline{\resizebox{90mm}{!}{\includegraphics{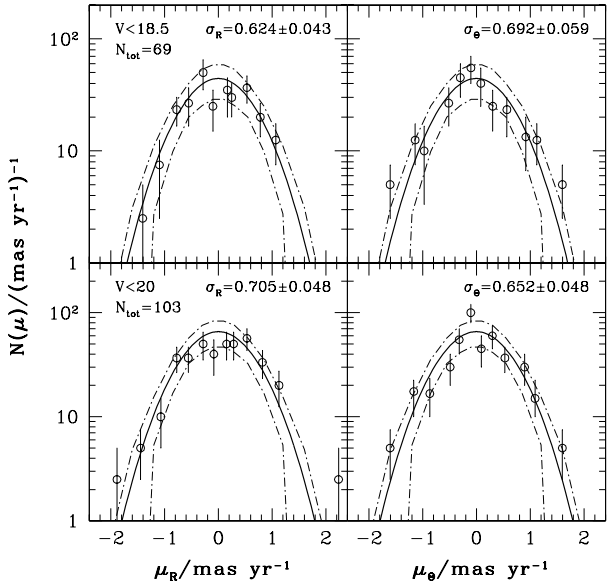}\hfil}}
\caption{
One-dimensional velocity distributions within 3\arcsec\ of the cluster center
for the proper-motion samples to which black-hole models are fit in Figures
\ref{fig:muvdens1} and \ref{fig:muvdens2}.
Left-hand panels show $N(\mu_R)$ for stars in the two samples of
Fig.~\ref{fig:nmuinner}, while right-hand panels display $N(\mu_\Theta)$ for
the same samples.
\label{fig:nmuinner1D}}
\end{figure}

Finally, Figure \ref{fig:nmuinner1D} shows the one-dimensional distributions
of the separate radial and tangential components of proper motion for stars
with $R<3\arcsec$ in the same two samples as before. The observed
distributions are obtained in the same way as the $N(\mu_{\rm tot})$
distributions we have focused on to this point; in particular, the errorbars
in Fig.~\ref{fig:nmuinner1D} are obtained by the same bootstrap procedure
described in \S\ref{subsec:twodvel}. For a given velocity sample, the
error-convolved model
curves for $N(\mu_R)$ (left-hand panels of Fig.~\ref{fig:nmuinner1D}) are
identical to those for $N(\mu_\Theta)$ (right-hand panels), due to the assumed
isotropy. They are calculated as in \S\ref{subsubsec:modconvol},
only using the appropriate one-dimensional Gaussian distribution of velocity
errors rather than equation (\ref{eq:bierrdist}) and with the function
$N_1(\mu_x|R/r_0)$ from equation (\ref{eq:onednorm}) taking the place of
$N_2(\mu_x,\mu_y|R/r_0)$ in equations (\ref{eq:twodconv}) and
(\ref{eq:twodave}).
As in Fig.~\ref{fig:nmuinner}, we use $\sigma_0$ and $r_0$ from the first
lines of Tables \ref{tab:bhfit18.5} and \ref{tab:bhfit20} to compare the
models and the data.

The excess of the observed velocity dispersion over that expected in a
\citet{king66} model with no black hole is perhaps more in evidence in these
plots than in Fig.~\ref{fig:nmuinner}. The broader wings
of the tangential $N(\mu_\Theta)$
distributions for both samples are particularly apparent, although it is also
clear that the discrepancies are at the $\sim$1-$\sigma$ level. The
distributions of radial proper motion in the left-hand panels may show more
interesting features, but again these have low significance. Specifically,
there appears to be an asymmetry in $N(\mu_R)$ in both velocity
samples. However, in the brighter sample with $V<18.5$ this is due to the
presence of just two stars moving radially inwards with
$\mu_R<-1.3$ mas yr$^{-1}$, vs.~none moving outwards with the corresponding
$\mu_R>+1.3$ mas yr$^{-1}$. In the fainter sample, the
asymmetry reflects the presence of four stars with $\mu_R<-1.3$ mas yr$^{-1}$
vs.~two with $\mu_R>1.3$ mas yr$^{-1}$. Similarly, the very
high observed dispersion $\sigma_R$ in the second sample
(which drives the non-zero best fit of $M_{\bullet}$
in Table \ref{tab:bhfit20} and Fig.~\ref{fig:muvdens2}) is due in large part
to a single star with $\mu_R\simeq-1.9$ mas yr$^{-1}$ and
one with $\mu_R\simeq+2.2$ mas yr$^{-1}$---neither of which is expected
in a regular \citeauthor{king66} (or Gaussian) model for a sample of this
size, but either of which could simply be
a statistical oddity.
It will be challenging indeed to build more
comprehensive velocity data sets, in this or any other globular cluster,
which might be able to overcome inherent statistical limitations such as
these and discern unambiguously the subtle kinematical effects of any
central mass concentration with $M_{\bullet}\sim 10^3\,M_\odot$ or less.

\section{Summary}
\label{sec:summary}

We have used multiple HST/WFPC2 and ACS images of the inner regions of 47
Tucanae (NGC 104), obtained between 1995 and 2002, to derive proper motions
for 14,366 stars within $R<100\arcsec$ of the cluster center---the largest
velocity data set yet collected for any globular cluster. At the same
time, we have used a single set of dithered ACS images to construct an
unbiased list of photometry and astrometrically calibrated positions for
nearly 130,000 stars in a $\simeq 3\arcmin \times 3\arcmin$
central field which contains the proper-motion sample. This ``master'' star
list is presented in Table \ref{tab:master} above (\S\ref{subsec:master}),
while the data going into our proper-motion determinations are in Table
\ref{tab:posdata} (\S\ref{subsec:pmsample}). Both of these catalogues can
be downloaded in their entirety through the electronic edition of
the {\it Astrophysical Journal}. Supplementary online material is also
available: an image ({\tt 47TucMaster.fits}) of a circular region
$R\le 150\arcsec$
around the cluster center, which contains the area covered by the master star
list of Table \ref{tab:master}; and an SM macro ({\tt pmdat.mon}) for
extracting and
plotting position vs.~time data from Table \ref{tab:posdata} for any star in
the proper-motion catalogue.

Section \ref{sec:hstdata} described in detail the basic procedures
and quality checks which we used to obtain calibrated astrometry and
relative position shifts (from local coordinate transformations) for many
thousands of stars with multi-epoch imaging
from different HST instruments with different pointings, fields of view,
and exposure times. Along the way, we used the unbiased master star list to
estimate the coordinates of the cluster center, finding our result to be in
good agreement with previous values from the literature (see Table
\ref{tab:basic}).

In \S\ref{sec:velsamples} we defined the subset of stars for which we derive
{\it relative} proper motions, in which the mean velocity of the entire
cluster is zero
by definition. We applied standard weighted, linear least-squares fitting to
find the plane-of-sky velocities and uncertainties, and to estimate the
quality of a straight-line fit to each component of motion. We found that
some $\simeq$10\% of the 14,366 proper-motion vectors we obtained were
either too uncertain, or not linear to high enough confidence, to include them
in subsequent kinematical analyses. However, these stars were retained in the
velocity catalogue of Table \ref{tab:posdata}.
Although we have not examined the issue
in this paper, in some cases the poor quality of linear proper-motion fits
may reflect truly nonlinear motion (rather than simple measurement errors or
noise) and thus be of physical interest. Section \ref{sec:velsamples} also
gave a brief description of a set of radial-velocity data---some already in
the literature, and some as yet unpublished---which we eventually used only
in order to estimate the distance to 47 Tuc.

Section \ref{sec:space} used stars in our unbiased master list again (not the
velocity sample, which has a highly irregular and biased spatial sampling) to
define the surface-density profile of stars with the main-sequence turn-off
mass in 47 Tuc (those with magnitudes $V\le 17.65$, or ${\rm F475W}\la 17.8$,
corresponding to $m_*\simeq0.85$--$0.9\,M_\odot$)
from $R\simeq70\arcsec$ to $R=0$.
After joining this profile on to ground-based $V$-band surface photometry from
the literature we fit the cluster structure to $R\approx1000\arcsec$ with
a single-mass, isotropic \citet{king66} model for a lowered isothermal
sphere. Our best-fit core radius, central surface brightness, and central
concentration are listed in Table \ref{tab:basic} and are in good
agreement with other estimates in the literature. This model does not
fit the density profile of 47 Tuc from $R>1000\arcsec$ to the tidal radius at
$R\sim3000\arcsec$; a realistic stellar mass function and/or velocity
anisotropy must be taken into account to explain the outer halo
structure. However, it does accurately describe the $R<100\arcsec$ field
covered by our proper-motion data, and thus we have used it as a convenient
point of reference for interpretation of some of the observed kinematics.
In particular, we have derived the {\it projected} one- and two-dimensional
velocity distributions as a function of clustercentric position in any
isotropic \citeauthor{king66} model (\S\ref{subsec:modvel} and Appendix
\ref{sec:kingveldist}), which could be profitably applied to other
large velocity samples in star clusters.

In \S\ref{sec:veldist} we presented the one- and two-dimensional velocity
distributions for our proper-motion sample and found them to be very regular
overall. For speeds less than the nominal central escape velocity in 47 Tuc
($\mu_{\rm max}\simeq2.6$ mas yr$^{-1}$, corresponding to $\simeq50$ km
s$^{-1}$ for a distance of 4 kpc), all distributions are well described by the
single-mass, isotropic \citeauthor{king66} model functions that we have
derived, after convolution with the velocity measurement errors and the stars'
spatial distribution. In turn, these model velocity distributions are nearly
Gaussian in the core regions of 47 Tuc, since this massive cluster defines a
rather deep potential well. We compared the observed and model distributions
as a function of radius for the brightest stars with the smallest proper-motion
uncertainties, showing that the velocity dispersion decreases slightly
more slowly with clustercentric radius than expected for a
single-mass \citeauthor{king66} model (\S\ref{subsec:twodvel}, Figure
\ref{fig:nmurad}). We argued that this implies that the
mass distribution in the inner few core radii of 47 Tuc is dominated by stars
less massive than the main-sequence turn-off mass, rather than by heavy
remnants such as neutron stars or massive white dwarfs. This is consistent
with other suggestions, from independent modeling and observations, that the
total mass of such remnants is quite low in this cluster.

In \S\ref{subsec:highpm} we took especial care to examine the properties of
stars with apparent proper-motion speeds above the nominal central escape
velocity of 2.6 mas yr$^{-1}$. Only a few dozen such stars are found,
amounting to $\la0.3\%$ of our total sample. Of these, fewer than ten, or
$<0.1\%$, can be said unequivocally not to have been scattered to high
velocities by measurement errors, and it is still possible that some of these 
may be interloping field stars. Thus, we have not found any evidence for a
significant population of very fast-moving, ``nonthermal'' stars in the core
of 47 Tuc, such as had been suggested on the basis of some earlier
radial-velocity surveys involving much smaller samples. We emphasize, however,
that we have not examined in any detail the $\sim5\%$ of stars in our
catalogue which are not fit well by straight-line proper motions. It is
conceivable that some legitimate nonlinear motions among these stars could be
of physical interest in connection with the close encounters and scattering
processes expected in this dense stellar environment.

Section \ref{sec:veldisp} looked at a few aspects of the proper-motion
velocity dispersion only. In \S\ref{subsec:strag} we showed that the velocity
dispersion of 18 blue stragglers within $R<20\arcsec$ (one core radius) is
smaller than the dispersion of comparably bright stars on the cluster
red-giant branch, by a factor of about $\sqrt{2}$. Together with other studies
in the literature showing that blue stragglers are more centrally concentrated
than the giants in 47 Tuc, this is consistent with the blue stragglers being
on average about twice as massive as normal main-sequence turn-off stars.

In \S\ref{subsec:kinmags} we presented the run of proper-motion dispersion
with clustercentric radius for stars in various magnitude ranges, compared it
with the expected behavior in a single-mass \citet{king66} model, and showed
quantitatively that the average velocity anisotropy in the inner
$R<100\arcsec\simeq5$ core radii of 47 Tuc is---as expected---negligible.

Section \ref{subsec:distance} compared the
line-of-sight velocity-dispersion profile defined by 419 bright giants
($V<14$) within $R<105\arcsec$ with the proper-motion dispersion profile of
6,336 of our stars with $V<18.5$ (and, thus, masses comparable to the
giants). This led to a kinematic estimate of the distance to the cluster:
$D=4.02\pm0.35$ kpc. While this result was obtained by ignoring the rotation
of 47 Tuc, on the small spatial scales probed by our kinematics the rotation
is quite weak and should not affect our estimate of $D$ by more than 2--3
percent at most (\S\ref{subsec:norotation}).
Our distance is consistent with results in the literature from fitting to the
white-dwarf cooling sequence. We also used a \citeauthor{king66} model to
extrapolate the velocity dispersions to $R=0$, giving improved estimates of
the central $\sigma_\mu$ and $\sigma_z$ (see Table \ref{tab:basic}).

Finally, \S\ref{subsec:bhmods} focused on the kinematics and
density structure in the innermost few arcseconds of 47 Tuc. We fit the
proper-motion velocity-dispersion profile with simplistic models based on
those of \citet{king66} but allowing for the presence of a central point mass. 
We found no clear need here for an ``intermediate-mass'' black
hole of the type recently claimed to have been detected in other
globular clusters. Nevertheless, one with a mass
$M_{\bullet}\la1000$--$1500\,M_\odot$ could be accommodated by the data
at the 1-$\sigma$ level. It would thus seem difficult to use these data alone
either to prove or to disprove whether the well known correlation, between
supermassive black-hole mass and nuclear velocity dispersion in galaxies,
extends to globular clusters.

The overarching conclusion from this work is that the core of 47 Tucanae is
remarkably unsurprising in its dynamical structure. The velocity distribution
is essentially isotropic and close to Gaussian, with radial and (secondarily)
mass dependences which---aside from a few relatively small details which can be
at least qualitatively understood---roughly match the expectations for standard
\citeauthor{king66}-type models of dynamically relaxed, lowered-isothermal
spheres. Our work was aimed expressly at an investigation of the core of
this cluster, and our results have emphasized that any truly new insight
into star-cluster dynamics
from large proper-motion samples may be more likely to come
from surveys of more off-center fields, where the potential is weaker,
relaxation may not be so advanced, and velocity anisotropy in particular
could be much more pronounced.

\acknowledgments

We are very grateful to the referee, Craig Heinke, for an exceptionally helpful
report.
JA and GM also thank Ivan King for discussions and support in the early stages
of this project.
DEM and GM thank the Space Telescope Science Institute, where a large portion
of this work was done, and the Director's Discretionary Fund at STScI for
financial support.
JA acknowledges support from NASA through HST grants AR-7993 and GO-9443
from the Space Telescope Science Institute, which is operated by AURA, Inc.,
under NASA contract NAS 5-26555.
GM acknowledges partial support from the Swiss National Science Foundation
(SNSF).
The National Science Foundation of the US provided a CAREER grant to KG
(AST 03-49095) and a grant to CP (AST 00-98650).
DM was supported by grant No.~15010003 from FONDAP Center for Astrophysics,
and by a Fellowship from the John Simon Guggenheim Foundation.

{\it Facility:} \facility{HST (WFPC2,ACS)}


\begin{appendix}
\section{Correcting Velocity Dispersion for Measurement and Transformation
         Errors}
\label{sec:errcor}

This Appendix addresses the need to interpret our observed velocity
dispersions in terms of the true dispersion (of the intrinsic stellar velocity
distribution) and the additional contribution made by our measurement
errors. There are two sources of error. The first, which is random, relates to
the effect that simple errors in position measurements have on the observed
proper
motions. The second, which is systematic, comes from our use of local
transformations to bring a star's coordinates at any epoch into our
master-frame system (see \S\ref{subsubsec:local}).

Recall that by applying standard linear regression to infer proper motions
from our measurements of displacement vs.\ time, we have assumed that the
position measurement error for each star at each epoch is Gaussian
distributed with a mean of 0, in which case the errorbars on the fitted
$\mu_\alpha$ and $\mu_\delta$ represent the dispersions of Gaussian
distributions of RA- and Dec-velocity measurement error. The errorbars on
the two velocity components need not be the same for any one star
and are in general
different from one star to the next, but the mean measurement error is always
assumed to be 0. Thus, the distribution of measurement error
in either component of velocity for a set of any ${\cal N}$ stars from our
sample is given by the normalized probability density
\begin{equation}
\label{eq:errdist}
P(\delta\mu) = \frac{1}{{\cal N}}\sum_{i=1}^{{\cal N}}
               \frac{1}{\sqrt{2\pi} \Delta_i}
               \exp\left[-\frac{(\delta\mu)^2}{2\Delta_i^2}\right]\ ,
\end{equation}
where $\delta\mu$ represents the error in either the RA or the Dec
component of the observed proper motion and $\Delta_i$ is the uncertainty
in that component for the $i^{\rm th}$ star in the sample, as given by our
least-squares fitting.

Any observed velocity $\mu_{\rm obs}$ can then be viewed as the sum
\begin{equation}
\label{eq:muobs}
\mu_{\rm obs} = \mu_{\rm cor} + \delta\mu \ ,
\end{equation}
where $\mu_{\rm cor}$ is a random variable drawn from the (a priori unknown)
error-free velocity distribution as a function of stellar position and
magnitude in 47 Tuc, and the measurement error $\delta\mu$ is a random
variable drawn from the distribution
(\ref{eq:errdist}). If the measurement errors are uncorrelated with the
velocities themselves (as the lack of correlation between the observed
$\mu_\alpha$ and the least-squares uncertainties $\Delta_\alpha$ in Figure
\ref{fig:errmua} of \S\ref{subsubsec:pmerrors}
implies is the case), then the expectation value of the
observed RA or Dec velocity for a set of ${\cal N}$ stars is simply
\begin{equation}
\label{eq:expmu}
E(\mu_{\rm obs}) = E(\mu_{\rm cor}) + E(\delta\mu) = E(\mu_{\rm cor})\ ,
\end{equation}
since $E(\delta\mu)$ vanishes by virtue of equation (\ref{eq:errdist}). Thus,
an unbiased estimate of the mean velocity with no measurement error is
provided by the usual
\begin{equation}
\label{eq:meanmu}
\langle\mu\rangle_{\rm cor} = \langle\mu\rangle_{\rm obs} =
    \frac{1}{{\cal N}} \sum_{i=1}^{{\cal N}} \mu_{i,\,\rm{obs}}\ .
\end{equation}

For uncorrelated random variables, the variances are similarly additive,
regardless of the form of the underlying distribution(s) of the variables.
Thus,
\begin{equation}
\label{eq:varmua}
{\rm Var}(\mu_{\rm obs}) = {\rm Var}(\mu_{\rm cor}) + {\rm Var}(\delta\mu)\ ,
\end{equation}
or
\begin{equation}
\label{eq:varmub}
E[(\mu_{\rm obs}-\langle\mu\rangle_{\rm obs})^2] =
E[(\mu_{\rm cor}-\langle\mu\rangle_{\rm cor})^2] +
E[(\delta\mu)^2] \ .
\end{equation}
Switching now to more standard notation, we write $\sigma_{\rm obs}^2 \equiv
{\rm Var}(\mu_{\rm obs})$ and  $\sigma_{\rm cor}^2 \equiv
{\rm Var}(\mu_{\rm cor})$; use equation (\ref{eq:errdist}) to evaluate
${\rm Var}(\delta\mu)$; and finally use the sample standard deviation as an
unbiased estimate of $\sigma_{\rm obs}$ to obtain
\begin{equation}
\label{eq:sigmu}
\sigma_{\rm cor}^2 =
    \frac{1}{{\cal N}-1} \left[\sum_{i=1}^{{\cal N}} (\mu_{i,\,{\rm obs}} -
           \langle\mu\rangle_{\rm obs})^2\right]
  - \frac{1}{{\cal N}}\sum_{i=1}^{{\cal N}} \Delta_i^2 \ .
\end{equation}
That is, the proper estimator for the error-corrected velocity dispersion
$\sigma_{\rm cor}$ of any ${\cal N}$ stars with unequal velocity uncertainties
is obtained by subtracting the rms errorbar in quadrature from the usual
standard deviation of the observed velocities. The only assumption behind
equation (\ref{eq:sigmu}) is that the velocity {\it errors} are normally
distributed, and thus it holds true for any form of the actual
velocity distribution. In particular, the latter need not be
Gaussian---indeed, it need not be specified at all.

There remains one further correction to be applied to $\sigma_{\rm cor}$
in order to reach an estimate of the true velocity
dispersion: As was mentioned in \S\ref{subsubsec:local}, the use of local
transformations to map stellar positions at our ten epochs into the
master-frame coordinate system introduces an unavoidable error in the
observed velocity dispersion, which is completely separate from the
inflation of dispersion due to measurement error.

Suppose that, as in \S\ref{subsubsec:local}, we use the positions of the
${\cal N}_{\rm trans}$ nearest neighbors of a star at some epoch to define
the local transformation of that star into the reference-frame coordinate
system. Suppose further that there are no position-measurement errors in any
frame at any epoch. Then the formal error in all local transformations is
zero, and the mean and dispersion of the observed velocities are
automatically $\langle\mu\rangle_{\rm cor}$ and $\sigma_{\rm cor}$ in the
notation of the preceding discussion. How do these relate to the true
moments of the intrinsic velocity distribution?

To address this question we look, without loss of generality, at a single
component (any component) of position in the master-frame system, measured
for a given target star and for its nearest ${\cal N}_{\rm trans}$ neighbors
at any
two epochs separated by $\Delta t$ in time. Call these positions $x_0$ for the
target star and $\{x_j; j=1, \dots, {\cal N}_{\rm trans}\}$ for the neighbor
stars.
By applying local coordinate transformations, the proper motion
$\mu_{\rm cor}$ we derive for a star is really measuring the change in its
position relative to the centroid of its nearest neighbors at the two epochs.
That is, if
\begin{equation}
\label{eq:centroid}
x_{\rm cen} \equiv \frac{1}{{\cal N}_{\rm trans}}
   \sum_{j=1}^{{\cal N}_{\rm trans}} x_j
\end{equation}
denotes the centroid of the neighbor stars, then
\begin{equation}
\label{eq:relmu}
\mu_{\rm cor}\,\Delta t
                = (x_0^\prime-x_{\rm cen}^\prime) - (x_0-x_{\rm cen})
                = (x_0^\prime - x_0) - (x_{\rm cen}^\prime - x_{\rm cen}) \ ,
\end{equation}
where the primes identify the second-epoch positions. Then, noting that the
expectation value of the position change for any star is $E(x^\prime - x)
= \Delta t\,E(\mu_{\rm true})$ (where $\mu_{\rm true}$ refers to the true
distribution of absolute proper motions),
equations (\ref{eq:relmu}) and (\ref{eq:centroid}) together imply that
\begin{equation}
\label{eq:relmean}
E(\rm \mu_{\rm cor}) = 0 \ .
\end{equation}
This only says, quite sensibly, that on average the velocity of a star relative
to its nearest neighbors will be 0. This is, of course, exactly what the true
situation is, so that even though the local transformation scheme leads to an
error in every observed stellar velocity, the {\it average} error is 0.

By contrast, the relationship between the observed dispersion
$\sigma_{\rm cor}$ and the second moment $\sigma_{\rm true}$ of
the true velocity distribution is nontrivial.
From equation (\ref{eq:relmu}) we have
\begin{equation}
\label{eq:relvar}
\begin{array}{lll}
(\Delta t)^2\,{\rm Var}(\mu_{\rm cor}) & = &
     {\rm Var}(x_0^\prime - x_0) + {\rm Var}(x_{\rm cen}^\prime-x_{\rm cen})
     \\
   & = & (\Delta t)^2\,{\rm Var}(\mu_{\rm true}) +
         (\Delta t)^2\,{\rm Var}(\mu_{\rm cen}) \ ,
\end{array}
\end{equation}
where
 $\mu_{\rm cen}\equiv
   (1/{\cal N}_{\rm trans})\sum_{j=1}^{{\cal N}_{\rm trans}}
   (x_j^\prime - x_j)/\Delta t$
from equation (\ref{eq:centroid}). Clearly,
${\rm Var}(\mu_{\rm true}) = \sigma_{\rm true}^2$, while the variance of
$\mu_{\rm cen}$ follows from its definition as the sample mean for the
${\cal N}_{\rm trans}$ neighbors of our target star:
${\rm Var}(\mu_{\rm cen}) = \sigma_{\rm true}^2/{\cal N}_{\rm trans}$.
Finally, then, equation (\ref{eq:relvar}) yields 
\begin{equation}
\label{eq:sigcor}
\sigma_{\rm cor}^2 =
   \left(1+\frac{1}{{\cal N}_{\rm trans}}\right) \sigma_{\rm true}^2 \ .
\end{equation}

Thus, from the combination of equations (\ref{eq:meanmu}) and
(\ref{eq:relmean}) we {\it expect} to find
\begin{equation}
\label{eq:meantrue}
\langle\mu\rangle_{\rm obs} \equiv
    \frac{1}{{\cal N}} \sum_{i=1}^{{\cal N}} \mu_{i,{\rm obs}} = 0
\end{equation}
for the mean relative proper motion (in any direction) of any set of
stars; and by combining equations (\ref{eq:sigmu}) and
(\ref{eq:sigcor}) we {\it calculate} 
\begin{equation}
\label{eq:sigtrue}
\sigma_{\rm true}^2 = 
    \left(1+\frac{1}{{\cal N}_{\rm trans}}\right)^{-1}
    \left\{
    \frac{1}{{\cal N}-1} \left[\sum_{i=1}^{{\cal N}} (\mu_{i,\,{\rm obs}} -
           \langle\mu\rangle_{\rm obs})^2\right]
  - \frac{1}{{\cal N}}\sum_{i=1}^{{\cal N}} \Delta_i^2
    \right\}
\end{equation}
to estimate their true one-dimensional velocity dispersion.

Equation (\ref{eq:sigtrue}), with ${\cal N}_{\rm trans}=45$ following the
discussion in \S\ref{subsubsec:local}, is used repeatedly in our analyses of
\S\ref{sec:veldist} and \S\ref{sec:veldisp}, to find
the intrinsic RA and Dec or radial and azimuthal velocity dispersions for
various subsets of our proper-motion sample.

\section{Projected Velocity Distributions in King (1966) Models}
\label{sec:kingveldist}

It is well known that the velocity distribution at any radius in a
\citet{king66} model cluster
takes the form of a ``lowered Maxwellian,'' i.e., a Gaussian distribution in
$v=(v_x^2+v_y^2+v_z^2)^{1/2}$ minus a position-dependent constant related to
the local escape speed. When comparing such models to data, however, it is
necessary to integrate this simple distribution over all phase-space
coordinates for which we have no direct information. This will always include
at least a projection
along the line of sight, which is already nontrivial. In a case where only
proper-motion data are being considered, an additional integration over one
component of velocity is also required; while if only radial-velocity data are
being modeled, yet another integration over a second velocity component is
necessary. In this Appendix, we develop general expressions and
outline a computational scheme for these various marginalized velocity
distributions.

If $E=v^2/2+\phi(r)$ is the specific energy in a self-gravitating cluster of
equal-mass stars, with $\phi(r)$ a relative potential defined so that $\phi=0$
at some tidal radius to be determined, then the phase-space distribution
function of a \citet{king66} model \citep[see also][Section 4.4]{bt87} is
defined as
\begin{equation}
\label{eq:kingdef}
f(E) = \left\{
\begin{array}{lll}
\varrho(2\pi\sigma_0^2)^{-3/2}\left(e^{-E/\sigma_0^2} - 1\right)
   & ,  & \quad E<  0 \\
0  & ,  & \quad E\ge0 \ ,
\end{array}
\right.
\end{equation}
where $\sigma_0$ is a velocity scale and $\varrho$ is a normalization factor
related to the central space density. The ``lowering'' term $(-1)$ in the
top line allows for the possibility of a finite escape velocity:
this distribution function only allows for stars with $E<0$, and thus
\begin{equation}
\label{eq:vmax}
v< v_{\rm max}(r) \equiv \sqrt{-2\phi(r)}\ .
\end{equation}
The usual space density
$\rho(r)$ then follows from integrating $f(E)$ over all allowed velocities:
\begin{equation}
\label{eq:modrho}
\rho(r) = \int_{0}^{v_{\rm max}(r)} f(E)\, 4\pi v^2\,dv =
 \varrho\left\{e^{W(r)}\, {\rm erf}[\sqrt{W(r)}] -
   \sqrt{4W(r)/\pi}\,\left[1+2W(r)/3\right]\right\} \ ,
\end{equation}
where $W(r)\equiv -\phi(r)/\sigma_0^2$ is the dimensionless relative
potential. If $W_0$ is the (positive, but otherwise arbitrary) value of $W$ at
$r=0$, then equation \ref{eq:modrho} defines $\varrho$ in terms of $W_0$ and
the central density $\rho_0$. The full density and potential profiles 
are obtained from Poisson's equation in dimensionless form:
\begin{equation}
\label{eq:poisson}
\frac{d}{d\widetilde{r}}\left(\widetilde{r}^2\,\frac{dW}{d\widetilde{r}}\right)
   = -9\, \frac{\rho}{\rho_0}\, \widetilde{r}^2\ ,
\end{equation}
in which
\begin{equation}
\label{eq:kingr0}
\widetilde{r}\equiv r/r_0\ ,
   \qquad r_0^2 \equiv 9\sigma_0^2 / (4\pi G \rho_0)\ .
\end{equation}
Equation (\ref{eq:modrho}) is used to write $\rho(\widetilde{r})/\rho_0$ in
terms of $W(\widetilde{r})$ and $W_0$, after which equation
(\ref{eq:poisson}) is integrated numerically [subject to the specification
of $W_0$ and the additional boundary condition
$(dW/d\widetilde{r})_{\widetilde{r}=0}=0$] to find $W$ and $\rho/\rho_0$
as functions of $r/r_0$.
The radius at which $\rho(r)$ vanishes defines the tidal radius of the
cluster, and the concentration parameter $c\equiv \log(r_t/r_0)$ is a
one-to-one function of $W_0$ (the \citeauthor{king66} model fit to 47 Tuc
in Figure \ref{fig:density} of \S\ref{subsec:density}, with $c=2.01$,
has $W_0=8.6$).
Subsequent projection of $\rho(r)$ along the line of sight $z$ is
straightforward, yielding the surface density as a function of projected
radius $R=(r^2-z^2)^{1/2}$:
\begin{equation}
\label{eq:modsurf}
\frac{\Sigma(\widetilde{R})}{\rho_0 r_0} =
     2 \int_0^{\left(\widetilde{r}_t^2-\widetilde{R}^2\right)^{1/2}}
   \widetilde{\rho}\,(\widetilde{r}\,)\ d\widetilde{z}\ ,
\end{equation}
where $\widetilde{R}=R/r_0$, $\widetilde{z}=z/r_0$,
$\widetilde{\rho}=\rho/\rho_0$, and $\widetilde{r}=r/r_0$ is evaluated as
$(\widetilde{R}^2+\widetilde{z}^2)^{1/2}$ in the integrand.
While $\rho(r)$ and $\Sigma(R)$ are actually mass densities, for a
single-mass cluster they are directly proportional to the stellar
number-density profiles, and to the luminosity density or
surface-brightness profiles. It is equation (\ref{eq:modsurf}) that was fit to
the surface-brightness profile constructed in \S\ref{subsec:density} to
constrain $W_0$, $r_0$, and $\rho_0$ (and thus to derive $c$ and $\mu_{V,0}$)
for 47 Tuc.

All of the above is completely standard and has been repeated here simply to
give some basic definitions and to establish our notation. The usual next step
in kinematical applications of these models is to obtain the dimensionless
velocity-dispersion profile, $\sigma_r/\sigma_0$ vs.\ $r/r_0$, from the
definition
\begin{equation}
\label{eq:moddisp1}
\sigma_r^2(r)=\frac{1}{3\rho(r)}\,\int_0^{v_{\rm max}(r)}
v^2f(E)\,4\pi v^2 dv\ ,
\end{equation}
which implies (using eqs.~[\ref{eq:kingdef}] and [\ref{eq:vmax}], and
eq.~[\ref{eq:modrho}] evaluated at $r=0$)
\begin{equation}
\label{eq:modsig}
\frac{\sigma_r^2(\widetilde{r})}{\sigma_0^2} =
  \frac{1}{3\widetilde{\rho}(\widetilde{r})}\,\frac{1}{\sqrt{2\pi}}\,
   \int_0^{\sqrt{2W(\widetilde{r})}}
  \frac{{\rm exp}\left[W(\widetilde{r})-v^2/2\sigma_0^2\right]-1}
     {{\rm exp}(W_0)\,{\rm erf}\left(\sqrt{W_0}\right)
       - \sqrt{4W_0/\pi}\,[1+2W_0/3]}\ 
  \frac{v^4}{\sigma_0^4}\ \frac{dv}{\sigma_0}\ .
\end{equation}
After finding the profile
$\widetilde{\sigma}_r^2 = \sigma_r^2/\sigma_0^2$, it is
again straightforward to integrate it along the line of sight to obtain the
projected velocity dispersion
\begin{equation}
\label{eq:modsigz}
\frac{\sigma_z^2(\widetilde{R})}{\sigma_0^2} =
    2\left[\frac{\Sigma(\widetilde{R})}{\rho_0 r_0}\right]^{-1}
      \int_0^{(\widetilde{r}_t^2 - \widetilde{R}^2)^{1/2}}
        \widetilde{\rho}(\widetilde{r})\
          \widetilde{\sigma}_r^2(\widetilde{r})\ d\widetilde{z}
\end{equation}
for direct comparison with radial-velocity observations. More generally,
because of the assumption of velocity isotropy in these models, $\sigma_z$ is
also the dispersion of any component of proper motion as a function of
projected clustercentric radius. Note that neither equation (\ref{eq:modsig})
nor equation (\ref{eq:modsigz}) requires that the parameter $\sigma_0$
correspond exactly to
the actual velocity dispersion at the center of the cluster; and indeed,
$\sigma_r(r=0)\ne\sigma_0$ and $\sigma_z(R=0)\ne\sigma_0$ in general for these
models. (It is only in the limit of infinite $W_0$ that the central
dispersions tend to $\sigma_0$). Section \ref{sec:veldisp} compares
the predictions of equation (\ref{eq:modsigz}), for models appropriate
to the spatial structure of 47 Tuc, to the observed dispersion profiles of
our proper-motion and line-of-sight velocity samples.

We are additionally interested here in the full distribution of projected
velocities, not just the second moment; and we have data for the joint
distribution of velocities in two orthogonal directions on the plane of the
sky, not just a single component. Thus, consider first the distribution
of three-dimensional velocity implied by equation (\ref{eq:kingdef}). At any
fixed radius $r$, this can be written (using the definition of $v_{\rm max}$ in
eq.~[\ref{eq:vmax}]) as
\begin{equation}
\label{eq:fulldist}
n_3(v_x,v_y,v_z | r) = \frac{\varrho\ e^{v^2_{\rm max}/2\sigma_0^2}}
{(2\pi \sigma_0^2)^{3/2}}\,
    \left[e^{-(v_x^2+v_y^2+v_z^2)/2\sigma_0^2} -
          e^{-v^2_{\rm max}/2\sigma_0^2}\right]\ ,
    \qquad v_x^2+v_y^2+v_z^2\le v^2_{\rm max}
\end{equation}
with $v^2_{\rm max}/2\sigma_0^2 = W(r)$ already
known as a function of radius from the procedure outlined just above. Recall
that $\int\!\!\int\!\!\int n_3(v_x,v_y,v_z)\, dv_xdv_ydv_z = \rho(r)$ by
definition of the phase-space density $f$.

Marginalizing the distribution (\ref{eq:fulldist}) over the two
velocity components $v_x$ and $v_y$ (for example) gives the velocity
distribution at $r$ in the $v_z$ direction alone:
\begin{eqnarray}
\label{eq:onedista}
n_1(v_z | r) & = &
     \int\!\!\!\int_{-v_{\rm max}(r)}^{v_{\rm max}(r)}
     n_3(v_x,v_y,v_z | r)\, dv_xdv_y                      \nonumber \\
           & = &
     \frac{\varrho}{(2\pi\sigma_0^2)^{1/2}}\,
     \left[ e^{(v_{\rm max}^2-v_z^2)/2\sigma_0^2} - 
           \left(1 + \frac{v_{\rm max}^2-v_z^2}{2\sigma_0^2}\right)
     \right]\ ,
\end{eqnarray}
which is the number of stars per unit $z$-velocity and per unit spatial
volume, at radius $r$ in the cluster.
Alternatively, equations (\ref{eq:vmax}) and (\ref{eq:modrho}) can be used
to write $v^2_{\rm max}$ and $\varrho$ in terms of $W(r)$ and $\rho(r)$,
so that
\begin{equation}
\label{eq:onedist}
n_1(v_z | r) = \frac{\rho(r)}{\sqrt{2\pi} \sigma_0} \,
\frac{{\rm exp}[W(r)-v_z^2/2\sigma_0^2] - 1
       - [W(r)-v_z^2/2\sigma_0^2]}
     {{\rm exp}[W(r)]\,{\rm erf}[\sqrt{W(r)}]
       - \sqrt{4W(r)/\pi}\,[1+2W(r)/3]} \ .
\end{equation}
By virtue of the assumed isotropy, the
distributions $n_1(v_x | r)$ and $n_1(v_y | r)$ are formally identical to
equation (\ref{eq:onedist}). Clearly, $n_1(v_{\rm max} | r) = 0$ always.

For comparison with observations, equation (\ref{eq:onedist}) must be
averaged along the line of sight. Denoting this coordinate by
$z=(r^2-R^2)^{1/2}$ again, and normalizing to a density of one star per unit
area at projected radius $R$ on the plane of the sky, we define
\begin{equation}
\label{eq:oneproj}
N_1(v_z|R) \equiv \frac{2 \int_0^{(r_t^2-R^2)^{1/2}} n_1(v_z | r)\,dz}
   {2 \int_0^{(r_t^2-R^2)^{1/2}} \rho(r)\,dz}\ ,
\end{equation}
or, in terms of the model's dimensionless radius, potential and surface
density,
\begin{equation}
\label{eq:onednorm}
N_1(v_z | \widetilde{R}) = 
   \frac{1}{\sqrt{2\pi}\sigma_0}\,
   \frac{2}{\widetilde{\Sigma}(\widetilde{R})}
        \int_0^{(\widetilde{r}_t^2 - \widetilde{R}^2)^{1/2}}
    \frac{{\rm exp}[W(\widetilde{r})-v_z^2/2\sigma_0^2] - 1
       - [W(\widetilde{r})-v_z^2/2\sigma_0^2]}
     {{\rm exp}(W_0)\,{\rm erf}\left(\sqrt{W_0}\right)
       - \sqrt{4W_0/\pi}\,[1+2W_0/3]} \ d\widetilde{z}\ ,
\end{equation}
where $\widetilde{\Sigma}$ is the dimensionless $\Sigma(\widetilde{R})/\rho_0
r_0$ given by equation (\ref{eq:modsurf}) and, as also in that definition,
$\widetilde{R}=R/r_0$, $\widetilde{z}=z/r_0$, 
and $\widetilde{r}=r/r_0=(\widetilde{R}^2+\widetilde{z}^2)^{1/2}$.
Note that this distribution has dimensions of inverse velocity and is
normalized such that $\int\! N_1(v_z | \widetilde{R})\ dv_z = 1$
at any $\widetilde{R}$. Thus, $N_1(v_z|\widetilde{R})$ is the probability
that a single star at projected clustercentric radius $R/r_0$ will have a
line-of-sight velocity between $v_z$ and $(v_z+dv_z)$. As with the volume
probability distribution $n_1(v_z|r)$, the corresponding distribution for any
one component of velocity on the plane of the sky has the same form as equation
(\ref{eq:onednorm}).

The two-dimensional distribution of velocities projected onto any spatial
plane can similarly be
derived from equation (\ref{eq:fulldist}). Obviously proper motions are the
motivation for looking at this, so let us now write
$\mu_x\equiv v_x$ and $\mu_y\equiv v_y$. Then
integrating equation (\ref{eq:fulldist}) over all velocities $v_z$ (with
$|v_z|\le v_{\rm max}$ as usual) and defining the auxiliary variable
\begin{equation}
\label{eq:bigu}
U(r) \ \equiv \ 
   \frac{v_{\rm max}^2}{2\sigma_0^2} - \frac{\mu_x^2+\mu_y^2}{2\sigma_0^2}
          \ = \ 
    W(r) - \frac{\mu_x^2+\mu_y^2}{2\sigma_0^2}
\end{equation}
gives
\begin{equation}
\label{eq:twodist}
n_2(\mu_x,\mu_y | r) = \frac{\rho(r)}{2\pi\sigma_0^2}\,
    \frac{{\rm exp}[U(r)]\,
          {\rm erf}[\sqrt{U(r)}]
           - \sqrt{4U(r)/\pi}}
         {{\rm exp}[W(r)]\,{\rm erf}[\sqrt{W(r)}]
           - \sqrt{4W(r)/\pi}\,[1+2W(r)/3]}
\end{equation}
for the number of stars per unit velocity element $d\mu_xd\mu_y$, per unit
spatial volume, at radius $r$ in the cluster. Through $U(r)$, this
distribution is a function
only of the total proper-motion speed, $\mu_{\rm tot}=(\mu_x^2+\mu_y^2)^{1/2}$
(i.e., it is circularly symmetric in the
$\mu_x$--$\mu_y$ plane of phase space), again because of the assumed
velocity isotropy. It vanishes at any radius
$r$ if $\mu_{\rm tot}^2/2\sigma_0^2\ge W(r)$. Averaging along
the line of sight as in equation (\ref{eq:oneproj})---including the
normalization there to one star per unit area on the sky---then gives the
observable 
\begin{equation}
\label{eq:twodnorm}
N_2(\mu_x,\mu_y | \widetilde{R}) = \frac{1}{2\pi\sigma_0^2}\,
                         \frac{2}{\widetilde{\Sigma}(\widetilde{R})}\,
 \int_{0}^{(\widetilde{r}_t^2-\widetilde{R}^2)^{1/2}}
    \frac{{\rm exp}\left[U(\widetilde{r})\right]\,
          {\rm erf}\left[\sqrt{U(\widetilde{r})}\right]
           - \sqrt{4U(\widetilde{r})/\pi}}
         {{\rm exp}(W_0)\,{\rm erf}\left(\sqrt{W_0}\right)
           - \sqrt{4W_0/\pi}\,[1+2W_0/3]} \ d\widetilde{z}\ ,
\end{equation}
for the dimensionless model quantities $\widetilde{\Sigma}$, $\widetilde{R}$,
$\widetilde{z}$, and $\widetilde{r}$ all defined as above. Note that 
$N_2(\mu_x,\mu_y | \widetilde{R})$ has units of inverse velocity squared
(it is analogous to a spatial surface density), and it is normalized so that
its integral over all allowed $\mu_x$ and $\mu_y$ (at any fixed $R/r_0$)
is always unity. Finally, note that $\mu_x$ and $\mu_y$ could, in fact, be
{\it any} two orthogonal components of velocity without changing any of the
results here. In particular, the joint distribution of radial and tangential
proper motions, $N_2(\mu_R, \mu_\Theta | \widetilde{R})$, has exactly the
same functional form as equation (\ref{eq:twodnorm}).

Unfortunately, equations (\ref{eq:onednorm}) and (\ref{eq:twodnorm}) can only
be evaluated by numerical integration. To do this, for any given $W_0$ we
first compute look-up tables for the dimensionless potential
$W(\widetilde{r})$ and surface-density profile
$\widetilde{\Sigma}(\widetilde{R})$. Then we define a series of dimensionless
projected radii ranging over $0\le R/r_0\le r_t/r_0$, where
the upper limit $r_t/r_0$ is determined by the value of $W_0$. Next,
at each point in this radius sequence we define a series of dimensionless
velocity values covering the interval $0\le u^2/2\sigma_0\le W(R/r_0)$,
where $u^2$ can be associated either with $v_z^2$ or with
$(\mu_x^2+\mu_y^2)$. Finally, we perform the integrals (\ref{eq:onednorm})
and (\ref{eq:twodnorm}) at each pair of fixed $(R/r_0, u^2/2\sigma_0^2)$,
ultimately obtaining two-dimensional look-up tables
for both $N_1(v_z | \widetilde{R})$ and $N_2(\mu_x,\mu_y | \widetilde{R})$.

Application to a real velocity sample further requires convolution both with
the observed spatial distribution of the stars in the sample and with the
distribution of their velocity-measurement errors.
Section \ref{subsec:modvel} describes how we have done this for
$N_2(\mu_x,\mu_y | \widetilde{R})$ in particular, and \S\S\ref{sec:veldist}
and \ref{subsec:bhmods} compare the results to our observed proper-motion
distributions.

\end{appendix}

\clearpage


\clearpage


\clearpage
\setcounter{table}{3}
\begin{landscape}
\begin{deluxetable}{ccccccccccccccc}
\tabletypesize{\scriptsize}
\tablecaption{129,733 Stars in the Master Frame \label{tab:master}}
\tablewidth{0pt}
\tablecolumns{15}
\tablehead{
\colhead{$\Delta$RA}  &  \colhead{$\Delta$Dec}  &  \colhead{F475W} &
\colhead{RA}  &  \colhead{Dec}  &
\colhead{RA}  &  \colhead{Dec}  &
\colhead{$m_{\rm max}$} & \colhead{OK} &
\colhead{$c$}  & \colhead{ID}                  &
\colhead{$x_{\rm meta}$}  & \colhead{$y_{\rm meta}$}      &
\colhead{$x_{\rm ref}$}   & \colhead{$y_{\rm ref}$}       \\
\multicolumn{2}{c}{[arcsec]} & \colhead{} & \colhead{[hours]} &
\colhead{[deg]}  & \colhead{[hh:mm:ss]}  & \colhead{[dd:mm:ss]} &
\colhead{}  & \colhead{} & \colhead{[\%]} & \colhead{} &
\multicolumn{2}{c}{[pixels]} & \multicolumn{2}{c}{[pixels]}  \\
 \colhead{(1)}  &  \colhead{(2)}  &  \colhead{(3)}  &  \colhead{(4)}  &
 \colhead{(5)}  &  \colhead{(6)}  &  \colhead{(7)}  &  \colhead{(8)}  &
 \colhead{(9)}  & \colhead{(10)}  & \colhead{(11)}  & \colhead{(12)}  &
\colhead{(13)}  & \colhead{(14)}  & \colhead{(15)}
}
\startdata
   88.1030 & -100.8748 & 21.51 & 0.4068860 & -72.109304 &
             00:24:24.789 & -72:06:33.49 & 25.39 & 1 & 97.51 &
             M000001 &  1238.94 &  983.50 &  4078.27 & 4167.41    \\
   85.7765 & -100.8247 & 21.75 & 0.4067457 & -72.109290 &
             00:24:24.284 & -72:06:33.44 & 26.90 & 1 & 95.67 &
             M000002 &  1285.47 &  984.51 &  4031.80 & 4172.30    \\
   84.9804 & -100.7607 & 23.83 & 0.4066977 & -72.109273 &
             00:24:24.111 & -72:06:33.38 & 29.44 & 1 & 86.33 &
             M000003 &  1301.39 &  985.79 &  4015.78 & 4173.04    \\
   87.6399 & -100.6749 & 19.11 & 0.4068580 & -72.109249 &
             00:24:24.688 & -72:06:33.29 & 27.79 & 1 & 99.56 &
             M000004 &  1248.20 &  987.50 &  4068.54 & 4164.60    \\
   86.0665 & -100.6557 & 23.00 & 0.4067632 & -72.109243 &
             00:24:24.347 & -72:06:33.27 & 25.31 & 1 & 92.87 &
             M000005 &  1279.67 &  987.89 &  4037.15 & 4168.19    \\
   86.9218 & -100.6459 & 23.29 & 0.4068147 & -72.109241 &
             00:24:24.533 & -72:06:33.26 & 26.41 & 1 & 93.39 &
             M000006 &  1262.56 &  988.08 &  4054.17 & 4165.84    \\
   82.3194 & -100.3598 & 19.76 & 0.4065373 & -72.109161 &
             00:24:23.534 & -72:06:32.98 & 27.28 & 1 & 98.35 &
             M000007 &  1354.61 &  993.80 &  3961.76 & 4171.78    \\
   86.2647 & -100.3566 & 19.76 & 0.4067751 & -72.109160 &
             00:24:24.390 & -72:06:32.97 & 24.29 & 1 & 98.95 &
             M000008 &  1275.71 &  993.87 &  4040.34 & 4161.74    \\
   87.9017 & -100.2980 & 23.65 & 0.4068738 & -72.109144 &
             00:24:24.745 & -72:06:32.91 & 25.16 & 1 & 92.68 &
             M000009 &  1242.97 &  995.04 &  4072.81 & 4156.43    \\
   81.5615 & -100.2525 & 19.87 & 0.4064916 & -72.109131 &
             00:24:23.369 & -72:06:32.87 & 24.61 & 1 & 97.97 &
             M000010 &  1369.77 &  995.95 &  3946.40 & 4171.56    \\
\enddata

\tablecomments{Table \ref{tab:master} is published in its entirety in the
electronic edition of the {\it Astrophysical Journal}. A portion is shown
here to illustrate its form and content.   \hfill\break
Key to columns: \hfill\break
{\bf Column (1)}---RA offset from the cluster center in
Table \ref{tab:center}, measured in arcsec at date 2002.26, from ACS images
taken as part of program GO-9028. \hfill\break
{\bf Column (2)}---Dec offset from the cluster center in
Table \ref{tab:center}, measured in arcsec at date 2002.26, from ACS images
taken as part of program GO-9028. \hfill\break
{\bf Column (3)}---Calibrated F475W magnitude.  \hfill\break
{\bf Column (4)}---Calibrated absolute right ascension (J2000) in fractional
hours.   \hfill\break
{\bf Column (5)}---Calibrated absolute declination (J2000) in fractional
degrees.   \hfill\break
{\bf Column (6)}---Calibrated absolute right ascension (J2000) in
(hour:minute:second) format.   \hfill\break
{\bf Column (7)}---Calibrated absolute declination (J2000) in
(degree:arcmin:arcsec) format.   \hfill\break
{\bf Column (8)}---Faintest magnitude a star could have at the given position
to be counted as ``OK.'' See text for details.  \hfill\break
{\bf Column (9)}---OK flag. Either OK=1 (for a bona fide stellar detection)
or OK=0 (for a likely PSF artifact). Set by the discrimination criterion
illustrated in Fig.~\ref{fig:PSFartifacts}.\hfill\break
{\bf Column (10)}---Percentage completeness: the fraction of the local
neighborhood (within 100 WFC pixels, or $\simeq5\arcsec$) where the
star could have fallen and still be found. Each entry in this table thus
corresponds to a completeness-corrected total ``number'' of $(100/c)$ stars
at the given position and magnitude in the master frame. \hfill\break
{\bf Column (11)}---Sequential ID number, running from M000001 to M129733.
\hfill\break
{\bf Column (12)}---Star's $x$ position, in pixels, in our
meta-image of the master frame ({\tt 47TucMaster.fits}, available in the
electronic edition of the {\it Astrophysical Journal}   \hfill\break
{\bf Column (13)}---Star's $y$ position, in pixels, in our
meta-image of the master frame ({\tt 47TucMaster.fits}, available in the
electronic edition of the {\it Astrophysical Journal}   \hfill\break
{\bf Column (14)}---Star's $x$ position, in pixels, in the
original ACS/GO-9028 image {\tt j8cd01a9q\_w.fits}.  \hfill\break
{\bf Column (15)}---Star's $y$ position, in pixels, in the
original ACS/GO-9028 image {\tt j8cd01a9q\_w.fits}.
}
\end{deluxetable}
\clearpage
\end{landscape}

\clearpage
\setcounter{table}{4}
\begin{landscape}
\begin{deluxetable}{lrrrrrrrrrr}
\tabletypesize{\scriptsize}
\tablecaption{Displacement and Proper-Motion Data for 14,366 Stars in the
              Core of 47 Tucanae \label{tab:posdata}}
\tablewidth{0pt}
\tablecolumns{11}
\tablehead{}
\startdata

DATASET &   MEYLANe1 &   MEYLANe2 &   GILLILU1 &   MEYLANe3 &   GILLILU2
        &   WFC-MEUR &   HRC-MEUR &   HRC-BOHL &   WFC-KING &   HRC-KING \\
   DATE &    1995.82 &    1997.84 &    1999.51 &    1999.82 &    2001.53
        &    2002.26 &    2002.26 &    2002.28 &    2002.52 &    2002.56 \\
M000872 &     8.3886 &   -88.5393 &     18.317 &     19.165 &      18.44
        &  0.4020806 & -72.105878 &            &            &            \\
M000872 &   7.18E-03 &        n/a &   1.21E-03 &   3.08E-03 &        n/a
        &   6.96E-05 &        n/a &        n/a &  -6.46E-05 &        n/a \\
M000872 &   3.11E-03 &        n/a &   2.12E-03 &   1.81E-03 &        n/a
        &   3.48E-04 &        n/a &        n/a &   2.98E-04 &        n/a \\
M000872 &   9.46E-03 &        n/a &   1.11E-03 &   1.78E-03 &        n/a
        &  -4.97E-06 &        n/a &        n/a &  -1.69E-04 &        n/a \\
M000872 &   2.91E-03 &        n/a &   1.63E-03 &   2.46E-03 &        n/a
        &   3.23E-04 &        n/a &        n/a &   1.44E-04 &        n/a \\
M000872 &  -9.50E-04 &   3.43E-04 &      0.264 &     0.8515 &  -9.89E-04
        &   3.13E-04 &      0.764 &     0.5141 &            &            \\
M000979 &     4.5367 &   -87.9300 &     18.649 &     19.517 &      18.61
        &  0.4018484 & -72.105709 &            &            &            \\
M000979 &   5.68E-03 &        n/a &  -6.36E-04 &  -2.59E-04 &  -1.77E-03
        &   3.98E-05 &        n/a &        n/a &   2.14E-04 &        n/a \\
M000979 &   3.55E-03 &        n/a &   1.84E-03 &   3.71E-03 &   2.33E-03
        &   1.99E-04 &        n/a &        n/a &   2.78E-04 &        n/a \\
M000979 &   3.48E-04 &        n/a &   2.08E-03 &   1.89E-04 &   2.84E-03
        &   1.19E-04 &        n/a &        n/a &  -1.59E-04 &        n/a \\
M000979 &   3.42E-03 &        n/a &   1.53E-03 &   2.58E-03 &   2.44E-03
        &   2.04E-04 &        n/a &        n/a &   3.03E-04 &        n/a \\
M000979 &  -2.27E-04 &   3.82E-04 &      0.799 &     0.5254 &  -4.19E-04
        &   3.43E-04 &      0.544 &     0.7034 &            &            \\ 
\enddata

\tablecomments{Table \ref{tab:posdata} is published in its entirety in the
electronic edition of the {\it Astrophysical Journal}.
A portion is shown here to illustrate its form and content.
The first two lines of the Table contain the names and dates of the 10
HST data sets we have used for position measurements (see Table
\ref{tab:datasets}). Following are 6 lines each for a total of 14,366 stars
which have calibrated $V$ magnitudes brighter than $V\la 20.5$,
colors $0\le (U-V)\le 3$, and proper motions obtained as described
in \S\ref{sec:velsamples} of the text. The lines recorded for each star contain
the following information: \hfill\break
{\bf Line 1:} Columns are: (1) the star's ID (corresponding to those in
Table \ref{tab:master}); (2) its right-ascension distance from the cluster
center (in arcseconds, positive Eastward, measured on the master frame);
(3) its distance in declination from the cluster center
(in arcseconds, positive Northward, measured on the master frame);
(4) its $V$ magnitude; (5)
its $U$ magnitude; (6) its F475W magnitude;
(7) its absolute RA position (J2000) at date 2002.26 (program GO-9028);
and (8) its absolute Dec position (J2000) at date 2002.26 (program GO-9028).
\hfill\break
{\bf Line 2:} Columns are: (1) the star's ID, followed by (2)--(11) its RA
displacement (in arcseconds), relative to the position in Line 1, in
each of the 10 HST data sets or ``epochs'' we have analyzed (following the
order in the top two lines of this table). An entry of `n/a' in any column
indicates that the star's position in that data set was not used in the
proper-motion measurement.
\hfill\break
{\bf Line 3:} Columns are: (1) the star's ID, followed by (2)--(11) the
uncertainties (in arcesonds) in the corresponding RA displacements in Line 2.
\hfill\break
{\bf Line 4:} Columns are (1) the star's ID, followed by (2)--(11) its Dec
displacement (in arcseconds), relative to the position in Line 1,
in each of the 10 HST data sets.
\hfill\break
{\bf Line 5:} Columns are: (1) the star's ID, followed by (2)--(11) the
uncertainties (in arcesonds) in the corresponding Dec displacements in Line 4.
\hfill\break
{\bf Line 6:} Columns are: (1) the star's ID, followed by (2) its weighted
least-squares proper-motion $\mu_{\alpha}$ in the RA direction (in units of
arcsec yr$^{-1}$, positive Eastward); (3) the uncertainty in $\mu_{\alpha}$;
(4) the $\chi^2$ of the weighted least-squares straight-line fit
yielding $\mu_{\alpha}$; (5) the probability $P$ that this or a higher
$\chi_{\alpha}^2$ could occur by chance if the true RA-motion of the star
follows a straight line; (6) the star's weighted
least-squares proper-motion $\mu_{\delta}$ in the Dec direction (in units of
arcsec yr$^{-1}$, positive Northward); (7) the uncertainty in $\mu_{\delta}$;
(8) the $\chi^2$ of the weighted least-squares straight-line fit
yielding $\mu_{\delta}$; and (9) the probability $P$ that this or a higher
$\chi_{\delta}^2$ could occur by chance if the true Dec-motion of the star
follows a straight line.
}
\end{deluxetable}
\clearpage
\end{landscape}

\setcounter{table}{9}
\LongTables
\begin{deluxetable}{lrrrrrrrrrr}
\tabletypesize{\scriptsize}
\tablewidth{0pt}
\tablecaption{Proper-Motion Kinematics in the Core of 47 Tucanae
               \label{tab:kinmags}}
\tablecolumns{11}
\tablehead{
\multicolumn{7}{c}{~~}   &
\multicolumn{4}{c}{$\langle\mu_R\rangle=\langle\mu_\Theta\rangle \equiv 0$} \\
\cline{8-11}
\colhead{Annulus} & \colhead{$R_{\rm eff}$} & \colhead{${\cal N}$} &
\colhead{$\langle\mu_{\alpha}\rangle$} &
\colhead{$\langle\mu_{\delta}\rangle$} &
\colhead{$\sigma_{\alpha}$} & \colhead{$\sigma_{\delta}$} &
\colhead{$\sigma_{R}$} & \colhead{$\sigma_{\Theta}$} &
\colhead{$\sigma_\mu$} &
\colhead{$(\sigma_R^2-\sigma_\Theta^2)/\sigma_\mu^2$} \\
\colhead{[arcsec]} &\colhead{[arcsec]} & \colhead{} &
\colhead{[mas yr$^{-1}$]} & \colhead{[mas yr$^{-1}$]} & 
\colhead{[mas yr$^{-1}$]} & \colhead{[mas yr$^{-1}$]} & 
\colhead{[mas yr$^{-1}$]} & \colhead{[mas yr$^{-1}$]} & 
\colhead{[mas yr$^{-1}$]} & \colhead{}  \\
\colhead{(1)} &  \colhead{(2)} & \colhead{(3)} & \colhead{(4)} & 
\colhead{(5)} &  \colhead{(6)} & \colhead{(7)} & \colhead{(8)} & 
\colhead{(9)} & \colhead{(10)} & \colhead{(11)}
}
\startdata
\cutinhead{$14\leq V < 16.75$, $1.5\leq (U-V) <2.6$,
           \qquad {\it and} \qquad
           $16.75\leq V < 18.5$, $0.6\leq (U-V) <1.9$ ;
           $\Delta R=3^{\prime\prime}$}
\phantom{1}0.0--3.0 & 2.12 & 69 &
        $-0.087^{+0.084}_{-0.094}$ & $0.025^{+0.071}_{-0.072}$ &
        $0.713^{+0.057}_{-0.071}$ & $0.604^{+0.036}_{-0.051}$ &
        $0.624^{+0.042}_{-0.046}$ & $0.692^{+0.053}_{-0.066}$ &
        $0.659^{+0.036}_{-0.042}$ & $-0.21^{+0.22}_{-0.19}$ \\
\phantom{1}3.0--6.0 & 4.74 & 176 &
        $-0.081^{+0.044}_{-0.043}$ & $-0.045^{+0.044}_{-0.048}$ &
        $0.564^{+0.027}_{-0.034}$ & $0.601^{+0.031}_{-0.035}$ &
        $0.600^{+0.036}_{-0.039}$ & $0.569^{+0.028}_{-0.028}$ &
        $0.585^{+0.021}_{-0.023}$ & $0.10^{+0.16}_{-0.18}$ \\
\phantom{1}6.0--9.0 & 7.65 & 303 &
        $0.003^{+0.036}_{-0.036}$ & $-0.074^{+0.034}_{-0.035}$ &
        $0.611^{+0.027}_{-0.030}$ & $0.585^{+0.021}_{-0.025}$ &
        $0.618^{+0.027}_{-0.028}$ & $0.580^{+0.022}_{-0.024}$ &
        $0.599^{+0.017}_{-0.018}$ & $0.13^{+0.12}_{-0.12}$ \\
\phantom{1}9.0--12.0 & 10.61 & 382 &
        $-0.005^{+0.033}_{-0.032}$ & $-0.035^{+0.034}_{-0.032}$ &
        $0.616^{+0.024}_{-0.025}$ & $0.620^{+0.020}_{-0.023}$ &
        $0.640^{+0.022}_{-0.023}$ & $0.595^{+0.022}_{-0.022}$ &
        $0.618^{+0.016}_{-0.016}$ & $0.15^{+0.10}_{-0.10}$ \\
12.0--15.0 & 13.58 & 353 &
        $-0.030^{+0.035}_{-0.034}$ & $-0.048^{+0.035}_{-0.030}$ &
        $0.601^{+0.024}_{-0.027}$ & $0.612^{+0.023}_{-0.024}$ &
        $0.623^{+0.025}_{-0.029}$ & $0.590^{+0.024}_{-0.024}$ &
        $0.607^{+0.018}_{-0.020}$ & $0.11^{+0.11}_{-0.12}$ \\
15.0--18.0 & 16.57 & 245 &
        $0.033^{+0.039}_{-0.040}$ & $-0.064^{+0.034}_{-0.035}$ &
        $0.584^{+0.030}_{-0.035}$ & $0.561^{+0.024}_{-0.026}$ &
        $0.545^{+0.026}_{-0.028}$ & $0.601^{+0.030}_{-0.034}$ &
        $0.573^{+0.019}_{-0.021}$ & $-0.20^{+0.14}_{-0.14}$ \\
18.0--21.0 & 19.56 & 143 &
        $-0.025^{+0.059}_{-0.049}$ & $-0.010^{+0.053}_{-0.054}$ &
        $0.638^{+0.029}_{-0.036}$ & $0.633^{+0.034}_{-0.042}$ &
        $0.585^{+0.033}_{-0.032}$ & $0.679^{+0.034}_{-0.035}$ &
        $0.634^{+0.023}_{-0.024}$ & $-0.30^{+0.17}_{-0.16}$ \\
21.0--24.0 & 22.55 & 50 &
        $-0.003^{+0.068}_{-0.073}$ & $-0.029^{+0.097}_{-0.087}$ &
        $0.468^{+0.037}_{-0.048}$ & $0.616^{+0.105}_{-0.134}$ &
        $0.587^{+0.087}_{-0.118}$ & $0.491^{+0.052}_{-0.056}$ &
        $0.541^{+0.065}_{-0.075}$ & $0.35^{+0.25}_{-0.37}$ \\
24.0--27.0 & 25.54 & 47 &
        $0.008^{+0.114}_{-0.104}$ & $-0.162^{+0.090}_{-0.101}$ &
        $0.700^{+0.066}_{-0.083}$ & $0.624^{+0.051}_{-0.067}$ &
        $0.625^{+0.063}_{-0.075}$ & $0.703^{+0.061}_{-0.071}$ &
        $0.665^{+0.042}_{-0.051}$ & $-0.23^{+0.30}_{-0.31}$ \\
27.0--30.0 & 28.54 & 84 &
        $-0.022^{+0.080}_{-0.075}$ & $-0.047^{+0.064}_{-0.067}$ &
        $0.666^{+0.042}_{-0.053}$ & $0.557^{+0.044}_{-0.051}$ &
        $0.622^{+0.055}_{-0.053}$ & $0.599^{+0.046}_{-0.048}$ &
        $0.611^{+0.037}_{-0.036}$ & $0.08^{+0.23}_{-0.22}$ \\
30.0--33.0 & 31.54 & 119 &
        $-0.103^{+0.066}_{-0.058}$ & $-0.049^{+0.061}_{-0.062}$ &
        $0.628^{+0.038}_{-0.049}$ & $0.638^{+0.046}_{-0.053}$ &
        $0.643^{+0.041}_{-0.044}$ & $0.628^{+0.050}_{-0.058}$ &
        $0.635^{+0.032}_{-0.035}$ & $0.05^{+0.23}_{-0.21}$ \\
33.0--36.0 & 34.53 & 154 &
        $-0.007^{+0.057}_{-0.053}$ & $0.036^{+0.051}_{-0.049}$ &
        $0.631^{+0.029}_{-0.038}$ & $0.571^{+0.033}_{-0.039}$ &
        $0.656^{+0.034}_{-0.034}$ & $0.539^{+0.036}_{-0.039}$ &
        $0.600^{+0.025}_{-0.025}$ & $0.39^{+0.17}_{-0.16}$ \\
36.0--39.0 & 37.53 & 162 &
        $-0.006^{+0.048}_{-0.050}$ & $-0.022^{+0.051}_{-0.054}$ &
        $0.597^{+0.036}_{-0.044}$ & $0.624^{+0.034}_{-0.041}$ &
        $0.567^{+0.035}_{-0.034}$ & $0.649^{+0.039}_{-0.042}$ &
        $0.609^{+0.026}_{-0.028}$ & $-0.27^{+0.18}_{-0.16}$ \\
39.0--42.0 & 40.53 & 180 &
        $-0.025^{+0.046}_{-0.046}$ & $0.038^{+0.047}_{-0.046}$ &
        $0.597^{+0.033}_{-0.039}$ & $0.591^{+0.034}_{-0.040}$ &
        $0.561^{+0.032}_{-0.033}$ & $0.623^{+0.038}_{-0.040}$ &
        $0.593^{+0.025}_{-0.026}$ & $-0.21^{+0.18}_{-0.18}$ \\
42.0--45.0 & 43.53 & 209 &
        $-0.020^{+0.042}_{-0.048}$ & $0.004^{+0.040}_{-0.043}$ &
        $0.641^{+0.030}_{-0.035}$ & $0.583^{+0.028}_{-0.029}$ &
        $0.601^{+0.031}_{-0.035}$ & $0.620^{+0.030}_{-0.034}$ &
        $0.611^{+0.020}_{-0.023}$ & $-0.06^{+0.16}_{-0.16}$ \\
45.0--48.0 & 46.52 & 229 &
        $-0.046^{+0.042}_{-0.037}$ & $-0.022^{+0.044}_{-0.045}$ &
        $0.562^{+0.025}_{-0.029}$ & $0.671^{+0.030}_{-0.039}$ &
        $0.565^{+0.029}_{-0.031}$ & $0.667^{+0.034}_{-0.034}$ &
        $0.618^{+0.023}_{-0.022}$ & $-0.33^{+0.15}_{-0.15}$ \\
48.0--51.0 & 49.52 & 259 &
        $-0.032^{+0.040}_{-0.039}$ & $-0.027^{+0.038}_{-0.035}$ &
        $0.586^{+0.023}_{-0.028}$ & $0.554^{+0.024}_{-0.024}$ &
        $0.599^{+0.030}_{-0.030}$ & $0.539^{+0.021}_{-0.023}$ &
        $0.570^{+0.018}_{-0.019}$ & $0.21^{+0.13}_{-0.12}$ \\
51.0--54.0 & 52.52 & 244 &
        $-0.024^{+0.040}_{-0.038}$ & $-0.013^{+0.040}_{-0.037}$ &
        $0.550^{+0.027}_{-0.036}$ & $0.577^{+0.022}_{-0.027}$ &
        $0.566^{+0.028}_{-0.030}$ & $0.559^{+0.026}_{-0.029}$ &
        $0.563^{+0.020}_{-0.021}$ & $0.02^{+0.14}_{-0.14}$ \\
54.0--57.0 & 55.52 & 278 &
        $-0.030^{+0.038}_{-0.036}$ & $-0.062^{+0.041}_{-0.036}$ &
        $0.578^{+0.026}_{-0.031}$ & $0.586^{+0.022}_{-0.031}$ &
        $0.601^{+0.028}_{-0.027}$ & $0.564^{+0.026}_{-0.029}$ &
        $0.583^{+0.019}_{-0.019}$ & $0.13^{+0.14}_{-0.14}$ \\
57.0--60.0 & 58.52 & 269 &
        $-0.059^{+0.032}_{-0.036}$ & $-0.039^{+0.035}_{-0.042}$ &
        $0.535^{+0.023}_{-0.023}$ & $0.596^{+0.033}_{-0.039}$ &
        $0.608^{+0.030}_{-0.032}$ & $0.524^{+0.025}_{-0.029}$ &
        $0.567^{+0.022}_{-0.023}$ & $0.30^{+0.13}_{-0.13}$ \\
60.0--63.0 & 61.52 & 277 &
        $-0.033^{+0.037}_{-0.037}$ & $-0.046^{+0.039}_{-0.033}$ &
        $0.565^{+0.027}_{-0.030}$ & $0.569^{+0.025}_{-0.030}$ &
        $0.528^{+0.023}_{-0.024}$ & $0.604^{+0.029}_{-0.029}$ &
        $0.567^{+0.019}_{-0.018}$ & $-0.27^{+0.12}_{-0.13}$ \\
63.0--66.0 & 64.52 & 277 &
        $-0.014^{+0.033}_{-0.035}$ & $0.041^{+0.039}_{-0.038}$ &
        $0.539^{+0.022}_{-0.026}$ & $0.579^{+0.023}_{-0.033}$ &
        $0.595^{+0.031}_{-0.033}$ & $0.521^{+0.024}_{-0.024}$ &
        $0.559^{+0.019}_{-0.020}$ & $0.26^{+0.14}_{-0.15}$ \\
66.0--69.0 & 67.52 & 260 &
        $-0.040^{+0.036}_{-0.038}$ & $-0.005^{+0.036}_{-0.043}$ &
        $0.570^{+0.027}_{-0.035}$ & $0.605^{+0.027}_{-0.032}$ &
        $0.614^{+0.032}_{-0.027}$ & $0.558^{+0.030}_{-0.031}$ &
        $0.587^{+0.022}_{-0.019}$ & $0.19^{+0.15}_{-0.13}$ \\
69.0--72.0 & 70.52 & 222 &
        $-0.028^{+0.041}_{-0.034}$ & $0.012^{+0.041}_{-0.036}$ &
        $0.548^{+0.024}_{-0.029}$ & $0.571^{+0.032}_{-0.042}$ &
        $0.561^{+0.029}_{-0.030}$ & $0.556^{+0.036}_{-0.034}$ &
        $0.559^{+0.025}_{-0.024}$ & $0.02^{+0.16}_{-0.16}$ \\
72.0--75.0 & 73.52 & 224 &
        $0.016^{+0.042}_{-0.039}$ & $-0.025^{+0.041}_{-0.039}$ &
        $0.567^{+0.027}_{-0.032}$ & $0.546^{+0.030}_{-0.034}$ &
        $0.579^{+0.034}_{-0.035}$ & $0.532^{+0.023}_{-0.025}$ &
        $0.556^{+0.020}_{-0.020}$ & $0.17^{+0.15}_{-0.16}$ \\
75.0--78.0 & 76.51 & 219 &
        $-0.038^{+0.043}_{-0.043}$ & $0.036^{+0.046}_{-0.044}$ &
        $0.584^{+0.030}_{-0.035}$ & $0.607^{+0.030}_{-0.035}$ &
        $0.581^{+0.027}_{-0.031}$ & $0.610^{+0.032}_{-0.032}$ &
        $0.595^{+0.021}_{-0.022}$ & $-0.10^{+0.14}_{-0.16}$ \\
78.0--81.0 & 79.51 & 227 &
        $-0.004^{+0.037}_{-0.037}$ & $-0.032^{+0.037}_{-0.042}$ &
        $0.528^{+0.027}_{-0.029}$ & $0.543^{+0.027}_{-0.031}$ &
        $0.526^{+0.027}_{-0.030}$ & $0.544^{+0.029}_{-0.032}$ &
        $0.535^{+0.019}_{-0.022}$ & $-0.07^{+0.16}_{-0.17}$ \\
81.0--84.0 & 82.51 & 212 &
        $-0.020^{+0.038}_{-0.041}$ & $0.105^{+0.042}_{-0.041}$ &
        $0.545^{+0.030}_{-0.033}$ & $0.551^{+0.030}_{-0.038}$ &
        $0.542^{+0.030}_{-0.035}$ & $0.562^{+0.028}_{-0.032}$ &
        $0.552^{+0.021}_{-0.024}$ & $-0.07^{+0.15}_{-0.16}$ \\
84.0--87.0 & 85.51 & 165 &
        $-0.022^{+0.046}_{-0.043}$ & $0.042^{+0.041}_{-0.044}$ &
        $0.540^{+0.028}_{-0.039}$ & $0.487^{+0.025}_{-0.031}$ &
        $0.521^{+0.035}_{-0.033}$ & $0.505^{+0.035}_{-0.035}$ &
        $0.514^{+0.025}_{-0.024}$ & $0.06^{+0.19}_{-0.19}$ \\
87.0--90.0 & 88.51 & 129 &
        $-0.054^{+0.051}_{-0.048}$ & $-0.012^{+0.047}_{-0.041}$ &
        $0.528^{+0.039}_{-0.050}$ & $0.477^{+0.034}_{-0.037}$ &
        $0.545^{+0.037}_{-0.045}$ & $0.456^{+0.029}_{-0.029}$ &
        $0.502^{+0.026}_{-0.028}$ & $0.35^{+0.16}_{-0.19}$ \\
90.0--93.0 & 91.51 & 93 &
        $0.116^{+0.063}_{-0.066}$ & $0.011^{+0.061}_{-0.062}$ &
        $0.578^{+0.051}_{-0.064}$ & $0.568^{+0.039}_{-0.044}$ &
        $0.667^{+0.061}_{-0.058}$ & $0.466^{+0.037}_{-0.040}$ &
        $0.575^{+0.039}_{-0.038}$ & $0.69^{+0.21}_{-0.21}$ \\
93.0--96.0 & 94.51 & 36 &
        $-0.175^{+0.120}_{-0.123}$ & $-0.007^{+0.089}_{-0.090}$ &
        $0.680^{+0.076}_{-0.108}$ & $0.482^{+0.043}_{-0.071}$ &
        $0.586^{+0.067}_{-0.069}$ & $0.599^{+0.078}_{-0.101}$ &
        $0.593^{+0.053}_{-0.062}$ & $-0.04^{+0.40}_{-0.34}$ \\
96.0--99.0 & 97.51 & 28 &
        $-0.038^{+0.101}_{-0.097}$ & $0.052^{+0.107}_{-0.100}$ &
        $0.490^{+0.047}_{-0.078}$ & $0.494^{+0.046}_{-0.076}$ &
        $0.525^{+0.054}_{-0.057}$ & $0.438^{+0.071}_{-0.091}$ &
        $0.483^{+0.043}_{-0.046}$ & $0.36^{+0.47}_{-0.40}$ \\
\cutinhead{$18.5\leq V < 19$, $0.6\leq (U-V)<1.9$ ; 
   $\Delta R=5^{\prime\prime}$}
\phantom{1}0.0--5.0 & 3.54 & 53 &
        $0.015^{+0.086}_{-0.092}$ & $-0.168^{+0.086}_{-0.085}$ &
        $0.666^{+0.043}_{-0.053}$ & $0.638^{+0.048}_{-0.070}$ &
        $0.687^{+0.048}_{-0.052}$ & $0.624^{+0.062}_{-0.061}$ &
        $0.656^{+0.038}_{-0.034}$ & $0.19^{+0.25}_{-0.26}$ \\
\phantom{1}2.5--7.5 & 5.59 & 119 &
        $-0.034^{+0.054}_{-0.062}$ & $-0.126^{+0.061}_{-0.062}$ &
        $0.632^{+0.040}_{-0.050}$ & $0.636^{+0.039}_{-0.044}$ &
        $0.589^{+0.036}_{-0.036}$ & $0.684^{+0.039}_{-0.044}$ &
        $0.638^{+0.028}_{-0.029}$ & $-0.30^{+0.16}_{-0.15}$ \\
\phantom{1}5.0--10.0 & 7.91 & 182 &
        $-0.059^{+0.052}_{-0.044}$ & $-0.132^{+0.045}_{-0.045}$ &
        $0.622^{+0.031}_{-0.039}$ & $0.604^{+0.028}_{-0.035}$ &
        $0.608^{+0.032}_{-0.034}$ & $0.631^{+0.034}_{-0.034}$ &
        $0.620^{+0.024}_{-0.025}$ & $-0.07^{+0.14}_{-0.15}$ \\
\phantom{1}7.5--12.5 & 10.31 & 209 &
        $-0.015^{+0.045}_{-0.045}$ & $-0.137^{+0.043}_{-0.045}$ &
        $0.639^{+0.027}_{-0.032}$ & $0.587^{+0.030}_{-0.035}$ &
        $0.648^{+0.028}_{-0.030}$ & $0.590^{+0.029}_{-0.031}$ &
        $0.620^{+0.020}_{-0.021}$ & $0.19^{+0.14}_{-0.13}$ \\
10.0--15.0 & 12.75 & 213 &
        $0.028^{+0.041}_{-0.047}$ & $-0.089^{+0.043}_{-0.043}$ &
        $0.632^{+0.028}_{-0.033}$ & $0.598^{+0.028}_{-0.030}$ &
        $0.631^{+0.025}_{-0.030}$ & $0.604^{+0.030}_{-0.033}$ &
        $0.617^{+0.020}_{-0.024}$ & $0.09^{+0.11}_{-0.13}$ \\
12.5--17.5 & 15.21 & 176 &
        $0.006^{+0.043}_{-0.043}$ & $-0.018^{+0.046}_{-0.048}$ &
        $0.581^{+0.028}_{-0.034}$ & $0.594^{+0.026}_{-0.029}$ &
        $0.566^{+0.025}_{-0.025}$ & $0.605^{+0.030}_{-0.030}$ &
        $0.586^{+0.021}_{-0.020}$ & $-0.13^{+0.12}_{-0.13}$ \\
15.0--20.0 & 17.68 & 129 &
        $-0.094^{+0.051}_{-0.049}$ & $-0.048^{+0.054}_{-0.055}$ &
        $0.548^{+0.035}_{-0.036}$ & $0.584^{+0.031}_{-0.039}$ &
        $0.515^{+0.032}_{-0.034}$ & $0.618^{+0.036}_{-0.040}$ &
        $0.569^{+0.025}_{-0.027}$ & $-0.36^{+0.17}_{-0.17}$ \\
17.5--22.5 & 20.16 & 83 &
        $-0.132^{+0.064}_{-0.068}$ & $-0.157^{+0.068}_{-0.072}$ &
        $0.570^{+0.040}_{-0.052}$ & $0.599^{+0.044}_{-0.055}$ &
        $0.590^{+0.046}_{-0.049}$ & $0.606^{+0.045}_{-0.054}$ &
        $0.598^{+0.033}_{-0.034}$ & $-0.05^{+0.24}_{-0.22}$ \\
20.0--25.0 & 22.64 & 35 &
        $-0.123^{+0.106}_{-0.116}$ & $-0.180^{+0.105}_{-0.123}$ &
        $0.614^{+0.055}_{-0.088}$ & $0.650^{+0.052}_{-0.082}$ &
        $0.704^{+0.066}_{-0.082}$ & $0.571^{+0.058}_{-0.070}$ &
        $0.641^{+0.041}_{-0.048}$ & $0.41^{+0.30}_{-0.34}$ \\
22.5--27.5 & 25.12 & 20 &
        $-0.301^{+0.158}_{-0.153}$ & $-0.130^{+0.143}_{-0.141}$ &
        $0.605^{+0.099}_{-0.157}$ & $0.585^{+0.053}_{-0.106}$ &
        $0.651^{+0.097}_{-0.117}$ & $0.589^{+0.084}_{-0.101}$ &
        $0.620^{+0.061}_{-0.071}$ & $0.20^{+0.47}_{-0.49}$ \\
25.0--30.0 & 27.61 & 32 &
        $-0.185^{+0.103}_{-0.112}$ & $0.038^{+0.107}_{-0.111}$ &
        $0.571^{+0.059}_{-0.095}$ & $0.548^{+0.063}_{-0.097}$ &
        $0.536^{+0.098}_{-0.107}$ & $0.590^{+0.062}_{-0.076}$ &
        $0.564^{+0.048}_{-0.053}$ & $-0.19^{+0.50}_{-0.49}$ \\
27.5--32.5 & 30.10 & 42 &
        $-0.044^{+0.100}_{-0.103}$ & $0.126^{+0.093}_{-0.100}$ &
        $0.595^{+0.047}_{-0.075}$ & $0.537^{+0.057}_{-0.075}$ &
        $0.512^{+0.077}_{-0.096}$ & $0.615^{+0.070}_{-0.085}$ &
        $0.566^{+0.048}_{-0.054}$ & $-0.37^{+0.44}_{-0.45}$ \\
30.0--35.0 & 32.60 & 52 &
        $0.204^{+0.086}_{-0.101}$ & $0.099^{+0.075}_{-0.081}$ &
        $0.640^{+0.060}_{-0.083}$ & $0.441^{+0.048}_{-0.067}$ &
        $0.605^{+0.070}_{-0.075}$ & $0.522^{+0.061}_{-0.083}$ &
        $0.565^{+0.049}_{-0.061}$ & $0.30^{+0.37}_{-0.33}$ \\
32.5--37.5 & 35.09 & 75 &
        $0.097^{+0.076}_{-0.077}$ & $-0.055^{+0.064}_{-0.066}$ &
        $0.620^{+0.051}_{-0.071}$ & $0.482^{+0.043}_{-0.059}$ &
        $0.633^{+0.062}_{-0.064}$ & $0.467^{+0.043}_{-0.055}$ &
        $0.556^{+0.041}_{-0.042}$ & $0.59^{+0.26}_{-0.25}$ \\
35.0--40.0 & 37.58 & 103 &
        $-0.110^{+0.068}_{-0.064}$ & $-0.047^{+0.072}_{-0.076}$ &
        $0.598^{+0.039}_{-0.045}$ & $0.679^{+0.053}_{-0.073}$ &
        $0.661^{+0.046}_{-0.052}$ & $0.621^{+0.051}_{-0.047}$ &
        $0.641^{+0.035}_{-0.034}$ & $0.13^{+0.20}_{-0.22}$ \\
37.5--42.5 & 40.08 & 125 &
        $-0.066^{+0.061}_{-0.066}$ & $0.015^{+0.063}_{-0.064}$ &
        $0.640^{+0.037}_{-0.044}$ & $0.648^{+0.054}_{-0.055}$ &
        $0.640^{+0.042}_{-0.052}$ & $0.646^{+0.043}_{-0.047}$ &
        $0.643^{+0.028}_{-0.033}$ & $-0.02^{+0.21}_{-0.22}$ \\
40.0--45.0 & 42.57 & 110 &
        $0.017^{+0.070}_{-0.064}$ & $0.032^{+0.069}_{-0.064}$ &
        $0.632^{+0.046}_{-0.065}$ & $0.618^{+0.044}_{-0.059}$ &
        $0.634^{+0.056}_{-0.055}$ & $0.610^{+0.056}_{-0.059}$ &
        $0.622^{+0.041}_{-0.040}$ & $0.07^{+0.26}_{-0.23}$ \\
42.5--47.5 & 45.07 & 124 &
        $-0.020^{+0.063}_{-0.060}$ & $0.023^{+0.056}_{-0.057}$ &
        $0.621^{+0.045}_{-0.053}$ & $0.573^{+0.043}_{-0.051}$ &
        $0.635^{+0.041}_{-0.052}$ & $0.553^{+0.044}_{-0.055}$ &
        $0.595^{+0.031}_{-0.040}$ & $0.27^{+0.22}_{-0.22}$ \\
45.0--50.0 & 47.57 & 158 &
        $-0.038^{+0.057}_{-0.054}$ & $-0.013^{+0.049}_{-0.053}$ &
        $0.596^{+0.031}_{-0.041}$ & $0.564^{+0.034}_{-0.042}$ &
        $0.611^{+0.036}_{-0.039}$ & $0.544^{+0.035}_{-0.034}$ &
        $0.579^{+0.024}_{-0.026}$ & $0.23^{+0.18}_{-0.18}$ \\
47.5--52.5 & 50.06 & 145 &
        $-0.003^{+0.051}_{-0.058}$ & $0.001^{+0.057}_{-0.062}$ &
        $0.564^{+0.036}_{-0.043}$ & $0.625^{+0.037}_{-0.042}$ &
        $0.582^{+0.040}_{-0.037}$ & $0.603^{+0.038}_{-0.041}$ &
        $0.593^{+0.029}_{-0.028}$ & $-0.07^{+0.19}_{-0.18}$ \\
50.0--55.0 & 52.56 & 147 &
        $-0.011^{+0.052}_{-0.048}$ & $0.037^{+0.051}_{-0.051}$ &
        $0.545^{+0.036}_{-0.040}$ & $0.544^{+0.035}_{-0.043}$ &
        $0.506^{+0.029}_{-0.033}$ & $0.577^{+0.041}_{-0.048}$ &
        $0.543^{+0.028}_{-0.030}$ & $-0.26^{+0.20}_{-0.17}$ \\
52.5--57.5 & 55.06 & 151 &
        $-0.071^{+0.046}_{-0.055}$ & $0.071^{+0.043}_{-0.051}$ &
        $0.583^{+0.028}_{-0.043}$ & $0.522^{+0.034}_{-0.040}$ &
        $0.575^{+0.037}_{-0.039}$ & $0.535^{+0.040}_{-0.043}$ &
        $0.556^{+0.026}_{-0.028}$ & $0.14^{+0.21}_{-0.21}$ \\
55.0--60.0 & 57.55 & 164 &
        $0.021^{+0.054}_{-0.055}$ & $0.050^{+0.049}_{-0.046}$ &
        $0.612^{+0.039}_{-0.047}$ & $0.542^{+0.040}_{-0.047}$ &
        $0.614^{+0.042}_{-0.045}$ & $0.538^{+0.040}_{-0.041}$ &
        $0.577^{+0.029}_{-0.032}$ & $0.26^{+0.19}_{-0.20}$ \\
57.5--62.5 & 60.05 & 173 &
        $0.075^{+0.047}_{-0.045}$ & $0.025^{+0.048}_{-0.050}$ &
        $0.533^{+0.035}_{-0.045}$ & $0.555^{+0.039}_{-0.046}$ &
        $0.577^{+0.044}_{-0.046}$ & $0.511^{+0.037}_{-0.040}$ &
        $0.545^{+0.030}_{-0.032}$ & $0.24^{+0.20}_{-0.20}$ \\
60.0--65.0 & 62.55 & 161 &
        $-0.014^{+0.050}_{-0.049}$ & $-0.001^{+0.056}_{-0.050}$ &
        $0.557^{+0.039}_{-0.042}$ & $0.568^{+0.033}_{-0.039}$ &
        $0.574^{+0.036}_{-0.037}$ & $0.547^{+0.046}_{-0.047}$ &
        $0.561^{+0.030}_{-0.031}$ & $0.10^{+0.20}_{-0.21}$ \\
62.5--67.5 & 65.05 & 162 &
        $0.029^{+0.054}_{-0.052}$ & $-0.042^{+0.046}_{-0.050}$ &
        $0.569^{+0.038}_{-0.048}$ & $0.520^{+0.031}_{-0.039}$ &
        $0.508^{+0.035}_{-0.035}$ & $0.578^{+0.047}_{-0.043}$ &
        $0.544^{+0.032}_{-0.029}$ & $-0.25^{+0.20}_{-0.19}$ \\
65.0--70.0 & 67.55 & 159 &
        $0.137^{+0.045}_{-0.053}$ & $0.006^{+0.048}_{-0.050}$ &
        $0.513^{+0.033}_{-0.042}$ & $0.531^{+0.031}_{-0.036}$ &
        $0.514^{+0.036}_{-0.038}$ & $0.543^{+0.035}_{-0.035}$ &
        $0.529^{+0.025}_{-0.025}$ & $-0.11^{+0.18}_{-0.20}$ \\
67.5--72.5 & 70.04 & 162 &
        $0.058^{+0.048}_{-0.049}$ & $0.071^{+0.050}_{-0.052}$ &
        $0.556^{+0.031}_{-0.041}$ & $0.612^{+0.034}_{-0.041}$ &
        $0.600^{+0.037}_{-0.039}$ & $0.571^{+0.036}_{-0.044}$ &
        $0.586^{+0.025}_{-0.026}$ & $0.10^{+0.19}_{-0.19}$ \\
70.0--75.0 & 72.54 & 157 &
        $-0.004^{+0.051}_{-0.050}$ & $0.044^{+0.052}_{-0.053}$ &
        $0.538^{+0.027}_{-0.037}$ & $0.619^{+0.037}_{-0.040}$ &
        $0.588^{+0.038}_{-0.038}$ & $0.569^{+0.038}_{-0.043}$ &
        $0.579^{+0.025}_{-0.026}$ & $0.06^{+0.20}_{-0.20}$ \\
72.5--77.5 & 75.04 & 153 &
        $-0.055^{+0.048}_{-0.051}$ & $0.010^{+0.047}_{-0.049}$ &
        $0.554^{+0.033}_{-0.045}$ & $0.551^{+0.028}_{-0.042}$ &
        $0.585^{+0.038}_{-0.041}$ & $0.516^{+0.031}_{-0.033}$ &
        $0.552^{+0.023}_{-0.027}$ & $0.25^{+0.19}_{-0.19}$ \\
75.0--80.0 & 77.54 & 143 &
        $-0.067^{+0.051}_{-0.050}$ & $-0.063^{+0.054}_{-0.054}$ &
        $0.545^{+0.036}_{-0.049}$ & $0.566^{+0.043}_{-0.052}$ &
        $0.605^{+0.042}_{-0.044}$ & $0.504^{+0.037}_{-0.037}$ &
        $0.557^{+0.029}_{-0.028}$ & $0.36^{+0.21}_{-0.20}$ \\
77.5--82.5 & 80.04 & 138 &
        $-0.028^{+0.051}_{-0.048}$ & $-0.017^{+0.058}_{-0.056}$ &
        $0.490^{+0.039}_{-0.049}$ & $0.595^{+0.046}_{-0.059}$ &
        $0.567^{+0.042}_{-0.043}$ & $0.519^{+0.043}_{-0.048}$ &
        $0.543^{+0.032}_{-0.034}$ & $0.18^{+0.22}_{-0.20}$ \\
80.0--85.0 & 82.54 & 145 &
        $-0.077^{+0.047}_{-0.046}$ & $0.040^{+0.052}_{-0.050}$ &
        $0.464^{+0.042}_{-0.046}$ & $0.546^{+0.042}_{-0.043}$ &
        $0.500^{+0.032}_{-0.037}$ & $0.515^{+0.044}_{-0.047}$ &
        $0.508^{+0.030}_{-0.032}$ & $-0.06^{+0.20}_{-0.21}$ \\
82.5--87.5 & 85.04 & 115 &
        $-0.035^{+0.054}_{-0.052}$ & $-0.013^{+0.056}_{-0.059}$ &
        $0.470^{+0.036}_{-0.041}$ & $0.540^{+0.050}_{-0.055}$ &
        $0.506^{+0.045}_{-0.060}$ & $0.503^{+0.041}_{-0.041}$ &
        $0.504^{+0.031}_{-0.035}$ & $0.01^{+0.23}_{-0.29}$ \\
85.0--90.0 & 87.54 & 95 &
        $0.135^{+0.064}_{-0.064}$ & $-0.094^{+0.067}_{-0.064}$ &
        $0.559^{+0.040}_{-0.058}$ & $0.524^{+0.050}_{-0.063}$ &
        $0.612^{+0.062}_{-0.075}$ & $0.481^{+0.032}_{-0.037}$ &
        $0.550^{+0.039}_{-0.042}$ & $0.47^{+0.23}_{-0.27}$ \\
87.5--92.5 & 90.03 & 85 &
        $0.063^{+0.074}_{-0.072}$ & $-0.096^{+0.062}_{-0.058}$ &
        $0.570^{+0.052}_{-0.062}$ & $0.459^{+0.041}_{-0.055}$ &
        $0.589^{+0.062}_{-0.067}$ & $0.439^{+0.036}_{-0.044}$ &
        $0.520^{+0.039}_{-0.043}$ & $0.57^{+0.25}_{-0.26}$ \\
90.0--95.0 & 92.53 & 66 &
        $-0.015^{+0.067}_{-0.066}$ & $-0.078^{+0.073}_{-0.067}$ &
        $0.460^{+0.046}_{-0.065}$ & $0.493^{+0.057}_{-0.092}$ &
        $0.482^{+0.048}_{-0.057}$ & $0.468^{+0.061}_{-0.079}$ &
        $0.475^{+0.042}_{-0.047}$ & $0.06^{+0.41}_{-0.36}$ \\
92.5--97.5 & 95.03 & 46 &
        $0.113^{+0.071}_{-0.071}$ & $-0.117^{+0.094}_{-0.091}$ &
        $0.404^{+0.046}_{-0.074}$ & $0.545^{+0.080}_{-0.105}$ &
        $0.427^{+0.053}_{-0.062}$ & $0.538^{+0.093}_{-0.106}$ &
        $0.486^{+0.057}_{-0.065}$ & $-0.46^{+0.44}_{-0.38}$ \\
95.0--100.0 & 97.53 & 23 &
        $-0.022^{+0.145}_{-0.129}$ & $0.145^{+0.117}_{-0.122}$ &
        $0.619^{+0.156}_{-0.293}$ & $0.450^{+0.086}_{-0.136}$ &
        $0.605^{+0.182}_{-0.277}$ & $0.456^{+0.087}_{-0.117}$ &
        $0.536^{+0.114}_{-0.135}$ & $0.55^{+0.45}_{-1.15}$ \\
\cutinhead{$19\leq V < 19.5$, $0.6\leq (U-V) <1.9$ ;
   $\Delta R=10^{\prime\prime}$}
\phantom{1}0.0--10.0 & 7.07 & 184 &
        $0.053^{+0.057}_{-0.054}$ & $-0.201^{+0.051}_{-0.056}$ &
        $0.728^{+0.033}_{-0.040}$ & $0.673^{+0.034}_{-0.043}$ &
        $0.705^{+0.040}_{-0.044}$ & $0.722^{+0.041}_{-0.043}$ &
        $0.714^{+0.029}_{-0.032}$ & $-0.05^{+0.16}_{-0.17}$ \\
\phantom{1}5.0--15.0 & 11.18 & 353 &
        $-0.018^{+0.039}_{-0.035}$ & $-0.113^{+0.037}_{-0.037}$ &
        $0.666^{+0.022}_{-0.025}$ & $0.647^{+0.025}_{-0.031}$ &
        $0.630^{+0.026}_{-0.028}$ & $0.690^{+0.025}_{-0.030}$ &
        $0.661^{+0.018}_{-0.020}$ & $-0.18^{+0.11}_{-0.11}$ \\
10.0--20.0 & 15.81 & 337 &
        $-0.088^{+0.042}_{-0.035}$ & $-0.051^{+0.035}_{-0.038}$ &
        $0.668^{+0.022}_{-0.027}$ & $0.619^{+0.026}_{-0.026}$ &
        $0.618^{+0.025}_{-0.026}$ & $0.674^{+0.027}_{-0.030}$ &
        $0.646^{+0.018}_{-0.019}$ & $-0.17^{+0.13}_{-0.12}$ \\
15.0--25.0 & 20.62 & 156 &
        $-0.120^{+0.062}_{-0.059}$ & $-0.027^{+0.058}_{-0.052}$ &
        $0.706^{+0.038}_{-0.047}$ & $0.638^{+0.037}_{-0.040}$ &
        $0.648^{+0.040}_{-0.040}$ & $0.703^{+0.041}_{-0.042}$ &
        $0.676^{+0.029}_{-0.027}$ & $-0.16^{+0.18}_{-0.19}$ \\
20.0--30.0 & 25.50 & 74 &
        $-0.120^{+0.082}_{-0.082}$ & $0.028^{+0.089}_{-0.081}$ &
        $0.617^{+0.061}_{-0.079}$ & $0.585^{+0.049}_{-0.072}$ &
        $0.657^{+0.072}_{-0.082}$ & $0.541^{+0.058}_{-0.064}$ &
        $0.602^{+0.042}_{-0.047}$ & $0.38^{+0.34}_{-0.35}$ \\
25.0--35.0 & 30.41 & 109 &
        $-0.095^{+0.066}_{-0.075}$ & $-0.042^{+0.076}_{-0.065}$ &
        $0.608^{+0.060}_{-0.069}$ & $0.559^{+0.050}_{-0.069}$ &
        $0.607^{+0.068}_{-0.085}$ & $0.561^{+0.054}_{-0.063}$ &
        $0.584^{+0.041}_{-0.048}$ & $0.16^{+0.34}_{-0.37}$ \\
30.0--40.0 & 35.36 & 166 &
        $0.040^{+0.060}_{-0.060}$ & $-0.051^{+0.055}_{-0.055}$ &
        $0.613^{+0.043}_{-0.052}$ & $0.563^{+0.041}_{-0.048}$ &
        $0.547^{+0.049}_{-0.052}$ & $0.625^{+0.052}_{-0.053}$ &
        $0.587^{+0.035}_{-0.035}$ & $-0.26^{+0.27}_{-0.25}$ \\
35.0--45.0 & 40.31 & 219 &
        $0.062^{+0.054}_{-0.056}$ & $0.055^{+0.047}_{-0.047}$ &
        $0.653^{+0.043}_{-0.056}$ & $0.504^{+0.038}_{-0.045}$ &
        $0.560^{+0.039}_{-0.041}$ & $0.608^{+0.045}_{-0.053}$ &
        $0.584^{+0.030}_{-0.034}$ & $-0.16^{+0.23}_{-0.21}$ \\
40.0--50.0 & 45.28 & 276 &
        $0.017^{+0.047}_{-0.046}$ & $0.067^{+0.041}_{-0.043}$ &
        $0.628^{+0.037}_{-0.044}$ & $0.528^{+0.040}_{-0.038}$ &
        $0.608^{+0.042}_{-0.043}$ & $0.552^{+0.040}_{-0.043}$ &
        $0.581^{+0.030}_{-0.031}$ & $0.19^{+0.20}_{-0.20}$ \\
45.0--55.0 & 50.25 & 303 &
        $0.034^{+0.044}_{-0.041}$ & $0.054^{+0.043}_{-0.039}$ &
        $0.575^{+0.030}_{-0.035}$ & $0.576^{+0.030}_{-0.038}$ &
        $0.621^{+0.037}_{-0.041}$ & $0.527^{+0.039}_{-0.038}$ &
        $0.576^{+0.025}_{-0.027}$ & $0.32^{+0.19}_{-0.22}$ \\
50.0--60.0 & 55.23 & 320 &
        $0.052^{+0.038}_{-0.044}$ & $0.075^{+0.044}_{-0.041}$ &
        $0.558^{+0.035}_{-0.035}$ & $0.628^{+0.039}_{-0.045}$ &
        $0.634^{+0.034}_{-0.043}$ & $0.556^{+0.035}_{-0.040}$ &
        $0.596^{+0.022}_{-0.029}$ & $0.26^{+0.17}_{-0.20}$ \\
55.0--65.0 & 60.21 & 365 &
        $0.026^{+0.037}_{-0.041}$ & $0.062^{+0.039}_{-0.038}$ &
        $0.552^{+0.035}_{-0.037}$ & $0.617^{+0.036}_{-0.041}$ &
        $0.597^{+0.036}_{-0.041}$ & $0.576^{+0.030}_{-0.032}$ &
        $0.586^{+0.022}_{-0.025}$ & $0.07^{+0.16}_{-0.18}$ \\
60.0--70.0 & 65.19 & 383 &
        $0.046^{+0.039}_{-0.038}$ & $0.011^{+0.038}_{-0.039}$ &
        $0.558^{+0.033}_{-0.040}$ & $0.550^{+0.031}_{-0.033}$ &
        $0.569^{+0.032}_{-0.030}$ & $0.538^{+0.029}_{-0.033}$ &
        $0.554^{+0.022}_{-0.024}$ & $0.11^{+0.17}_{-0.16}$ \\
65.0--75.0 & 70.18 & 358 &
        $0.080^{+0.039}_{-0.035}$ & $0.013^{+0.037}_{-0.036}$ &
        $0.543^{+0.029}_{-0.036}$ & $0.530^{+0.033}_{-0.038}$ &
        $0.576^{+0.035}_{-0.038}$ & $0.497^{+0.032}_{-0.033}$ &
        $0.538^{+0.026}_{-0.024}$ & $0.29^{+0.16}_{-0.17}$ \\
70.0--80.0 & 75.17 & 324 &
        $0.010^{+0.036}_{-0.037}$ & $-0.011^{+0.041}_{-0.037}$ &
        $0.522^{+0.031}_{-0.037}$ & $0.541^{+0.031}_{-0.032}$ &
        $0.541^{+0.035}_{-0.039}$ & $0.520^{+0.030}_{-0.033}$ &
        $0.531^{+0.024}_{-0.026}$ & $0.08^{+0.16}_{-0.18}$ \\
75.0--85.0 & 80.16 & 315 &
        $-0.002^{+0.038}_{-0.037}$ & $-0.026^{+0.041}_{-0.039}$ &
        $0.513^{+0.033}_{-0.035}$ & $0.570^{+0.035}_{-0.040}$ &
        $0.541^{+0.031}_{-0.038}$ & $0.541^{+0.036}_{-0.036}$ &
        $0.541^{+0.027}_{-0.029}$ & $0.00^{+0.14}_{-0.16}$ \\
80.0--90.0 & 85.15 & 265 &
        $0.033^{+0.041}_{-0.040}$ & $0.030^{+0.039}_{-0.040}$ &
        $0.524^{+0.040}_{-0.049}$ & $0.513^{+0.040}_{-0.049}$ &
        $0.484^{+0.038}_{-0.039}$ & $0.549^{+0.044}_{-0.052}$ &
        $0.518^{+0.032}_{-0.034}$ & $-0.25^{+0.23}_{-0.20}$ \\
85.0--95.0 & 90.14 & 175 &
        $-0.091^{+0.059}_{-0.056}$ & $-0.042^{+0.046}_{-0.043}$ &
        $0.600^{+0.054}_{-0.061}$ & $0.404^{+0.039}_{-0.048}$ &
        $0.458^{+0.043}_{-0.046}$ & $0.564^{+0.058}_{-0.064}$ &
        $0.514^{+0.038}_{-0.039}$ & $-0.41^{+0.29}_{-0.27}$ \\
90.0--100.0 & 95.13 & 91 &
        $-0.193^{+0.077}_{-0.088}$ & $-0.066^{+0.071}_{-0.072}$ &
        $0.676^{+0.060}_{-0.081}$ & $0.507^{+0.053}_{-0.067}$ &
        $0.599^{+0.058}_{-0.053}$ & $0.619^{+0.060}_{-0.067}$ &
        $0.609^{+0.045}_{-0.042}$ & $-0.06^{+0.28}_{-0.25}$ \\
95.0--105.0 & 100.12 & 24 &
        $-0.059^{+0.177}_{-0.162}$ & $0.193^{+0.157}_{-0.137}$ &
        $0.710^{+0.134}_{-0.242}$ & $0.586^{+0.088}_{-0.140}$ &
        $0.614^{+0.110}_{-0.139}$ & $0.677^{+0.111}_{-0.142}$ &
        $0.646^{+0.088}_{-0.107}$ & $-0.19^{+0.49}_{-0.48}$ \\
\cutinhead{$19.5\leq V < 20$, $0.6\leq (U-V) <1.9$ ;
   $\Delta R=10^{\prime\prime}$}
\phantom{1}0.0--10.0 & 7.07 & 149 &
        $0.043^{+0.060}_{-0.063}$ & $0.022^{+0.063}_{-0.067}$ &
        $0.671^{+0.048}_{-0.055}$ & $0.673^{+0.057}_{-0.063}$ &
        $0.704^{+0.059}_{-0.064}$ & $0.635^{+0.053}_{-0.055}$ &
        $0.670^{+0.040}_{-0.041}$ & $0.21^{+0.24}_{-0.24}$ \\
\phantom{1}5.0--15.0 & 11.18 & 237 &
        $0.036^{+0.051}_{-0.055}$ & $0.004^{+0.048}_{-0.053}$ &
        $0.719^{+0.040}_{-0.054}$ & $0.676^{+0.040}_{-0.051}$ &
        $0.749^{+0.045}_{-0.053}$ & $0.640^{+0.047}_{-0.051}$ &
        $0.697^{+0.032}_{-0.037}$ & $0.31^{+0.20}_{-0.22}$ \\
10.0--20.0 & 15.81 & 207 &
        $-0.050^{+0.059}_{-0.055}$ & $0.001^{+0.052}_{-0.057}$ &
        $0.726^{+0.044}_{-0.058}$ & $0.682^{+0.048}_{-0.055}$ &
        $0.725^{+0.057}_{-0.060}$ & $0.680^{+0.055}_{-0.063}$ &
        $0.703^{+0.038}_{-0.044}$ & $0.13^{+0.21}_{-0.23}$ \\
15.0--25.0 & 20.62 & 89 &
        $-0.215^{+0.075}_{-0.076}$ & $0.055^{+0.082}_{-0.070}$ &
        $0.582^{+0.050}_{-0.068}$ & $0.617^{+0.080}_{-0.105}$ &
        $0.599^{+0.085}_{-0.104}$ & $0.630^{+0.079}_{-0.083}$ &
        $0.615^{+0.055}_{-0.064}$ & $-0.10^{+0.41}_{-0.46}$ \\
25.0--35.0 & 30.41 & 69 &
        $-0.208^{+0.094}_{-0.095}$ & $-0.040^{+0.097}_{-0.097}$ &
        $0.399^{+0.095}_{-0.157}$ & $0.365^{+0.099}_{-0.180}$ &
        $0.343^{+0.110}_{-0.183}$ & $0.449^{+0.104}_{-0.135}$ &
        $0.399^{+0.073}_{-0.092}$ & $-0.53^{+0.88}_{-0.96}$ \\
30.0--40.0 & 35.36 & 105 &
        $-0.078^{+0.087}_{-0.079}$ & $-0.091^{+0.076}_{-0.083}$ &
        $0.548^{+0.078}_{-0.116}$ & $0.442^{+0.084}_{-0.122}$ &
        $0.519^{+0.090}_{-0.106}$ & $0.475^{+0.085}_{-0.101}$ &
        $0.498^{+0.058}_{-0.066}$ & $0.18^{+0.54}_{-0.58}$ \\
35.0--45.0 & 40.31 & 150 &
        $0.040^{+0.066}_{-0.071}$ & $0.053^{+0.069}_{-0.064}$ &
        $0.463^{+0.079}_{-0.104}$ & $0.479^{+0.088}_{-0.107}$ &
        $0.517^{+0.079}_{-0.096}$ & $0.414^{+0.086}_{-0.109}$ &
        $0.468^{+0.068}_{-0.075}$ & $0.43^{+0.53}_{-0.52}$ \\
40.0--50.0 & 45.28 & 195 &
        $0.063^{+0.056}_{-0.056}$ & $0.136^{+0.067}_{-0.062}$ &
        $0.386^{+0.079}_{-0.108}$ & $0.563^{+0.068}_{-0.090}$ &
        $0.520^{+0.066}_{-0.073}$ & $0.459^{+0.080}_{-0.090}$ &
        $0.490^{+0.052}_{-0.063}$ & $0.25^{+0.45}_{-0.41}$ \\
45.0--55.0 & 50.25 & 221 &
        $0.035^{+0.052}_{-0.060}$ & $0.092^{+0.058}_{-0.060}$ &
        $0.419^{+0.063}_{-0.075}$ & $0.535^{+0.053}_{-0.067}$ &
        $0.460^{+0.061}_{-0.073}$ & $0.503^{+0.057}_{-0.076}$ &
        $0.482^{+0.040}_{-0.050}$ & $-0.18^{+0.40}_{-0.41}$ \\
50.0--60.0 & 55.23 & 243 &
        $0.035^{+0.049}_{-0.046}$ & $0.033^{+0.058}_{-0.055}$ &
        $0.377^{+0.063}_{-0.072}$ & $0.552^{+0.066}_{-0.077}$ &
        $0.372^{+0.067}_{-0.072}$ & $0.552^{+0.058}_{-0.068}$ &
        $0.471^{+0.046}_{-0.049}$ & $-0.75^{+0.37}_{-0.34}$ \\
55.0--65.0 & 60.21 & 266 &
        $0.048^{+0.046}_{-0.047}$ & $0.089^{+0.053}_{-0.058}$ &
        $0.366^{+0.053}_{-0.075}$ & $0.595^{+0.068}_{-0.079}$ &
        $0.460^{+0.068}_{-0.069}$ & $0.531^{+0.062}_{-0.072}$ &
        $0.496^{+0.050}_{-0.052}$ & $-0.29^{+0.35}_{-0.32}$ \\
60.0--70.0 & 65.19 & 262 &
        $0.024^{+0.047}_{-0.049}$ & $0.094^{+0.049}_{-0.056}$ &
        $0.418^{+0.056}_{-0.071}$ & $0.506^{+0.070}_{-0.086}$ &
        $0.507^{+0.069}_{-0.076}$ & $0.422^{+0.061}_{-0.072}$ &
        $0.466^{+0.046}_{-0.054}$ & $0.36^{+0.38}_{-0.42}$ \\
65.0--75.0 & 70.18 & 236 &
        $0.094^{+0.056}_{-0.044}$ & $0.112^{+0.050}_{-0.052}$ &
        $0.322^{+0.067}_{-0.086}$ & $0.458^{+0.069}_{-0.086}$ &
        $0.438^{+0.077}_{-0.086}$ & $0.372^{+0.066}_{-0.077}$ &
        $0.406^{+0.051}_{-0.053}$ & $0.32^{+0.54}_{-0.58}$ \\
70.0--80.0 & 75.17 & 221 &
        $0.073^{+0.047}_{-0.045}$ & $0.036^{+0.056}_{-0.051}$ &
        $0.237^{+0.076}_{-0.147}$ & $0.428^{+0.054}_{-0.079}$ &
        $0.307^{+0.084}_{-0.102}$ & $0.382^{+0.065}_{-0.077}$ &
        $0.347^{+0.052}_{-0.053}$ & $-0.43^{+0.68}_{-0.74}$ \\
75.0--85.0 & 80.16 & 234 &
        $0.045^{+0.054}_{-0.052}$ & $-0.038^{+0.050}_{-0.047}$ &
        $0.360^{+0.065}_{-0.096}$ & $0.310^{+0.085}_{-0.115}$ &
        $0.000^{+0.177}_{-0.000}$ & $0.488^{+0.069}_{-0.079}$ &
        $0.345^{+0.057}_{-0.048}$ & $-2.00^{+0.45}_{-0.00}$ \\
80.0--90.0 & 85.15 & 211 &
        $0.034^{+0.051}_{-0.048}$ & $0.009^{+0.047}_{-0.053}$ &
        $0.341^{+0.074}_{-0.101}$ & $0.323^{+0.082}_{-0.123}$ &
        $0.081^{+0.169}_{-0.081}$ & $0.458^{+0.073}_{-0.087}$ &
        $0.329^{+0.074}_{-0.053}$ & $-1.88^{+0.81}_{-0.12}$ \\
85.0--95.0 & 90.14 & 134 &
        $-0.039^{+0.064}_{-0.060}$ & $-0.044^{+0.063}_{-0.065}$ &
        $0.160^{+0.132}_{-0.160}$ & $0.246^{+0.104}_{-0.246}$ &
        $0.246^{+0.132}_{-0.246}$ & $0.146^{+0.134}_{-0.146}$ &
        $0.202^{+0.109}_{-0.102}$ & $0.96^{+0.04}_{-2.67}$ \\
\enddata
\end{deluxetable}

\end{document}